\documentclass[12pt]{iopart}

\usepackage{epsfig}
\begin{document}

\title[Finite-size effect on two-particle
production]{Finite-size effect on two-particle
production in continuous and discrete spectrum}

\author{R Lednick\'y}

\address{Joint Institute for Nuclear Research,
Dubna, Moscow Region, 141980, Russia\\
Institute of Physics ASCR,
Na Slovance 2, 18221 Prague 8, Czech Republic}
\ead{lednicky@fzu.cz}

\begin{abstract}
The formalism allowing one to account for
the effect of a finite space-time extent of particle production
region is given. Its applications to
the lifetime measurement of hadronic atoms produced by
a high-energy beam in a thin target,
as well as to
the femtoscopy techniques
widely used to measure space-time characteristics
of the production processes, are discussed.
Particularly, it is found that the neglect of
the finite-size effect on the pionium
lifetime measurement in the experiment DIRAC at CERN
could lead to the lifetime overestimation comparable with the
$10\%$ statistical error.
The theoretical systematic errors arising in the calculation
of the finite-size effect due to the neglect
of non-equal emission times in the pair center-of-mass system,
the space-time coherence and the
residual charge are shown to be negligible.
\end{abstract}

\maketitle

\section{Introduction}
The determination,
on a percent level accuracy, of the breakup probability of the
$\pi^+\pi^-$ atoms produced by a high-energy beam in a thin target
is of principle importance for a precise lifetime measurement of
these atoms in the experiment DIRAC at CERN [1-4].
This experiment aims to measure the lifetime $\tau_{10}$
of the $\pi^+\pi^-$
atoms in the ground state with 10$\%$ precision.
As this lifetime of order $10^{-15}$ s
is determined by the probability of the
annihilation $\pi^+\pi^- \rightarrow \pi^0\pi^0$:
$1/\tau_{10} \sim |a_0^0-a_0^2|^2$, the DIRAC measurement
enables to determine the absolute
value of the difference $a_0^0-a_0^2$ of the
s-wave isoscalar and isotensor $\pi\pi$ scattering lengths
to $5\%$. This represents a factor of 4 improvement of
the accuracy achieved in previous studies \cite{nag79},
and is comparable with the precision of
the most recent experiments E865 at BNL \cite{pis01}
and NA48/2 at CERN \cite{na48}. The former
is based on a study of $K_{e4}$ decays and yields the
statistical error of $6\%$ in $a_0^0$; this measurement
essentially exploits other experimental data
together with dispersion relations (Roy equations),
the systematic and theoretical errors being estimated
on the level of several percent.
The latter studied the cusp effect
at the $\pi^+\pi^-$ threshold
in the distribution of the $2\pi^0$ effective mass
in $K^{\pm}\to \pi^{\pm}\pi^0\pi^0$ decays and
yields $|a_0^0-a_0^2|$ with a few percent statistical
precision and $\sim 5\%$ theoretical error.
Both these measurements are in agreement with the
preliminary DIRAC result based on $\sim 40\%$ of the collected
statistics \cite{ade05} as well as with the prediction
of the standard chiral perturbation theory \cite{cgl01}.

It should be stressed that the theoretical prediction for the
difference $a_0^0-a_0^2$ depends on the structure of the QCD vacuum.
Thus, on the standard assumption of a strong quark condensate one has
$a_0^0-a_0^2=0.374\pm 0.006$ fm \cite{cgl01}. With the decreasing
condensate this difference increases and can be up to 25 $\%$
larger \cite{kne95}.
The precise measurements of the $\pi\pi$ scattering lengths
thus submit the understanding of chiral symmetry
breaking of QCD to a crucial test.

The method of the lifetime measurement is based on the production
of $\pi^+\pi^-$ atoms in a thin target
and subsequent detection of
highly correlated $\pi^+\pi^-$ pairs leaving the target as a result
of the breakup of a part of the $\pi^+\pi^-$ atoms
which did not decay within the target \cite{nem85}.
Clearly, the breakup probability is a unique function of the
target geometry and material, the Lorentz factor and the
ground-state lifetime of the $\pi^+\pi^-$ atom. The analysis shows that,
to achieve the accuracy of 10$\%$ in the lifetime, the
breakup probability, in more or less optimal conditions,
should be measured to 4$\%$ \cite{ade95}.

There are two methods \cite{ade04} --
extrapolation and subtraction ones --
which can be used to measure the breakup probability $P_{\rm br}$
(or a combination of the breakup probabilities in
different targets)
defined as the ratio of the number $N_A^{\rm br}$ of breakup
atoms to the number $N_A$ of the atoms produced in the target:
\begin{equation}
P_{\rm br}=N_A^{\rm br}/N_A .
\label{pbr}
\end{equation}
The extrapolation method requires the calculation
of the number of produced
$\pi^+\pi^-$ atoms $N_A$ based on the theory of the final-state
interaction (FSI) in discrete and
continuous spectrum \cite{nem85,lyu88,ll82}.
This calculation, as well as the determination of $N_A^{\rm br}$,
is not required in the
subtraction method which exploits the data taken
on at least three different targets made out of the same material
but consisting of a different number of layers of the same total
thickness. However, this method needs a factor of 7 larger
statistics \cite{ade95} and cannot yield the required precision
within the approved time-scale of the experiment DIRAC.

The FSI effect on $\pi^+\pi^-$ production
is sensitive to the space-time extent of the pion
production region mainly through
the distance $r^{*}$ between the $\pi^+$ and
$\pi^-$ production points in their center-of-mass (c.m.)
system.
In \cite{nem85}, only the Coulomb FSI was considered and
the $r^*$-dependence was treated in an approximate way,
dividing the pion emitters into short-lived (SL)
and long-lived (LL) ones.
It was assumed that $r^*=0$ for pion pairs arising solely from the
SL emitters and characterized by the distances $r^*$
much smaller than the Bohr radius $|a|$ of the $\pi^+\pi^-$ system
($a=-387.5$ fm), otherwise $r^*=\infty$.

The finite-size correction to such calculated number of
non-atomic $\pi^+\pi^-$ pairs
in the region of very small relative momenta in the pair c.m. system,
$Q\ll 1/r^*$,
is determined by the three dimensionless
combinations $r^*/a$, $f_0/r^*$ and $f_0/a$
of $r^*$, $a$  and the s-wave $\pi^+\pi^-$ scattering length:
$f_0=\frac 13 (2a_0^0+a_0^2)\approx 0.2$ fm.
Typically $\langle r^*\rangle^{\mbox{\tiny SL}} \sim$ 10 fm
so that the correction is dominated by the strong interaction
effect and can amount up to $\sim 10\%$.

Fortunately,
due to a small binding energy $\epsilon_b\sim (m_\pi a^2)^{-1}$,
the finite-size correction to the production probability
in discrete spectrum at
$r^*\ll |a|$
is nearly the same as that in continuous spectrum at zero energy.
Since the calculated number $N_A$ of produced atoms is approximately
determined by the measured number of non-atomic $\pi^+\pi^-$ pairs
and the ratio of weighted means of the finite-size correction factors
corresponding to the production in discrete
and continuous spectrum, the finite-size correction
would cancel out in $N_A$,
up to $\Or((r^*/a)^2)$ and $\Or(f_0/a)$,
provided we could measure the number of non-atomic pairs
in the region of very small
$Q\ll 1/r^*$ \cite{led98,av99}.

At small values of $Q$ and $r^*$,
the relative correction to the number of non-atomic pairs
is positive (due to the effect of the strong FSI $\sim 2f_0/r^*$)
and changes sign at $r^* \sim$ 10 fm
(due to negative finite-size effect of the Coulomb FSI
$\sim 2r^*/a$).
It appears that
for $r^* < 20$ fm the correction shows a quasi-linear
behaviour in $Q$ up to $\sim 50$ MeV/$c$, with almost a universal
negative slope. For larger distances $r^*$, the slope becomes
positive and has a non-trivial $Q$-dependence.
If the pions were produced at small distances $r^*$
of several fm, one could safely
neglect the non-universal correction $\Or((r^*/a)^2)$
and use the quasi-linear $Q$-dependence of the
correction factor to interpolate to $Q=0$.
However, there is a non-negligible tail of the distances
$r^* > 10$ fm due to particle rescatterings and resonances
(particularly, $r^{*}\sim 30$ fm in the case when one of the
two pions comes from the $\omega$-meson decay).
In the experiment DIRAC, the finite-size correction
can lead to about percent underestimation of $N_A$
and -- to several percent overestimation
of $N_A^{\rm br}$. As a result, the corresponding
overestimation of the extracted lifetime
can be comparable with the
$10\%$ statistical error and should be taken into account.

We discuss how to diminish the systematic error
due to the finite-size effect on the lifetime measurement
of hadronic atoms, using
the correlation data on identical charged pions (containing the
information about the distances $r^*$ between the
pion production points in the same experiment)
together with the complete phase-space
simulations within transport models.

The formalism accounting for the finite space-time separation
of particle emitters is also in the basis of the
correlation measurements of the space-time characteristics
of particle production -
so called {\it particle interferometry}
or {\it correlation femtoscopy}
(see reviews
\cite{pod89,lor89,boa90,wh99,cso02,led04,lis05}).
In fact, the femtoscopic correlations due to
the Coulomb FSI between the emitted electron or
positron and the residual nucleus
in beta-decay are known for more than 70 years;
the sensitivity of the differential decay rate to the nucleus charge
and radius is taken into account in terms of the Fermi function
which can be considered as
an analogue of the correlation function in multiparticle production
(see \cite{led07} for a discussion of the similarity
and difference of femtoscopic correlations in beta-decay and multiparticle production).
The femtoscopic correlations due to the quantum statistics (QS)
of produced identical particles were observed almost 50 years
ago as an enhanced production of pairs of identical
pions with small opening angles (GGLP effect). The basics of
the modern correlation femtoscopy were settled by Kopylov and
Podgoretsky in early seventieth of the last century;
they also pointed out a striking analogy between
the femtoscopic momentum correlations of identical particles and
the spectroscopic space-time correlations of photons (HBT effect),
the latter allowing one to measure the spectral width of the
light source as well as
the angular radii of distant (stellar) objects
(see \cite{led07} and references therein).
Besides the space-time characteristics of particle production,
the femtoscopic correlations yield also a valuable information on
low-energy strong interaction between specific particles
which can hardly be achieved
by other means (see \cite{led04} and subsection 4.3.5).

The paper is organized now as follows.
In sections 2 and 4 we give the basics of the theory of
two-particle correlations due to the FSI and QS effects.
Particularly, the formalism and assumptions behind the
correlation femtoscopy are discussed in section 4.
Sections 5 and 6 deal with the one- and two-channel
wave functions in the continuous and discrete spectrum.
In sections 3 and 7, we apply the developed formalism
to estimate the finite-size effect on the pionium
lifetime measurement in the experiment DIRAC at CERN.
The results are summarized in section 8.
In Appendices A and B we consider the effect of non-equal
times and derive the analytical expression for the
normalization effect of the short-range interaction on
the wave function of a hadronic atom, modifying the
usual $n^{-3/2}$ dependence of the pure Coulomb wave
function on the main quantum number $n$.
The reader interested mainly in practical application
of the formalism to the lifetime measurement of hadronic atoms
can start reading from sections 3 and 7 and
consult the rest of the paper to clarify the
eventual questions.

\section{Formalism}

The production of two particles at small relative momenta is strongly
influenced by their mutual FSI and,
for identical particles, also by QS. One can
separate the FSI effect from the production amplitude provided
a sufficiently long two-particle interaction time in the final state
as compared with the
characteristic time of the production process.
This condition requires the magnitude of the relative three-momentum
${\bf q}^* = {\bf p}_1^*-{\bf p}_2^*\equiv 2{\bf k}^*\equiv {\bf Q}$
in the two-particle c.m. system much smaller
than several hundreds MeV/$c$ -- the momentum transfer typical for
particle production. For a two-particle bound state the
momentum $k^*$ in this condition has to be substituted by
$(2\mu\epsilon_b)^{1/2}$, where $\epsilon_b$ is the binding energy
and $\mu=m_1m_2/(m_1+m_2)$ is the reduced mass.

Consider first the differential cross section for the production
of a pair of non-identical
particles 1 and 2 with the four-momenta
$p_i=\{\omega_i,{\bf p}_i\}$ and the
Lorentz factors $\gamma_i=\omega_i/m_i$. It can be
expressed through the invariant
production amplitudes $T(p_1,p_2;\alpha)$ in the form
\begin{equation}
\label{1} (2\pi)^6\gamma_1\gamma_2
\frac{{\rm d}^6\sigma}{{\rm d}^3{\bf p}_1{\rm d}^3{\bf p}_2} =
\sum\limits_{\alpha}|T(p_1,p_2;\alpha )|^2,
\end{equation}
where the sum is done over $\alpha=\{S,M,\alpha'\}$, i.e.
the total spin
$S$ of the pair and
its projection
$M$
(which is equivalent to the sum over
helicities of the two particles)
and the quantum numbers $\alpha'$ of other produced particles,
including integration over their momenta with the
energy--momentum conservation taken into account.

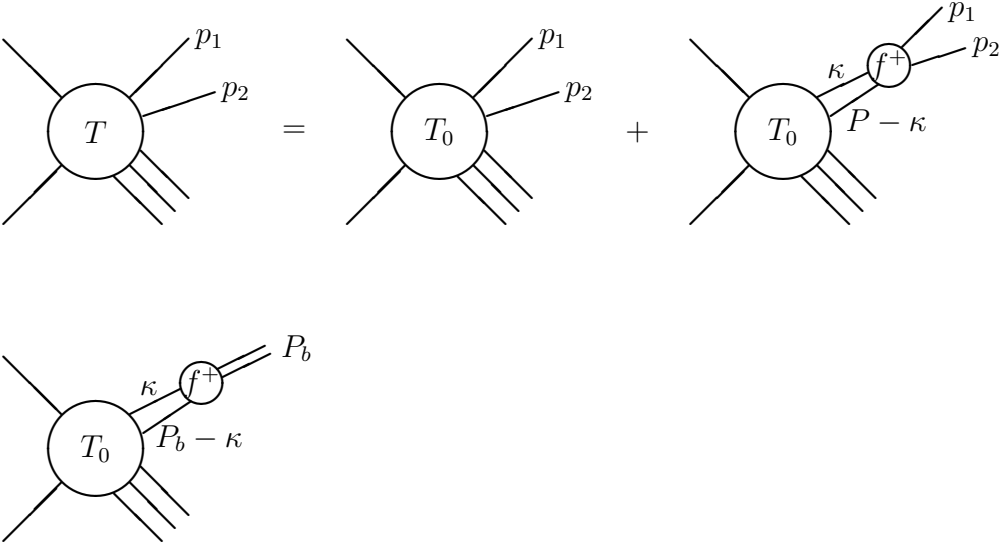
\begin{figure}
\begin{picture}(400,240)(0,0)
\thicklines
\newsavebox{\dia}
\savebox{\dia}{
\put(40,180){\circle{35}}
\put(5,145){\line(1,1){22}}
\put(5,215){\line(1,-1){22}}
\put(70,150){\line(-1,1){17}}
\put(65,145){\line(-1,1){18}}
\put(75,155){\line(-1,1){18}}
}
\put(40,180){\makebox(0,0){$T$}}
\put(115,180){\makebox(0,0){$=$}}
\multiput(0,0)(130,0){3}{\usebox{\dia}}
\multiput(170,180)(130,0){2}{\makebox(0,0){$T_0$}}
\put(245,180){\makebox(0,0){$+$}}
\multiput(83,215)(130,0){2}{\makebox(0,0){$p_1$}}
\multiput(93,195)(130,0){2}{\makebox(0,0){$p_2$}}
\multiput(75,215)(130,0){2}{\line(-1,-1){22}}
\multiput(85,195)(130,0){2}{\line(-3,-1){27}}
\put(340,205){\circle{17}}
\put(340,205){\makebox(0,0){$f^+$}}
\put(345,212){\line(1,1){15}}
\put(349,205){\line(3,1){20}}
\put(368,225){\makebox(0,0){$p_1$}}
\put(377,212){\makebox(0,0){$p_2$}}
\put(339,184){\makebox(0,0){$P-\kappa$}}
\put(320,203){\makebox(0,0){$\kappa$}}
\put(318,186){\line(3,2){18}}
\put(313,193){\line(2,1){19}}

\put(40,60){\circle{35}}
\put(5,25){\line(1,1){22}}
\put(5,95){\line(1,-1){22}}
\put(70,30){\line(-1,1){17}}
\put(65,25){\line(-1,1){18}}
\put(75,35){\line(-1,1){18}}
\put(40,60){\makebox(0,0){$T_0$}}
\put(80,85){\circle{17}}
\put(80,85){\makebox(0,0){$f^+$}}
\put(86,90){\line(2,1){18}}
\put(88,87){\line(2,1){18}}
\put(116,98){\makebox(0,0){$P_b$}}
\put(79,64){\makebox(0,0){$P_b-\kappa$}}
\put(60,83){\makebox(0,0){$\kappa$}}
\put(58,66){\line(3,2){18}}
\put(53,73){\line(2,1){19}}
\end{picture}
\caption{The diagrams describing production of
particles $1$ and $2$ in continuous and discrete spectrum.}
\label{fig1}
\end{figure}

We are interested in the pairs $(1,2)$
of the particles produced with a small relative velocity
in a process with an ordinary phase space
density of final-state particles
so that a main contribution to the double inclusive cross section
comes from
the configurations $(1,2,..,i,..)$ with large relative velocities
of the particles $1$ and $2$ with respect to
other produced particles $(i = 3,4,..n)$. Due to a sharp fall
of FSI with the increasing relative velocity, we can then neglect
the effect of FSI in all pairs $(1,i)$ and $(2,i)$ except $(1,2)$
and,
in accordance with the upper diagram in figure~\ref{fig1},
write the production amplitude as
(see however \cite{3body} and subsection 4.3.3 for the account of the
residual
Coulomb field)
\begin{equation}
\label{2} T(p_1,p_2;\alpha) = T_0(p_1,p_2;\alpha) + \Delta
T(p_1,p_2;\alpha).
\end{equation}
Here $T_0(p_1,p_2;\alpha)$ is the production amplitude in the case
of no FSI, and $\Delta T(p_1,p_2;\alpha)$ represents the
contribution of the FSI between particles $1$ and $2$,
described by the formula
\begin{equation}
\label{3} \Delta T(p_1,p_2;\alpha) = \frac{{\rm i}\sqrt{P^2}}{2\pi^3}
\int {\rm d}^4\kappa
\frac{T_0(\kappa ,P-\kappa ;\alpha)
f^{S*}(p_1,p_2;\kappa ,P-\kappa )} {(\kappa^2-m_1{}^2-{\rm i}0)[(P-\kappa
)^2-m_2{}^2-{\rm i}0]},
\end{equation}
where $P \equiv 2p = p_1 +p_2$,
$T_0(\kappa ,P-\kappa ;\alpha)$ is
the production amplitude analytically continued off mass-shell,
$f^{S}(p_1,p_2;\kappa ,P-\kappa )$ is the scattering
amplitude of particles $1$ and $2$ also analytically continued to
the unphysical region.
{
In the case of small ${k}^*$, we are interested in, the
central forces dominate so the scattering amplitude $f^S$ is
diagonal with respect to the total spin $S$ and doesn't depend
on its projections. Since most of the systems of our
interest ($\pi^+\pi^-$, $K^{\pm}\pi^{\mp}$, $\pi^-p$, $K^+K^-$,
$K^-p$, except for $\bar{p}p$) is described by a single
value of $S$, we will often skip it to simplify the notation.
}

Let us express the amplitude $T_0$ in a form of the Fourier integral
\begin{equation}
\label{4}
\eqalign{
T_0(p_1,p_2;\alpha) =
\int {\rm d}^4x_1\,{\rm d}^4x_2
{\rm e}^{-{\rm i}p_1x_1-{\rm i}p_2x_2} {\mathcal T}(x_1,x_2;\alpha)
\cr
=\int {\rm d}^4x {\rm e}^{-{\rm i}\tilde qx/2}\tau_P(x;\alpha),
}
\end{equation}
where the last expression arises after the
integration over the pair c.m. four-coordinate
$X= [(p_1P)x_1 +(p_2P)x_2]/P^2$
(${\rm d}^4x_1{\rm d}^4x_2={\rm d}^4X{\rm d}^4x$)
based on the separation
of the phase factors due to the free c.m. and relative motions:
${\rm e}^{-{\rm i}p_1x_1-{\rm i}p_2x_2}=
{\rm e}^{-{\rm i}PX}{\rm e}^{-{\rm i}\tilde qx/2}$.
Here the relative four-coordinate
$x\equiv \{t,{\bf r}\}=x_1-x_2$ and the generalized relative
four-momentum
$\widetilde{q}=q-P(qP)/P^2$, $q = p_1-p_2$; note that
$qP = m_1{}^2-m_2{}^2$.
Apparently, the function ${\mathcal T}(x_1,x_2;\alpha)$ represents
the production amplitude of particles $1$ and $2$ at the space-time
points $x_1$ and $x_2$, respectively. It should be stressed that
the representation (\ref{4}) concerns virtual particles as well.
Inserting now in (\ref{3}) the representation (\ref{4}) with
the substitutions $p_1 \rightarrow \kappa$,
$p_2 \rightarrow P-\kappa $, we get
\begin{equation}
\label{5}
\eqalign{
T(p_1,p_2;\alpha) =
\int {\rm d}^4x_1\,{\rm d}^4x_2 \Psi ^{S(-)}_{p_1,p_2}(x_1,x_2)
{\mathcal T}(x_1,x_2;\alpha)
\cr
=\int {\rm d}^4x \psi ^{S(-)}_{\widetilde{q}}(x)
\tau_P(x;\alpha),
}
\end{equation}
where
\begin{equation}
\label{5'}
\Psi ^{S(-)}_{p_1,p_2}(x_1,x_2)=
\left[\Psi ^{S(+)}_{p_1,p_2}(x_1,x_2)\right]^*=
\left[{\rm e}^{{\rm i}PX}\psi ^{S(+)}_{\widetilde{q}}(x)\right]^*
\end{equation}
coincides with the Bethe--Salpeter amplitude in continuous
spectrum \cite{Sch55}.
The second equality in (\ref{5}), similar to the one in (\ref{4}),
merely arises after the
integration over the pair c.m. coordinate $X$ as a consequence
of the factorization of the free c.m. motion in the phase
factor ${\rm e}^{-{\rm i}PX}$. Thus, on the assumption of the
quasi-free propagation of the low-mass two-particle system,
the momentum dependence of the two-particle amplitude is
determined by the convolution of the reduced production amplitude
\begin{equation}
\label{tau}
\tau_P(x;\alpha) =
\int {\rm d}^4X{\rm e}^{-{\rm i}PX} {\mathcal T}(x_1,x_2;\alpha)
\end{equation}
and the reduced Bethe--Salpeter amplitude
$\psi ^{S(-)}_{\widetilde{q}}(x)$, the latter depending only
on the relative four-coordinate
$x$ and the generalized relative four-momentum
$\widetilde{q}$.
Using (\ref{2})-(\ref{5}),
we can write
\begin{equation}
\label{6a}
\psi ^{S(+)}_{\widetilde{q}}(x)={\rm e}^{{\rm i}\tilde qx/2}+
\Delta\psi ^{S(+)}_{\widetilde{q}}(x),
\end{equation}
where the correction to the
plane wave is
\begin{eqnarray}
\label{7}
\Delta\psi ^{S(+)}_{\widetilde{q}}(x) =
\frac{\sqrt{P^2}}{2\pi^3{\rm i}}
{\rm e}^{-{\rm i}Px(1+Pq/P^2)/2}
\nonumber\\
\cdot\int {\rm d}^4\kappa
\frac{{\rm e}^{{\rm i}\kappa x}f^S(p_1,p_2;\kappa
,P-\kappa )} {(\kappa^2-m_1{}^2+{\rm i}0)[(P-\kappa )^2-m_2{}^2+{\rm i}0]}.
\end{eqnarray}
In the two-particle c.m. system, where ${\bf P} = 0$,
$\tilde q = \{0,2{\bf k}^*\}$, $x = \{t^*,{\bf r}^*\}$, the
amplitude $\psi ^{S(+)}_{\widetilde{q}}(x)$
at $t^* \equiv t_1^*-t_2^* = 0$ coincides with
a stationary solution $\psi ^{S}_{-{\bf k}^*}({\bf r}^*)$ of the
scattering problem having at large
${r}^*=|{\bf r}^*|$ the asymptotic form of a
superposition of the plane and outgoing spherical
waves \cite{Lan77}.

We see that one and the same production amplitude
${\mathcal T}(x_1,x_2;\alpha)$ or
$\tau_P(x;\alpha)$, corresponding to the
space-time representation,
enters into relations (\ref{4}) and (\ref{5}).
The effect of FSI manifests itself in the
fact that the role of the functional basis, which the
asymptotic two-particle state is
projected on, is transferred from the plane waves to
Bethe--Salpeter amplitudes
$\Psi ^{S(-)}_{p_1,p_2}(x_1,x_2)$ or
$\psi ^{S(-)}_{\widetilde{q}}(x)$.

Equation (\ref{5}) is valid also for identical particles $1$ and $2$
provided the substitution of the non-symmetrized Bethe--Salpeter
amplitudes
$\Psi ^{S(-)}_{p_1,p_2}(x_1,x_2)$ by their properly symmetrized
combinations satisfying the requirements of QS:
\begin{equation}
\label{12} \Psi ^{S(-)}_{p_1,p_2}(x_1,x_2) \rightarrow
\frac 1{\sqrt{2}} \left[\Psi ^{S(-)}_{p_1,p_2}(x_1,x_2)
+(-1)^S\Psi ^{S(-)}_{p_2,p_1}(x_1,x_2)\right].
\end{equation}
In this case $m_1 = m_2$, $\widetilde q = q$ and $X = (x_1+x_2)/2$.
Similar to the case of non-identical particles,
the assumption of the ordinary phase space density
of the final-state particles allows one to account for the
FSI and QS effects only in a given pair of identical particles
produced with a small relative velocity.

After substituting the representation (5) into (\ref{1}),
the double inclusive cross section takes on the form
\begin{eqnarray}
\label{16}
(2\pi)^6\gamma_1\gamma_2\frac{{\rm d}^6\sigma}
{{\rm d}^3{\bf p}_1{\rm d}^3{\bf p}_2}
\nonumber\\
=
\sum\limits_{S}
\int {\rm d}^4x_1\,{\rm d}^4x_2\,{\rm d}^4x_1'\,{\rm d}^4x_2'
\mbox{{\large $\rho$}}_{PS}(x_1,x_2;x_1',x_2')
\Psi ^{S(-)}_{p_1,p_2}(x_1,x_2)
\Psi ^{S(-)*}_{p_1,p_2}(x_1',x_2')
\nonumber\\
=
\sum\limits_{S}
\int {\rm d}^4x\,{\rm d}^4x'
\rho_{PS}(x;x')
\psi ^{S(-)}_{\widetilde{q}}(x)
\psi ^{S(-)*}_{\widetilde{q}}(x'),
\end{eqnarray}
where the functions
\begin{equation}
\label{17}
\eqalign{
\mbox{{\large $\rho$}}_{PS}(x_1,x_2;x_1',x_2')=\sum\limits_{M,\alpha '}
{\mathcal T}(x_1,x_2;S,M,\alpha ')
{\mathcal T}^{*}(x_1',x_2';S,M,\alpha ')
\cr
\rho_{PS}(x;x')=\sum\limits_{M,\alpha '}
\tau_P(x;S,M,\alpha ')
\tau_P^{*}(x';S,M,\alpha ')
\equiv
\int {\rm d}^4X\,{\rm d}^4X'
\cr
\cdot
{\rm e}^{-{\rm i}P(X-X')}
\mbox{{\large $\rho$}}_{PS}
\!\left(
X+\frac{p_2P}{P^2}x,X-\frac{p_1P}{P^2}x;
X'+\frac{p_2P}{P^2}x',X'-\frac{p_1P}{P^2}x'
\right)
}
\end{equation}
represent elements of the
unnormalized two-particle space-time
density matrices;
the density matrix $\mbox{{\large $\rho$}}_{PS}(x_1,x_2;x_1',x_2')$
depends on the pair four-momentum $P$ due to the
account of the energy--momentum conservation
in the sum $\sum_{\alpha'}$.
{
On the assumption of an instantaneous emission in the two-particle
c.m. system ($t_1^*=t_2^*$), the second expression in (\ref{16})
reduces to the
ansatz used in \cite{nem85,av99}.
}

Switching off the FSI and QS effects, for example, by mixing
particles from different events with similar global characteristics,
one can define the reference differential cross section
\begin{eqnarray}
\label{16ref}
(2\pi)^6\gamma_1\gamma_2\frac{{\rm d}^6\sigma_0}
{{\rm d}^3{\bf p}_1{\rm d}^3{\bf p}_2}
\nonumber\\
=
\sum\limits_{S}
\int {\rm d}^4x_1\,{\rm d}^4x_2\,{\rm d}^4x_1'\,{\rm d}^4x_2'
\mbox{{\large $\rho$}}_{PS}(x_1,x_2;x_1',x_2')
{\rm e}^{-{\rm i}p_1(x_1-x_1')-{\rm i}p_2(x_2-x_2')}
\nonumber\\
=
\sum\limits_{S}
\int {\rm d}^4x\,{\rm d}^4x'
\rho_{PS}(x;x')
{\rm e}^{-{\rm i}\widetilde{q}(x-x')/2}
\end{eqnarray}
and rewrite (\ref{16}) as
\begin{equation}
\label{16aver}
\eqalign{
\frac{{\rm d}^6\sigma}{{\rm d}^3{\bf p}_1{\rm d}^3{\bf p}_2}
\cr
=\frac{{\rm d}^6\sigma_0}{{\rm d}^3{\bf p}_1{\rm d}^3{\bf p}_2}
\sum\limits_{S}
{\cal G}_S(p_1,p_2)
\left\langle
\Psi ^{S(-)}_{p_1,p_2}(x_1,x_2)
\Psi ^{S(-)*}_{p_1,p_2}(x_1',x_2')
\right\rangle'_{p_1p_2S}
\cr
=\frac{{\rm d}^6\sigma_0}{{\rm d}^3{\bf p}_1{\rm d}^3{\bf p}_2}
\sum\limits_{S}
{\cal G}_S(p_1,p_2)
\left\langle
\psi ^{S(-)}_{\widetilde{q}}(x)
\psi ^{S(-)*}_{\widetilde{q}}(x')
\right\rangle'_{\widetilde{q}PS},
}
\end{equation}
where we have introduced the quasi-averages of the bilinear
products of the Bethe--Salpeter amplitudes:
\begin{equation}
\label{quasi-aver}
\eqalign{
\left\langle
\Psi ^{S(-)}_{p_1,p_2}(x_1,x_2)
\Psi ^{S(-)*}_{p_1,p_2}(x_1',x_2')
\right\rangle'_{p_1p_2S}
\cr
=\frac{\int {\rm d}^4x_1\,{\rm d}^4x_2\,{\rm d}^4x_1'\,{\rm d}^4x_2'
\mbox{{\large $\rho$}}_{PS}(x_1,x_2;x_1',x_2')
\Psi ^{S(-)}_{p_1,p_2}(x_1,x_2)
\Psi ^{S(-)*}_{p_1,p_2}(x_1',x_2')}
{\int {\rm d}^4x_1\,{\rm d}^4x_2\,{\rm d}^4x_1'\,{\rm d}^4x_2'
\mbox{{\large $\rho$}}_{PS}(x_1,x_2;x_1',x_2')
{\rm e}^{-{\rm i}p_1(x_1-x_1')-{\rm i}p_2(x_2-x_2')}}
\cr
=
\left\langle
\psi ^{S(-)}_{\widetilde{q}}(x)
\psi ^{S(-)*}_{\widetilde{q}}(x')
\right\rangle'_{\widetilde{q}PS}
=\frac{\int {\rm d}^4x\,{\rm d}^4x'
\rho_{PS}(x;x')
\psi ^{S(-)}_{\widetilde{q}}(x)
\psi ^{S(-)*}_{\widetilde{q}}(x')}
{\int {\rm d}^4x\,{\rm d}^4x'
\rho_{PS}(x;x')
{\rm e}^{-{\rm i}\widetilde{q}(x-x')/2}}
}
\end{equation}
and the statistical factors ${\cal G}_S$ --
the population probabilities of the pair spin-$S$
states in the absence of the correlation effect:
\begin{equation}
\label{statfactors}
\eqalign{
{\cal G}_S(p_1,p_2)=
\cr
\frac{\int {\rm d}^4x_1\,{\rm d}^4x_2\,{\rm d}^4x_1'\,{\rm d}^4x_2'
\mbox{{\large $\rho$}}_{PS}(x_1,x_2;x_1',x_2')
{\rm e}^{-{\rm i}p_1(x_1-x_1')-{\rm i}p_2(x_2-x_2')}}
{\sum\limits_{S}
\int {\rm d}^4x_1\,{\rm d}^4x_2\,{\rm d}^4x_1'\,{\rm d}^4x_2'
\mbox{{\large $\rho$}}_{PS}(x_1,x_2;x_1',x_2')
{\rm e}^{-{\rm i}p_1(x_1-x_1')-{\rm i}p_2(x_2-x_2')}}
\cr
=\frac{\int {\rm d}^4x\,{\rm d}^4x'
\rho_{PS}(x;x')
{\rm e}^{-{\rm i}\widetilde{q}(x-x')/2}}
{\sum\limits_{S}
\int {\rm d}^4x\,{\rm d}^4x'
\rho_{PS}(x;x')
{\rm e}^{-{\rm i}\widetilde{q}(x-x')/2}}.
}
\end{equation}
Note that for unpolarized particles with spins $j_1$ and $j_2$
one has
\begin{equation}
\label{186''}
{\cal G}_S=(2S+1)[(2j_{1}+1)(2j_{2}+1)]^{-1} \qquad \sum_S {\cal G}_S =1.
\end{equation}
Generally, the spin factors are sensitive to particle polarization.
For example, if two spin-1/2 particles were emitted independently
with the polarizations ${\bf P}_1$ and ${\bf P}_2$, then
${\cal G}_0=(1-{\bf P}_1\cdot {\bf P}_2)/4$ and
${\cal G}_1=(3+{\bf P}_1\cdot {\bf P}_2)/4$.

The same procedure can be also applied to describe the production
of weakly bound two-particle systems, like deuterons or hadronic
atoms ($\pi^+\pi^-$ atoms, in particular). Due to a low binding
energy, as compared with the energy transfers at the initial
stage of the collision, there is practically no direct production
of such bound systems. Their dominant production mechanism
is thus due to the particle interaction in the final state.
The invariant production amplitude
$T_b(P_b;S,M,\alpha ')$ of a spin-$S$ bound
system $b=\{1+2\}$ is then
described by the lower diagram in figure~\ref{fig1}
corresponding to the second term in the upper
diagram with the free two-particle final
state substituted by the bound one.
Therefore, similar to (\ref{5}),
this amplitude is related to the Fourier
transforms ${\mathcal T}$ or $\tau_{P_b}$
of the off-mass-shell two-particle amplitude $T_0$:
\begin{equation}
\label{5a}
\eqalign{
T_b(P_b;S,M,\alpha ') =
\int {\rm d}^4x_1\,{\rm d}^4x_2 \Psi ^{S(-)}_{b,P_b}(x_1,x_2)
{\mathcal T}(x_1,x_2;S,M,\alpha ')
\cr
=\int {\rm d}^4x \psi ^{S(-)}_{b}(x)
\tau_{P_b}(x;S,M,\alpha ')
}
\end{equation}
and the corresponding differential cross section -
to the same two-particle space-time
density matrices $\mbox{{\large $\rho$}}_{PS}$ or $\rho_{PS}$
as enter into (\ref{16}), up to the substitution $P\to P_b$:
\begin{equation}
\label{16a}
\eqalign{
(2\pi)^3\gamma_b\frac{{\rm d}^3\sigma_b^S}{{\rm d}^3{\bf P}_b}
\cr
=
\int {\rm d}^4x_1\,{\rm d}^4x_2\,{\rm d}^4x_1'\,{\rm d}^4x_2'
\mbox{{\large $\rho$}}_{P_bS}(x_1,x_2;x_1',x_2')
\Psi ^{S(-)}_{b,P_b}(x_1,x_2)
\Psi ^{S(-)*}_{b,P_b}(x_1',x_2')
\cr
=\int {\rm d}^4x\,{\rm d}^4x'\rho_{P_bS}(x;x')
\psi ^{S(-)}_{b}(x)
\psi ^{S(-)*}_{b}(x').
}
\end{equation}
Here
$\Psi ^{S(-)}_{b,P_b}(x_1,x_2)
=[\Psi ^{S(+)}_{b,P_b}(x_1,x_2)]^*
=[{\rm e}^{{\rm i}P_bX}\psi ^{S(+)}_b(x)]^*$
is the Bethe--Salpeter amplitude for the bound system. At equal
emission times of the two particles in their c.m. system,
the amplitude
$\psi ^{S(+)}_b(x)$, describing their
relative motion, coincides with the usual
non-relativistic wave function in discrete spectrum
$\psi ^{S}_b({\bf r}^*)$.
Similar to (\ref{16aver}), one can also rewrite the
production cross section of the bound system through the
reference cross section in (\ref{16ref}) taken at
${\bf p}_1\doteq {\bf P}_b m_1/(m_1+m_2)$,
${\bf p}_2\doteq {\bf P}_b m_2/(m_1+m_2)$ and
$\gamma_1\doteq \gamma_2\doteq \gamma_b$
(i.e. $\widetilde{q}\doteq 0$, $P\doteq P_b$):
\begin{equation}
\label{16aaver}
\eqalign{
\frac{{\rm d}^3\sigma_b^S}{{\rm d}^3{\bf P}_b} =
(2\pi)^3\gamma_b
\frac{{\rm d}^6\sigma_0}{{\rm d}^3{\bf p}_1{\rm d}^3{\bf p}_2}
{\cal G}_S(p_1,p_2)
\left\langle
\Psi ^{S(-)}_{b,P_b}(x_1,x_2)
\Psi ^{S(-)*}_{b,P_b}(x_1',x_2')
\right\rangle'_{p_1p_2S}
\cr
=(2\pi)^3\gamma_b
\frac{{\rm d}^6\sigma_0}{{\rm d}^3{\bf p}_1{\rm d}^3{\bf p}_2}
{\cal G}_S(p_1,p_2)
\left\langle
\psi ^{S(-)}_{b}(x)
\psi ^{S(-)*}_{b}(x')
\right\rangle'_{0PS}.
}
\end{equation}

We see that the production of a weakly bound system $\{1+2\}$ is
closely related with the production of particles $1$ and $2$
in continuous spectrum at small kinetic energies in their
c.m. system.
This relation was first formulated \cite{mig55}
in connection with the production of non-relativistic deuterons
and then generalized \cite{ani80}
to the relativistic case and the inclusive production.
Similar relation was obtained,
in the limit of an instantaneous
emission from a point-like region, also for the case of the
production of pure Coulomb hadronic atoms \cite{nem85}.
A complete treatment of the production of weakly bound systems,
accounting for the finite-size effect,
can be found in \cite{lyu88}.

\section{Approximate description of the $\pi^+\pi^-$ production}

Following \cite{nem85}, let us first neglect the
$\pi^+\pi^-$ strong FSI and assume only two types of pion
emitters:
SL emitters (e.g., $\rho$- or $\Delta$-resonances)
characterized by small sizes or decay lengths on a fm level,
and LL emitters (e.g., $\eta$, $K_s$ or
$\Lambda$) with very large or macroscopic decay lengths.
Since the relative space-time distance between the emission
points $x$ enters in the pure Coulomb $\pi^+\pi^-$
amplitudes $\psi ^{(-){\rm coul}}_{{q}}(x)$
and $\psi ^{(-){\rm coul}}_b(x)$
scaled by the Bohr radius $a= -387.5$ fm,
one can put in (\ref{16}) and (\ref{16a})
$\psi ^{(-)}_{{q}}(x)\approx {\rm e}^{-{\rm i}\tilde qx/2}
\psi ^{(-){\rm coul}}_{{q}}(0)= {\rm e}^{{\rm i}{\bf k}^*{\bf r}^*}
[\psi ^{\rm coul}_{-{\bf k}^*}(0)]^*$ and
$\psi ^{(-)}_b(x)\approx \psi ^{(-){\rm coul}}_b(0)=
[\psi ^{\rm coul}_b(0)]^*$ for the fraction $\lambda$
of the pairs
with both pions from SL emitters ($r^*\ll |a|$), and
$\psi ^{(-){\rm coul}}_{{q}}(x)\approx
\exp[{\rm i}{\bf k}^*{\bf r}^*-
{\rm i}(k^* a)^{-1}\ln({\bf k}^*{\bf r}^*+k^*r^*)]$
(a plane wave amplitude with the phase modified by Coulomb interaction)
and
$\psi ^{(-){\rm coul}}_b(x)\approx
\psi ^{(-){\rm coul}}_b(\infty)=0$
for the remaining fraction $(1-\lambda)$ of the pairs
with at least one pion from a LL emitter ($r^*\gg |a|$).
As a result, (\ref{16}) and
(\ref{16a}) reduce to:
\begin{equation}
\label{16nem}
\frac{{\rm d}^6\sigma}{{\rm d}^3{\bf p}_1{\rm d}^3{\bf p}_2}
\approx
\frac{{\rm d}^6\sigma_0}{{\rm d}^3{\bf p}_1{\rm d}^3{\bf p}_2}
\left[\lambda \left|\psi ^{\rm coul}_{-{\bf k}^*}(0)
\right|^2+(1-\lambda)\right]
\end{equation}
\begin{equation}
\label{16anem}
\frac{{\rm d}^3\sigma_b}{{\rm d}^3{\bf P}_b}
\approx
(2\pi)^3\gamma_b\frac{{\rm d}^3\sigma_0}{{\rm d}^3{\bf p}_1
{\rm d}^3{\bf p}_2}
\lambda \left|\psi ^{\rm coul}_b(0)\right|^2,
\end{equation}
where $\sigma_0$ represents the production cross section of
the non-interacting pions and the expression for the
production of bound $\pi^+\pi^-$ system implies
${\bf p}_1\doteq {\bf p}_2\doteq {\bf P}_b/2$ and
$\gamma_1\doteq \gamma_2\doteq \gamma_b$.
The squares of the
non-relativistic Coulomb wave
functions at zero separation are well known:
\begin{equation}
\label{Ac}
\left|\psi ^{\rm coul}_{-{\bf k}^*}(0)\right|^2\equiv A_c(\eta)=
2\pi\eta[\exp(2\pi\eta)-1]^{-1} \qquad \eta=(k^*a)^{-1}
\end{equation}
\begin{equation}
\label{wfds0}
\left|\psi_{b}^{\rm coul}(0)\right|^2
=\delta_{l0}\left(\pi|a|^3n^3\right)^{-1}\qquad  b=\{nl\},
\end{equation}
where the Bohr radius $a=(\mu e_1e_2)^{-1}$ is negative
for $\pi^+\pi^-$ system due to the opposite signs of
$\pi^+$ and $\pi^-$ charges ($e_1=-e_2= e$).
The Coulomb penetration factor $A_c(\eta)$ (sometimes called
Gamow factor)
behaves
at small $Q=2k^* < Q_c$ as $Q_c/Q$ and at large $Q$ approaches
unity as $1+\frac12 Q_c/Q$, where $Q_c=4\pi/|a|=6.4$ MeV/$c$.
As for the bound $\pi^+\pi^-$ states $b=\{nl\}$,
only the s-wave states $\{n0\}$ are produced at zero separation
and their fractions with given main
quantum numbers $n$ are uniquely fixed by the $n^{-3}$ law
in (\ref{wfds0}).

The numbers $N_A$ and $N_A^{\rm br}$ of produced and breakup
$\pi^+\pi^-$ atoms, required to calculate the breakup
probability (\ref{pbr}), can then be obtained
in two steps \cite{nem85}. First, one simulates the
non-correlated two-pion spectrum
${{\rm d}^6N_0}/{{\rm d}^3{\bf p}_1{\rm d}^3{\bf p}_2}$
or, constructs it by mixing pions from different
events, and determines the overall normalization parameter $g$
and the fraction $\lambda$ or $\Lambda=\lambda g$ by fitting
the theoretical spectrum
\begin{equation}
\label{16nem'}
\eqalign{
\frac{{\rm d}^6N}{{\rm d}^3{\bf p}_1{\rm d}^3{\bf p}_2}
\approx
g\frac{{\rm d}^6N_0}{{\rm d}^3{\bf p}_1{\rm d}^3{\bf p}_2}
\left[\lambda A_c(\eta)+(1-\lambda)\right]
\cr
\equiv
\frac{{\rm d}^6N_0}{{\rm d}^3{\bf p}_1{\rm d}^3{\bf p}_2}
\left[\Lambda A_c(\eta)+\Lambda'\right]
}
\end{equation}
to the measured spectrum of the pion pairs;
to get rid of the pairs from the breakup of the
$\pi^+\pi^-$ atoms in the target,
the fit should be done in the region $Q>\sim 3$ MeV/$c$.
In the second step, one can use (\ref{16anem}) and
the fitted parameter
$\Lambda=\lambda g$ to calculate the
three-momentum distribution of the numbers of produced
atoms in given states $b=\{n0\}$:
\begin{equation}
\label{16anem'}
\frac{{\rm d}^3N_b}{{\rm d}^3{\bf P}_b}
\approx
(2\pi)^3\gamma_b\frac{{\rm d}^3N_0}{{\rm d}^3{\bf p}_1
{\rm d}^3{\bf p}_2}
\Lambda \left|\psi ^{\rm coul}_{n0}(0)\right|^2,
\end{equation}
where ${\bf p}_1\doteq {\bf p}_2\doteq {\bf P}_b/2$.
Then, taking into account that
$d^3{\bf p}_1d^3{\bf p}_2=\frac18 d^3{\bf P}d^3
\widetilde{\bf q}=\frac18 \gamma d^3{\bf P}d^3{\bf Q}$
and that at sufficiently small $Q<Q_0$ the non-correlated
two-pion spectrum is practically $Q$-independent
(up to a correction $\Or(Q_0^2/m_\pi^2)$),
one can calculate $N_A$
from the number $N_{0}(Q<Q_0)$
of simulated or mixed
non-correlated pion pairs with $Q<Q_0$:
\begin{equation}
\label{N_A}
N_A= \sum_b \int {\rm d}^3{\bf P}_b
\frac{{\rm d}^3N_b}{{\rm d}^3{\bf P}_b} \doteq
\frac{3\Lambda}{4\pi^2}N_{0}(Q<Q_0)
\left(\frac{4\pi}{|a|Q_0}\right)^3
\sum_n n^{-3}.
\end{equation}
One can also calculate $N_A$ from the number of correlated
non-atomic pairs,
\begin{equation}
\label{Ncc}
N^{\rm cna}_{\pi^+\pi^-}(Q<Q_0)\doteq
\frac{3\Lambda}{Q_0^3}N_{0}(Q<Q_0)
\int_0^{Q_0}{\rm d}Q\, Q^2 A_c(\eta),
\end{equation}
using so called $k$-factor \cite{ade05}:
\begin{equation}
\label{k-fac}
k(Q<Q_0)=N_A/N^{\rm cna}_{\pi^+\pi^-}(Q<Q_0)\doteq
\frac{(4\pi)^3}{4\pi^2|a|^3}\frac{\sum_n n^{-3}}
{\int_0^{Q_0}{\rm d}Q\, Q^2 A_c(\eta)};
\end{equation}
e.g., $k=0.615$, 0.263 and 0.140 for $Q_0=2$, 3
and 4 MeV/$c$ respectively.
As for the number of breakup atoms $N_A^{\rm br}$,
it is simply obtained by
subtracting the fitted numbers of correlated (cna)
and non-correlated (nc) non-atomic pion pairs
(see (\ref{16nem'})) from the
measured number of pion pairs:
\begin{equation}
\label{NA}
N^{\rm br}_A \doteq
N_{\pi^+\pi^-}(Q<Q_{\rm cut})
-N^{\rm cna}_{\pi^+\pi^-}(Q<Q_{\rm cut})
-N^{\rm nc}_{\pi^+\pi^-}(Q<Q_{\rm cut})
\end{equation}
\begin{equation}
\label{nc}
N^{\rm nc}_{\pi^+\pi^-}(Q<Q_{\rm cut})=
\Lambda' N_{0}(Q<Q_{\rm cut}).
\end{equation}
The value of $Q_{\rm cut}$ should be chosen sufficiently
large so that the interval $(0,Q_{\rm cut})$ contains
the signal from practically all atomic pairs. The
possible choice is $Q_{\rm cut}=Q_0=4$ MeV/$c$.
One can also choose a smaller value, correcting for the
loss of atomic pairs with the help of simulated
efficiency factor $\epsilon^{\rm br}_A$ \cite{ade05}.
One can obtain $N^{\rm br}_A$ also
in a more direct way using the
data from multi-layer targets \cite{ade04}.

Let us now consider the modification of
(\ref{16nem'}) and (\ref{16anem'}) due to the
strong FSI and finite space-time separation of the
particle emitters. Formally one can write
\begin{equation}
\label{16nem''}
\frac{{\rm d}^6N}{{\rm d}^3{\bf p}_1{\rm d}^3{\bf p}_2}
\doteq
\frac{{\rm d}^6N_0}{{\rm d}^3{\bf p}_1{\rm d}^3{\bf p}_2}
\left\{\Lambda \left[1+\delta({\bf k}^*)
\right]A_c(\eta)+\Lambda'\right\}
\end{equation}
\begin{equation}
\label{16anem''}
\frac{{\rm d}^3N_b}{{\rm d}^3{\bf P}_b}
\doteq
(2\pi)^3\gamma_b\frac{{\rm d}^3N_0}{{\rm d}^3{\bf p}_1
{\rm d}^3{\bf p}_2}
\Lambda (1+\delta_n)\left|\psi ^{\rm coul}_{n0}(0)\right|^2
\end{equation}
\begin{equation}
\label{N_A_new}
N_A\doteq
\frac{3\Lambda}{4\pi^2}N_{0}(Q<Q_0)
\left(\frac{4\pi}{|a|Q_0}\right)^3
\sum_n n^{-3}(1+\delta_n)
\end{equation}
\begin{equation}
\label{Ncc_new}
N^{\rm cna}_{\pi^+\pi^-}(Q<Q_0)\doteq
\frac{3\Lambda}{Q_0^3}N_{0}(Q<Q_0)
\int_0^{Q_0}{\rm d}Q\, Q^2 A_c(\eta)[1+\delta({k}^*)]
\end{equation}
\begin{equation}
\label{k-fac_new}
k(Q<Q_0) \doteq
\frac{(4\pi)^3}{4\pi^2|a|^3}\frac{\sum_n n^{-3}(1+\delta_n)}
{\int_0^{Q_0}{\rm d}Q\, Q^2 A_c(\eta)[1+\delta({k}^*)]},
\end{equation}
where the correction factors are determined by the averaging
of the bilinear products of the reduced Bethe--Salpeter
amplitudes over the distribution of the relative space-time
separations of the SL emitters:
\begin{equation}
\label{corcs}
1+\delta({\bf k}^*)\doteq
\left\langle \psi ^{(-)}_{\widetilde{q}}(x)
\psi ^{(-)*}_{\widetilde{q}}(x')
\right\rangle_{\widetilde{q}P}^{'\mbox{\tiny SL}}
[A_c(\eta)]^{-1}
\end{equation}
\begin{equation}
\label{cords}
1+\delta_n\doteq
\left\langle \psi ^{(-)}_{n0}(x)\psi ^{(-)*}_{n0}(x')
\right\rangle_{0P}^{'\mbox{\tiny SL}}
\left|\psi ^{\rm coul}_{n0}(0)\right|^{-2}.
\end{equation}
The averaging is defined in (\ref{quasi-aver})
with the reduced space-time density matrix substituted
by its part,
$\rho_{P}^{\mbox{\tiny SL}}$, related only with the
SL emitters.
Equations (\ref{corcs}) and (\ref{cords}) account
only for the elastic transition $\alpha \rightarrow \alpha$
and ignore a small contribution of the inelastic one
$\beta \rightarrow \alpha$, where $\alpha=\{\pi^+\pi^-\}$,
$\beta=\{\pi^0\pi^0\}$; see section 6 and
(\ref{corcspi}), (\ref{cordspi}) for the complete
treatment.

In fact, it can be argued \cite{led98,av99} that
\begin{equation}
\label{migdal}
\delta_n\approx \delta(0)
\end{equation}
provided the characteristic spatial separation of the pion SL
emitters in the two-pion c.m. system
is much less than the two-pion Bohr radius $|a|$.
This result immediately follows from the well known Migdal's
argument \cite{mig55}. Namely, since the particles in continuous
spectrum at zero kinetic energy and in discrete spectrum at
very small binding energy $\kappa^2/(2\mu) \rightarrow 0$
are described by practically the same wave equations,
the $r^*$-dependence of the corresponding wave functions
at a given orbital angular momentum should be the same for
the distances $r^*\ll \kappa^{-1}$ (i.e. $r^*\ll n|a|$
in the case of a hadronic atom with the main quantum number $n$).

One may see that the approximate equality (\ref{migdal}),
together with the assumption of a weak $k^*$-dependence of the correction
$\delta({\bf k}^*)$, justify the use of the approximate
equations (\ref{16nem'})-(\ref{k-fac}). In this approximation,
the finite-size correction merely reduces to the
rescaling $\Lambda\to \Lambda [1+\delta(0)]$.

We show in sections 5 and 6 that
(\ref{migdal}) is subject to a normalization
correction $\Or(f_0/a)\sim 0.3\%/n$ and other small
corrections $4\pi\Or(k^*_\beta(f_0^{\beta\alpha})^2/a)$
and $\Or(a^{-2})$, where $k^*_\beta = 35.5$ MeV/$c$ is the
momentum in the channel $\beta=\{\pi^0\pi^0\}$ at the
threshold transition to the channel $\alpha=\{\pi^+\pi^-\}$,
$f_0^{\beta\alpha}=\sqrt{2}(a_0^2-a_0^0)/3 \approx -0.2$ fm
is the transition amplitude.
Taking the normalization correction explicitly into account
(see (\ref{nemansatz1}), (\ref{Aaa1}) and (\ref{normapp1})):
\begin{eqnarray}
\label{norm_corr}
1+\delta_n = (1+\delta_n')
\nonumber\\
\cdot
\left\{1+ \phi(n)
\frac{2f_0}{n|a|}
\left[1+\Or\!\left(\left(k^*_\beta f_0^{\beta\alpha}\right)^2
\right)\right]
-4\pi^2\Or\!\left(\left(\frac{f_0}{a}\right)^2\right)
\right\},
\end{eqnarray}
where $\phi(n)\approx 3$ is defined in (\ref{phi}),
one can rewrite the approximate equality in
(\ref{migdal}) as
\begin{equation}
\label{migdal_mod}
\delta_n' = \delta(0)+
4\pi\Or\!\left(k^*_\beta(f_0^{\beta\alpha})^2/a\right)
+\Or(\langle r^{*2}/a^2\rangle^{\mbox{\tiny SL}}).
\end{equation}

The neglect of the corrections
$\delta_n$ and $\delta({\bf k}^*)$, i.e.
the use of the approximate equations
(\ref{16nem'})-(\ref{k-fac}) instead of
(\ref{16nem''})-(\ref{k-fac_new}),
leads to the systematic shift of the breakup
probability:
\begin{equation}
\label{DPbr}
\Delta P_{\rm br}/P_{\rm br}\doteq
-\Delta N_A/N_A + \Delta N^{\rm br}_A/N^{\rm br}_A .
\end{equation}
To estimate this shift, one can approximate
the correlation function
\begin{equation}
\label{Rcna}
{\cal R}(Q)\equiv \frac{{\rm d}N/{\rm d}Q}
{{\rm d}N_0/{\rm d}Q}=
A_c(\eta)[1+\delta({k}^*)],
\end{equation}
where $\delta({k}^*)=\langle\delta({\bf k}^*)\rangle$
is the finite-size correction averaged over the
pion three-momenta at a fixed $k^*=|{\bf k}^*|=Q/2$,
by
\begin{equation}
\label{Rapp}
\widetilde{{\cal R}}(Q)=
\widetilde{\Lambda}A_c(\eta)+\widetilde{\Lambda}'
\end{equation}
and use the fitted parameters $\widetilde{\Lambda}$,
$\widetilde{\Lambda}'$ to calculate $\Delta N_A/N_A$
and $\Delta N^{\rm br}_A/N^{\rm br}_A$:
\begin{equation}
\label{DNA}
-\frac{\Delta N_A}{N_A} =
\frac{\sum_n n^{-3}(1+\delta_n - \widetilde{\Lambda})}
{\sum_n n^{-3}(1+\delta_n)}
\end{equation}
\begin{equation}
\label{DNAbr}
\frac{\Delta N^{\rm br}_A}{N^{\rm br}_A}=
\frac{N^{\rm cna}_{\pi^+\pi^-}(Q<Q_{\rm cut})}
{N^{\rm br}_A(Q<Q_{\rm cut})}
\frac{\int_0^{Q_{\rm cut}}{\rm d}Q\, Q^2
[{\cal R}(Q)-\widetilde{{\cal R}}(Q)]}
{\int_0^{Q_{\rm cut}}{\rm d}Q\, Q^2
{\cal R}(Q)} .
\end{equation}
In the experiment DIRAC,
$N^{\rm cna}_{\pi^+\pi^-}/N^{\rm br}_A \doteq 16$
at $Q_{\rm cut}=4$ MeV/$c$ \cite{ade05} and,
for $Q_{\rm cut}< Q_c$, when $Q A_c(\eta)\approx const$, it
decreases approximately quadratically with decreasing
$Q_{\rm cut}$ provided the signal region contains
most of the atomic pion pairs, i.e. down to
$Q_{\rm cut}\approx 2$ MeV/$c$.

\begin{figure}
\epsfig{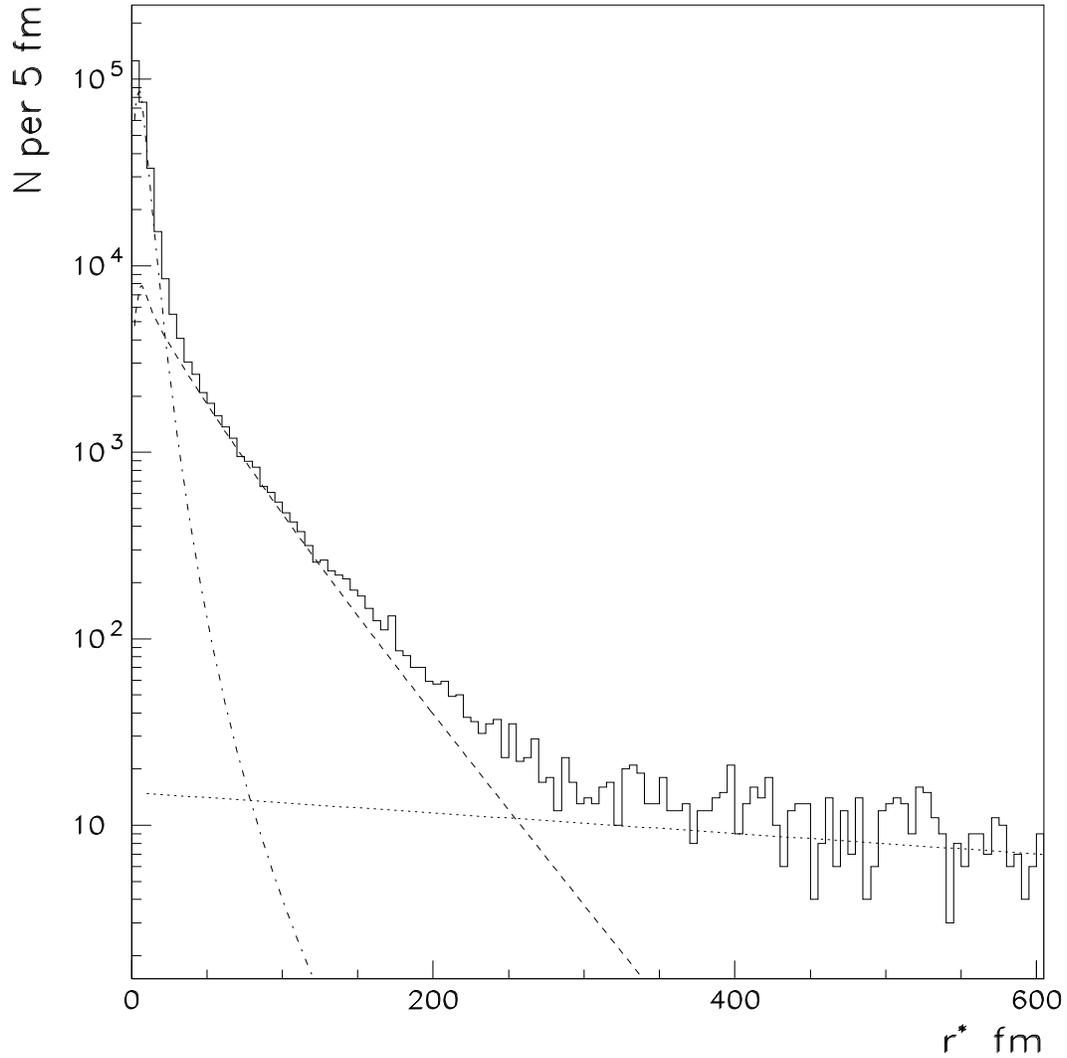}
\caption{The distribution of the relative distance $r^*$
between the pion production points in the pair c.m. system
simulated with the UrQMD transport code
\cite{bas98} for pNi interactions
at 24 GeV and the relative momenta in the pair c.m. system
$Q=2 k^* < 50$ MeV/$c$
in the conditions of the DIRAC experiment at CERN
\cite{smol}. The curves are the results of the fits to
short-distance, $\omega$ and $\eta'$ contributions
described in the text.}
\label{fig_rst}
\end{figure}

In the following, we perform an analytical and numerical study
of the corrections $\delta_n$ and $\delta(k^*)$
and their effect on the breakup probability $P_{\rm br}$.
Here we only mention that the condition $r^*\ll n|a|$ can be
violated for pion pairs containing pions from the decays of some
resonances such as $\omega$ and $\eta'$ with the decay lengths
of about 30 and 900 fm respectively. The corresponding exponential
tails are clearly seen in figure~\ref{fig_rst}, where the
$r^*$-distribution simulated with the UrQMD transport code
\cite{bas98} is shown for pion pairs produced in pNi interactions
at 24 GeV in the conditions of the DIRAC experiment at CERN
\cite{smol}.

\section{Correlation femtoscopy -- basic assumptions}

\subsection{Non-interacting non-identical particles: space-time coherence}
To clarify the meaning of the two-particle space-time density
matrix $\mbox{{\large $\rho$}}_S(x_1,x_2;x_1',x_2')$,
let us first neglect
the FSI effect and substitute the Bethe--Salpeter amplitudes
by the plane waves.
Changing in (\ref{16}) the
integration variables $x_i, x_i'$ by the new ones:
\begin{equation}
\label{new_var}
\bar x_i=\frac 12 (x_i+x_i') \qquad
\epsilon_i=x_i-x_i',
\end{equation}
we can rewrite the production cross section of two
non-identical particles as
\begin{equation}
\label{16'}
\eqalign{
(2\pi)^6\gamma_1\gamma_2\frac{{\rm d}^6\sigma_0}
{{\rm d}^3{\bf p}_1{\rm d}^3{\bf p}_2}
\cr
= \sum\limits_{S}
\int {\rm d}^4\bar x_1\,{\rm d}^4\bar x_2
G_S(\bar x_1,p_1;\bar x_2,p_2)
= \sum\limits_{S}\int {\rm d}^4\bar x
g_{PS}(\bar x,\widetilde{q}),
}
\end{equation}
where $\bar x=\bar x_1-\bar x_2$ and
the real functions
\begin{equation}
\label{16''}
\eqalign{
G_S(\bar x_1,p_1;\bar x_1,p_2)=
\cr
\int {\rm d}^4\epsilon_1\,{\rm d}^4\epsilon_2
{\rm e}^{-{\rm i}p_1\epsilon_1-{\rm i}p_2\epsilon_2}
\mbox{{\large $\rho$}}_{PS}\!\left(\bar x_1+\frac{\epsilon_1}{2},
\bar x_2+\frac{\epsilon_2}{2};
\bar x_1-\frac{\epsilon_1}{2},\bar x_2-\frac{\epsilon_2}{2}\right)
\cr
g_{PS}(\bar x;\widetilde{q})=
\int {\rm d}^4{\bar X}
G_S\!\left(\bar X+\frac{p_2P}{P^2}\bar x,p_1;
\bar X-\frac{p_1P}{P^2}\bar x,p_2\right)
\cr
=
\int {\rm d}^4\epsilon
{\rm e}^{-{\rm i}\widetilde{q}\epsilon/2}
\rho_{PS}\!\left(\bar x+\frac{\epsilon}{2},
\bar x-\frac{\epsilon}{2}\right).
}
\end{equation}
The function $G_S$, usually called emission function,
being a partial Fourier transform of the
space-time density matrix,
is closely related to the Wigner density, the latter collecting
all contributions due to free streaming of the emitted particles
to given space-time points through an integral over the emission
function (see (49) in \cite{als}).

It is clear from (\ref{16'}) and (\ref{16''}) that
more narrow is the width of the
diagonal of the space-time density matrix (the width of the
$\epsilon_i$-distribution), more wide is the distribution
of particle four-momenta.
In particular, the diagonal space-time density matrix
(i.e., zero width of the $\epsilon_i$-distribution) would
yield the uniform four-momentum distribution,
in correspondence with the infinite uncertainty in the
four-momenta of the particles localized at certain space-time
points.

Consider as an example the particle emission by
independent one-particle emitters of various types $A$
according to the one-particle production amplitudes
(see also \cite{led00})
\begin{equation}
\label{188}
\eqalign{
{\mathcal T}_A^{(1)}(x_1;x_A)
\sim v_A(x_A)\exp
\!\left[-\frac{({\bf x}_1-{\bf x}_A)^2}{2r_A{}^2}
-\frac{(x_{01}-x_{0A})^2}{2\tau_A{}^2}\right]
\cr \qquad
v_A(x_A)\sim \exp\!\left(-\frac{{\bf x}_A{}^2}{4r_0{}^2}-
\frac{x_{0A}{}^2}{4\tau_0{}^2}\right).
}
\end{equation}
These amplitudes correspond to the emitters at rest
with a Gaussian distribution of the emission points
$x_1=\{t_1,{\bf r}_1\}$ around the emitter centers
$x_A=\{t_A,{\bf r}_A\}$,
also distributed according to a Gaussian law.
In four-momentum representation,
\begin{equation}
\label{188m}
\eqalign{
{T}_A^{(1)}(p_1;x_A)\sim v_A(x_A)u_A(p_1)\exp(-ip_1x_A)
\cr \qquad
u_A(p_1)\sim
\exp
\!\left(-r_A{}^2{\bf p}_1{}^2/2-\tau_A{}^2p_{01}{}^2/2\right).
}
\end{equation}

Assuming further that the emitters are sufficiently heavy,
we can describe them classically.
The four-coordinates of the emitter centers $x_A$ can then
be considered as a part of the quantum numbers $\alpha '$,
the sum in (\ref{17}) thus containing the integration
over $x_A$. Performing this integration, we get
for the elements of the one-particle space-time
density matrix related to the emitter $A$:
\begin{equation}
\label{189}
\eqalign{
\rho_A^{(1)}(x_1,x_1')=\int {\rm d}^4x_A
{\mathcal T}_A^{(1)}(x_1;x_A){\mathcal T}_A^{(1)*}(x_1';x_A)
\cr
\sim \exp
\!\left(-\frac{\mbox{\boldmath $\epsilon$}_1{}^2}{4r_A{}^2}
-\frac{\epsilon_{01}{}^2}{4\tau_A{}^2}\right)
\exp\!\left(-\frac{{\bf \bar x}_1{}^2}{2r_0{}^2+r_A{}^2}-
\frac{\bar x_{01}{}^2}{2\tau_0{}^2+\tau_A{}^2}\right)
}
\end{equation}
and for the corresponding emission function:
\begin{equation}
\label{190}
\eqalign{
G_A^{(1)}(\bar x_1,p_1)=\int {\rm d}^4\epsilon_1
{\rm e}^{-ip_1\epsilon_1}
\rho_A^{(1)}\left(\bar x_1+\frac{\epsilon_1}{2},
\bar x_1-\frac{\epsilon_1}{2}\right)
\cr
\sim \left|u_A(p_1)\right|^2
\exp\!\left(-\frac{{\bf \bar x}_1{}^2}{2r_0{}^2+r_A{}^2}-
\frac{\bar x_{01}{}^2}{2\tau_0{}^2+\tau_A{}^2}\right).
}
\end{equation}
The contribution of the emitter $A$ to the single-particle
production cross section is
\begin{equation}
\label{contrib_A}
\eqalign{
(2\pi)^3\gamma_1\frac{{\rm d}^3\sigma_A}
{{\rm d}^3{\bf p}_1}=\int {\rm d}^4\bar x_1
G_A^{(1)}(\bar x_1,p_1)=
\int {\rm d}^4x_A
\left|{T}_A^{(1)}(p_1;x_A)\right|^2
\cr
\sim \left|u_A(p_1)\right|^2 \sim
\exp
\!\left(-r_A{}^2{\bf p}_1{}^2-\tau_A{}^2p_{01}{}^2\right).
}
\end{equation}
We may see that the emitter space-time dimensions $r_A$ and
$\tau_A$ determine both the space-time coherence of particle
production (the non-diagonality of the space-time
density matrix) and the distribution of particle four-momenta.
In particular case of the emitters
of a vanishing space-time extent: $r_A=\tau_A=0$ (no coherence)
any particle four-momenta are equally probable.

Note that for the emitter moving with a non-relativistic velocity
$\mbox{\boldmath$\beta$}_A$ and emitting a particle 1 with the
mean three-momentum ${\bf p}_A=m_1\mbox{\boldmath$\beta$}_A$,
the amplitude (\ref{188}) and the density matrix (\ref{189})
respectively acquire phase factors
${\rm e}^{-{\rm i}{\bf p}_A{\bf x}_1}$ and
${\rm e}^{-{\rm i}{\bf p}_A\mbox{\boldmath$\epsilon$}_1}$
and
the substitution ${\bf p}_1\rightarrow {\bf p}_1-{\bf p}_A$
has to be done in the amplitude $u_A(p_1)$.
After averaging over
the ${\bf p}_A$-distribution that decouples from the distribution
of other emitter characteristics in a Gaussian form of a width
$\Delta_0$, we still arrive at (\ref{189})-(\ref{contrib_A}),
up to a substitution
$r_A{}^2\rightarrow r_A{}^2/[2(r_A\Delta_0)^2+1]$
in the $\epsilon$- and momentum-dependent factors,
corresponding to a widening of the momentum distribution
due to the dispersion of the emitter velocities.

As for the actual values of the parameters
$r_A$ and $\tau_A$, we can estimate
them using the information about particle transverse momenta,
$p_t$,
which are much less influenced by the motion of the emitters
than the longitudinal ones.
Doing this for pions or kaons, we should however exclude the
low-$p_t$ region which is dominantly populated by the
decays of low-lying resonances. We can also use the
$p_t$-distributions of these resonances. In both cases
the $p_t{}^2$-slopes in the interactions
of elementary hadrons are of $\sim$3 (GeV/$c$)$^{-2}$
(see, e.g., \cite{rid88}), yielding
on average
$r_A{}^2, \tau_A{}^2\sim 0.1$ fm$^2$.
Somewhat larger values can be expected in heavy ion collisions
where a substantial part of the emitters can be associated
with the centers of the last rescatterings characterized by
sufficiently large momentum transfer.
It is important that the estimated values of
$r_A{}^2, \tau_A{}^2$ appear to be much smaller than the effective
values of the parameters $r_0{}^2$, $\tau_0{}^2$
obtained in femtoscopic measurements.
The latter being of about 1 fm$^2$
for pions produced at $p_t \sim \langle p_t\rangle$ in
hadron--hadron interactions and up to several tens fm$^2$ in
the collisions involving heavy nuclei.

\subsection{Non-interacting identical particles: QS correlations}
\label{3.2}
\subsubsection{Correlation function}
Consider the
production of non-interacting identical particles.
It should be noted that this consideration is not of academic
interest only. Thus for identical pions or kaons, the effect
of the strong FSI is usually small and the effect of the
Coulomb FSI can be in first approximation simply corrected for
(see \cite{slape98} and references therein).
The corrected correlation effect is then determined by the
QS symmetrization only, i.e. the Bethe--Salpeter amplitudes
have to be substituted by
properly symmetrized combinations of the plane waves
(see (\ref{12})). As a result of the interference of
these waves, there appears the additional term, not present
in (\ref{16'}):
\begin{equation}
\label{16a'}
\eqalign{
(2\pi)^6\gamma_1\gamma_2\frac{{\rm d}^6\sigma}
{{\rm d}^3{\bf p}_1{\rm d}^3{\bf p}_2}
\cr
=
\sum\limits_{S}
\int {\rm d}^4\bar x_1\,{\rm d}^4\bar x_2
\left[G_S(\bar x_1,p_1;\bar x_2,p_2)+
G_S(\bar x_1,p;\bar x_2,p)(-1)^S\cos(q\bar x)\right]
\cr
=\sum\limits_{S}\int {\rm d}^4\bar x
\left[g_{PS}(\bar x,q)+
g_{PS}(\bar x,0)(-1)^S\cos(q\bar x)\right] .
}
\end{equation}
Note that
the off-mass-shell four-momentum $p=(p_1+p_2)/2$ enters as
an argument of the emission function $G_S$
in the interference term.

It is convenient to define the correlation function
${\cal R}(p_1,p_2)$ as the ratio
of the double inclusive cross section ${\rm d}^6\sigma $
to the reference one ${\rm d}^6\sigma_0$
which would be observed in the case of absent
QS and FSI effects:
\begin{equation}
\label{CF-def}
{\cal R}(p_1,p_2)=\frac{{\rm d}^6\sigma(p_1,p_2)}
{{\rm d}^6\sigma_0(p_1,p_2)}.
\end{equation}
In the high-energy collisions involving nuclei, we can neglect
the kinematic constraints as well as rather weak dynamical
correlations and construct the reference distribution using
the particles from different events with similar topology.
In  case of a negligible FSI, there is no correlation
for non-identical particles: ${\cal R}(p_1,p_2)=1$, while
for identical particles the correlation arises due to the
interference effect:
\begin{equation}
\label{181}
\eqalign{
{\cal R}(p_1,p_2) = 1+
\frac{\sum\limits_{S}
\int {\rm d}^4\bar x_1\,{\rm d}^4\bar x_2
G_S(\bar x_1,p;\bar x_2,p)
(-1)^S\cos (q\bar x)}
{\sum\limits_{S}
\int {\rm d}^4\bar x_1\,{\rm d}^4\bar x_2
G_S(\bar x_1,p_1;\bar x_2,p_2)}
\cr
\equiv
1+\sum\limits_{S}{\cal G}_S(-1)^S
\left\langle\cos (q\bar x)\right\rangle''_{p_1p_2S}
\cr
= 1+
\frac{\sum\limits_{S}
\int {\rm d}^4\bar x
g_{PS}(\bar x,0)(-1)^S\cos (q\bar x)}
{\sum\limits_{S}
\int {\rm d}^4\bar x
g_{PS}(\bar x,q)}
\cr
\equiv
1+\sum\limits_{S}{\cal G}_S(-1)^S
\left\langle\cos (q\bar x)\right\rangle''_{qPS},
}
\end{equation}
where the quasi-averages satisfy the equalities
\begin{equation}
\label{equalities}
\eqalign{
\langle \cos(q\bar x)\rangle''_{p_1p_2S}=
\!\left\langle {\rm e}^{{\rm i}p_1(x_1-x_2')+
{\rm i}p_2(x_2-x_1')}\right\rangle'_{p_1p_2S}
\cr
=\left\langle \cos(q\bar x)\right\rangle''_{qPS}=
\!\left\langle {\rm e}^{{\rm i}q(x+x')/2}\right\rangle'_{qPS};
}
\end{equation}
the factors ${\cal G}_S$ represent the
population probabilities of the pair spin-$S$ states
out of the region of the correlation effect. They are
defined in (\ref{statfactors}) and can be expressed
through the emission functions as
\begin{equation}
\label{185}
\eqalign{
{\cal G}_S(p_1,p_2) = \frac{\int {\rm d}^4x_{1}\,{\rm d}^{4}x_{2}
G_S(x_1,p_1;x_2,p_2)}
{\sum_{S}\int {\rm d}^4x_{1}\,{\rm d}^{4}x_{2}
G_S(x_1,p_1;x_2,p_2)}
\cr
=
\frac{\int {\rm d}^4x g_{PS}(x,q)}
{\sum_{S}\int {\rm d}^4x g_{PS}(x,q)}
\qquad
\sum_{S} {\cal G}_S = 1.
}
\end{equation}
They can be also considered as the initial (QS switched off)
statistical factors.
For initially unpolarized spin-$j$ particles:
$\sum_S {\cal G}_S(-1)^S=(-1)^{2j}/(2j+1)$.

Assuming, for example, that for a (generally momentum dependent)
fraction $\lambda$ of the pairs the particles are emitted by
independent SL one-particle emitters described by the Gaussian
amplitudes (\ref{188}) or (\ref{188m}), while for the remaining fraction
$(1-\lambda)$, related to LL emitters
($\eta$, $K^0_s$, $\Lambda$, \dots),
the relative distances $r^*$ between the emission
points in the pair c.m. system are extremely large,
the correlation function
\begin{equation}
\label{181'a}
\eqalign{
{\cal R}(p_1,p_2)
\cr
= 1+\lambda\sum_S{\cal G}_S(-1)^S
\frac{\Re\sum_{A,B}u_A(p_1)u_B(p_2)u_A^*(p_2)u_B^*(p_1)
{\rm e}^{-iq(x_A-x_B)}}{\sum_{A,B}|u_A(p_1)u_B(p_2)|^2},
}
\end{equation}
where the sum $\sum_{A,B}$ is done over all characteristics of the
emitters. In the case of only one type of the SL emitters
that are at rest and differ only by the four-coordinates
$x_A$ of their centers, the amplitudes $u_A$ reduce to
a single universal amplitude $u$ and the sum merely
reduces to the averaging over $x_A$, i.e.
\begin{equation}
\label{181'}
\eqalign{
{\cal R}(p_1,p_2)=1+\lambda\sum_S{\cal G}_S(-1)^S
\langle\cos(q(x_A-x_B))\rangle
\cr
 = 1+\lambda\sum_S{\cal G}_S(-1)^S
\exp\!\left(-r_0{}^2{\bf q}^2-\tau_0{}^2q_0^2\right).
}
\end{equation}
We see that a characteristic feature of the correlation function
of identical particles is the presence of an interference maximum
or minimum at small $|{\bf q}|$, changing to a horizontal plateau
at sufficiently large $|{\bf q}|$, large compared with the
inverse characteristic
space-time distance between the particle emission points.

\subsubsection{Smoothness assumption}
In the simple model of only one type of the emitters contributing
to the observable interference effect and in the absence of the
relative emitter motion, the width of the low-$|{\bf q}|$
structure is solely determined by the characteristic space-time
distance between the one-particle emitters and does not depend on
the parameters $r_A$ and $\tau_A$, characterizing the space-time
extent of the emitters themselves -- see (\ref{181'}).
It means that the enlargement of the production
region related to the latter
($r_0{}^2 \rightarrow r_0{}^2+\frac 12r_A{}^2$,
$\tau_0{}^2 \rightarrow \tau_0{}^2+\frac 12\tau_A{}^2$)
is compensated in the correlation function due to the
different momentum arguments of the emission functions in the
numerator and denominator of (\ref{181}).
This is clearly seen when calculating the correlation function
directly from (\ref{1}), substituting
the production amplitude $T(p_1,p_2;\alpha)$ by the
symmetrized product of the Kopylov--Podgoretsky
one-particle amplitudes in momentum representation,
see (\ref{181'a}) and (\ref{181'}).
Of course, the independence of the interference effect
on the space-time extent of the emitters in this model
(assuming that the emitters decay according to
a single universal amplitude $u$
and differ by the four-coordinates of their centers only)
is justified only in the case of sufficiently small
overlap of the emitters to guarantee the assumption
of their independence.

Generally, even in the case of independent emitters,
the particles are emitted by moving emitters of
different types and the correlation function depends also on
their space-time extent $r_A, \tau_A$.
Particularly, for a Gaussian distribution of the mean emission
three-momentum
${\bf p}_A$ of a width $\Delta_0$, (\ref{181'}) is modified
by the substitution \cite{led00}
$r_0{}^2 \rightarrow r_0{}^2+r_A{}^2/[2+(r_A\Delta_0)^{-2}]$.
Usually, the effect of a finite space-time extent of the
one-particle emitters is negligible:
\begin{equation}
\label{smoothcon}
r_A{}^2/2 \ll r_0{}^2,~~ \tau_A{}^2/2 \ll \tau_0{}^2.
\end{equation}
Note that
these conditions guarantee sufficiently smooth four-momentum
dependence of the emission function
$G_S(\bar x_1,p_1;\bar x_2,p_2)$, such that
we can neglect its dependence on the four-momentum difference
$q$ in the region of the interference
effect characterized by the inverse space-time distance
between the particle production points. On this,
so called {\it smoothness} assumption, (\ref{181})
reduces to:
\begin{equation}
{\cal R}(p_1,p_2) \doteq 1+\sum\limits_{S}
(-1)^S {\cal G}_S\left\langle\cos (qx)\right\rangle_{qPS},
\label{183}
\end{equation}
where ${\cal G}_S$ are the normalized spin factors
defined in (\ref{statfactors})
and (\ref{185}), and
\begin{equation}
\label{184a}
\eqalign{
\langle\cos (qx)\rangle_{qPS}=
\frac{\int {\rm d}^4x_1\,{\rm d}^4x_2
G_S(x_1,p_1;x_2,p_2)
\cos (qx)}
{\int {\rm d}^4x_1\,{\rm d}^4x_2
G_S(x_1,p_1;x_2,p_2)}
\cr
=
\frac{\int {\rm d}^4x g_{PS}(x,q)\cos (qx)}
{\int {\rm d}^4x g_{PS}(x,q)}.
}
\end{equation}

Equation (\ref{183}) is valid up to a correction representing a
fraction of $r_A{}^2/r_0{}^2$, $\tau_A{}^2/\tau_0{}^2$.
This correction composes a few percent
for high-energy hadron--hadron collisions and a fraction of
percent for the collisions involving heavy nuclei.
Note that (\ref{183}) is often used to calculate the
correlation functions of non-interacting identical particles
with the help of various classical transport codes
(like RQMD, VENUS or UrQMD) \cite{bas98} -
the emission points are identified with the points of the last
collisions or the resonance decays.

At sufficiently small $Q$, one can calculate
the one-dimensional correlation function ${\cal R}(Q)$
using a more simple and faster procedure than the averaging
according to (\ref{183}). For this one can
exploit the fact that the
angular distribution of the vector ${\bf Q}$ becomes
isotropic at $Q\to 0$ and calculate
$\langle\cos (qx)\rangle\equiv\langle\cos ({\bf Qr}^*)\rangle$
by averaging over the uniform distribution of the cosine
of the angle between the vectors ${\bf Q}$
and ${\bf r}^*$ and, over the one-dimensional
$r^*$-distribution determined at $Q\to 0$.
We have checked the accuracy of this procedure
using the UrQMD simulation of the pNi interactions
at 24 GeV in the conditions of experiment DIRAC \cite{smol}
and determined the ${r}^*$-distributions
in various $Q$-intervals. It appears that the
$\pi^-\pi^-$ correlation functions calculated from
the ${r}^*$-distributions corresponding to
the intervals 50-100, 100-150 and 150-200 MeV/$c$
agree with that corresponding to the ${r}^*$-distribution
in the lowest $Q$-interval of 0-50 MeV/$c$
within 0.2-0.9$\%$, $\sim 3\%$ and $\sim 7\%$, respectively.
It should be noted that the increasing difference of the
correlation functions with the increasing lower
boundary of  the above $Q$-intervals
is not related with the violation of the
{\it smoothness} assumption but rather with the
approximate treatment of the angular dependence
of the vector ${\bf Q}$ and with the
$Q$-dependence of the
fractions of pairs containing pions from resonance decays.

\subsubsection{Femtoscopy with identical particles}
One can see from (\ref{181'}) that, due to the
on-shell constraint
$q_0 = {\bf v}{\bf q} \equiv vq_L$,
the correlation function at $v\tau_{0} > r_{0}$ substantially
depends on the direction of the vector ${\bf q}$
even in the case of
spherically symmetric spatial form of the production region.
Thus the transverse (${\bf q}\perp {\bf v}$) and longitudinal
(${\bf q}\parallel {\bf v}$)
correlation radii are
$r_T=r_0$ and $r_L=(r_0^2+v^2\tau_0^2)^{1/2}$ respectively.

The on-shell constraint makes the $q$-dependence
of the correlation function essentially three-dimensional
(particularly, in pair c.m. system,
$q x=-2{\bf k}^*{\bf r}^*$)
and thus makes impossible the unique Fourier reconstruction
of the space-time characteristics of the emission process.
However, within realistic models,
the directional and velocity
dependence of the correlation function
can be used to determine both
the duration of the emission and the form
of the emission region \cite{pod89},
as well as -- to reveal the details of the
production dynamics (such as collective flows;
see, e.g., \cite{pra84,mak87} and reviews
\cite{wh99,cso02,led04,lis05}).
For this, the correlation functions can be analyzed
in terms of the {out} (x), {side} (y) and {longitudinal} (z)
components of the relative momentum vector
${\bf q}=\{q_x,q_y,q_z\}$ \cite{pod83,bdh94};
the {out} and {side} denote the transverse components
of the vector ${\bf q}$, the {out} direction is
parallel to the transverse component of the pair three-momentum.
The corresponding correlation widths are
usually parameterized in terms
of the Gaussian correlation (interferometry) radii $r_i$,
e.g., for spin-0 bosons
\begin{equation}
{\cal R}(p_{1},p_{2})=
1+\lambda\exp(-r_x^2q_x^2-r_y^2q_y^2-r_z^2q_z^2
-2 r_{xz}^2q_xq_z),
\label{osl}
\end{equation}
and the radii dependence on pair rapidity and transverse momentum
is studied.
The correlation strength parameter $\lambda$
can differ from unity due to the contribution of LL
emitters, particle misidentification and coherence effects.
Equation~(\ref{osl}) assumes azimuthal symmetry of the
production process. Generally, e.g., in case of the
correlation analysis with respect to the reaction plane, all three
cross terms $q_iq_j$ contribute.

It is well known that particle correlations at high energies
usually measure only a small part of the space-time emission
volume, being only slightly sensitive to its increase related
to the fast longitudinal motion of particle emitters.
In fact, due to limited emitter decay momenta $p_{\rm dec}$
of few hundred MeV/$c$, the correlated particles with nearby
velocities are emitted by almost comoving emitters and so -
at nearby space-time points.
In other words, the maximal contribution of the relative motion
to the correlation radii in the two-particle c.m. system is
limited by the moderate emitter decay length $\tau p_{\rm dec}/m$.
The dynamical examples are resonances,
colour strings or hydrodynamic expansion.
To substantially eliminate the effect of the longitudinal
motion, the correlations can be analyzed in terms of the
invariant variable $Q = 2k^* \equiv (-\widetilde{q}^2)^{1/2}$
and the components of the  three-momentum difference in the
pair c.m. system (${\bf q}^*\equiv {\bf Q}= 2{\bf k}^*$) or in
the longitudinally comoving system (LCMS) \cite{lcms}.
In LCMS, each pair is emitted transverse to the reaction axis
so that the generalized relative three-momentum
$\widetilde{{\bf q}}$ coincides with ${\bf q}^*$,
except for the {\it out}-component
$\widetilde{q}_x=\gamma_tq_x^*$,
where $\gamma_t$ is the LCMS Lorentz factor of the pair.
Particularly, in the case of one-dimensional boost-invariant
expansion, the longitudinal correlation radius in LCMS
reads \cite{mak87}
\begin{equation}
r_z = (T/m_t)^{1/2}\tau,
\label{rz}
\end{equation}
where $T$ is the
freezeout temperature, $\tau$ is the proper freezeout time
and $m_t$ is the transverse particle mass.
In this model, the side radius measures the
transverse radius of the system while the square of the out radius gets an additional
contribution $(p_t/m_t)^2\Delta\tau^2$ due to the finite emission duration $\Delta\tau$.
The additional transverse expansion leads to a slight modification of the
$p_t$-dependence of the longitudinal radius and -- to a noticeable decrease of the
side radius and the spatial part of the out radius with $p_t$. Thus in the case of
a linear non-relativistic transverse flow velocity profile
$\beta_{\mathrm F}= \beta_0r_t/R$ of the expanding fireball with the
freeze-out transverse radius $R$, the side radius
\begin{equation}
       r_y\approx  R/ (1+m_t\beta_0^2/T)^{1/2}.
\label{ry}
\end{equation}
The decrease of the two-pion correlation radii
with increasing transverse mass (expansion) and
decreasing centrality (geometry)
has been demonstrated, e.g., in Au+Au collisions at
$\sqrt{s}_{NN}=200$ GeV \cite{star05}.

Since the freeze-out temperature and the transverse flow determine also the shapes
of the $m_t$ spectra, the simultaneous analysis of correlations and single particle
spectra for various particle species allows one to disentangle all the freeze-out
characteristics (see, e.g., \cite{wh99}).
Thus in heavy ion collisions
the correlation data show rather weak energy
dependence and point to the kinetic freeze-out temperature
somewhat below the pion mass, a strong transverse flow (with the mean transverse
flow velocity of about half the velocity of light), a short evolution time
of 8--10 fm/$c$ and a very short emission duration of about 2--3 fm/$c$
(see, e.g., a recent review \cite{lis05}).

\subsection{Interacting particles: FSI correlations}
\subsubsection{Production of interacting particles}
It is clear that the {\it smoothness} assumption allows one to
express the production cross section through the emission
function $G_S(x_1,p_1;x_2,p_2)$ also in the case of
interacting particles. Thus, separating the two-particle c.m.
motion in the phase factor
$\exp[iP(X-X')]\equiv \exp[i(p_1-\widetilde{q}/2)\epsilon_1
+{\rm i}(p_2+\widetilde{q}/2)\epsilon_2]$ and using the
{\it smoothness} assumption to neglect here $\widetilde{q}$
compared with $p_{1,2}$
\footnote
{
The account of $\widetilde{q}$ in the phase factor would
lead to the substitution of the particle four-momenta in
the emission function by their mean (off-mass-shell) values:
$p_i\rightarrow Pm_i/(m_1+m_2)$.
}
and substitute, in the amplitudes
$\psi ^{S(+)}_{\widetilde{q}}(x)$, the relative coordinates
$x=\bar x+(\epsilon_1-\epsilon_2)/2$ and
$x'=\bar x-(\epsilon_1-\epsilon_2)/2$
by their mean value $\bar x$, we can rewrite (\ref{16})
in a simple approximate form:
\begin{eqnarray}
\label{160}
(2\pi)^6\gamma_1\gamma_2\frac{{\rm d}^6\sigma}
{{\rm d}^3{\bf p}_1{\rm d}^3{\bf p}_2}
&\doteq& \sum\limits_{S}
\int {\rm d}^4x_1\,{\rm d}^4x_2
G_S(x_1,p_1;x_2,p_2)
\left|\psi ^{S(+)}_{\widetilde{q}}(x)\right|^2 \nonumber \\
&=& \sum\limits_{S}
\int {\rm d}^4x
g_{PS}(x,{\widetilde q})
\left|\psi ^{S(+)}_{\widetilde{q}}(x)\right|^2 \nonumber \\
&\equiv&
(2\pi)^6\gamma_1\gamma_2\frac{{\rm d}^6\sigma_0}
{{\rm d}^3{\bf p}_1{\rm d}^3{\bf p}_2}
\sum\limits_{S}
{\cal G}_S\left\langle \left|\psi ^{S(+)}_{\widetilde{q}}(x)\right|^2
\right\rangle_{\widetilde{q}PS},
\end{eqnarray}
where ${\rm d}^6\sigma_0$ is the production cross section of
non-interacting particles introduced in (\ref{16'}).
The averaging
$\langle \dots\rangle_{\widetilde{q}PS}$ and the initial
spin factors ${\cal G}_S$ are defined
in (\ref{184a}) and (\ref{185}).
The correlation function defined as the ratio
${\rm d}^6\sigma/{\rm d}^6\sigma_0$
then takes on the form:
\begin{equation}
\label{cf1}
{\cal R}(p_1,p_2) \doteq
\sum\limits_{S}
{\cal G}_S\left\langle \left|\psi ^{S(+)}_{\widetilde{q}}(x)\right|^2
\right\rangle_{\widetilde{q}PS}.
\end{equation}
Recall that for identical
particles the Bethe--Salpeter amplitudes
$\psi ^{S(+)}_{\widetilde{q}}(x)$
should be symmetrized according to (\ref{12}).

Note that for non-identical particles, one also arrives at
(\ref{160}) and (\ref{cf1}) using the approximate ansatz
$\Psi ^{S(+)}_{p_1,p_2}(x_1,x_2)\doteq
{\rm e}^{{\rm i}(p_1\epsilon_1+p_2\epsilon_2)}
\Psi ^{S(+)}_{p_1,p_2}(\bar x_1,\bar x_2)$
which becomes exact in the absence of FSI.
For identical particles, this ansatz, applied to the
non-symmetrized amplitudes $\widetilde{\Psi}^{S(+)}$,
leads to the correlation
function (see also (58) and (60) in \cite{als})
\begin{equation}
\label{cf2}
\eqalign{
{\cal R}(p_1,p_2)
\cr
\doteq
\sum\limits_{S}
{\cal G}_S\left[
\left\langle \left|\widetilde{\psi}^{S(+)}_{q}(x)\right|^2
\right\rangle_{qPS} + (-1)^S
\Re\!\left\langle \widetilde{\psi}^{S(+)}_{q}(x)
\widetilde{\psi}^{S(+)*}_{-q}(x)
\right\rangle''_{qPS}
\right],
}
\end{equation}
where $\widetilde{\psi}$ is the reduced non-symmetrized
Bethe--Salpeter amplitude ($\widetilde{\psi}^{S(+)}_{q}(x)=
{\rm e}^{{\rm i}qx/2}$ for non-interacting particles).
Clearly, the {\it smoothness} assumption allows one to put
$\langle\dots\rangle''_{qPS}\doteq \langle\dots\rangle_{qPS}$
and thus recover symmetrized equation (\ref{cf1}).

Similar to the case of non-interacting particles,
the relative correction to the {\it smoothness} approximations in
(\ref{160})-(\ref{cf2})
is determined by the ratios
$r_A{}^2/r_0{}^2$, $\tau_A{}^2/\tau_0{}^2$ -
the measures of the non-diagonality
of the space-time density matrix.
For identical particles, the correction arises mainly
from the simplified treatment of the symmetrization
effect and, according to subsection 4.2.2,
it is expected on a few per mil level for the
processes involving heavy nuclei.
For non-identical particles, the corrections
to the finite-size FSI contributions are of the same
order while, those to the complete correlation functions
are usually
substantially smaller, being scaled by the relative
finite-size contributions of the strong
and Coulomb FSI.
In case of $|f^S|\ll r^*\ll |a|$, we are interested in,
the corresponding strong and Coulomb FSI contributions are
of $2f^S/r^*$ and $2r^*/a$ respectively (see section 5).

Proceeding in a similar way with the production cross section
of a bound two-particle system, we arrive,
on the same conditions as in the case of continuous spectrum,
at the approximate form:
\begin{eqnarray}
\label{16aa}
(2\pi)^3\gamma_b\frac{{\rm d}^3\sigma_b^S}{{\rm d}^3{\bf P}_b}
&\doteq&
\int {\rm d}^4x_1\,{\rm d}^4x_2
G_S(x_1,p_1;x_2,p_2)
\left|\psi ^{S(+)}_{b}(x) \right|^2 \nonumber \\
&=&\int {\rm d}^4x g_{PS}(x,0)
\left|\psi ^{S(+)}_{b}(x) \right|^2 \nonumber \\
&\equiv&
(2\pi)^6\gamma_1\gamma_2\frac{{\rm d}^6\sigma_0}
{{\rm d}^3{\bf p}_1{\rm d}^3{\bf p}_2}
{\cal G}_S\left\langle \left|\psi ^{S(+)}_{b}(x)
\right|^2\right\rangle_{0PS},
\end{eqnarray}
where ${\bf p}_i = {\bf P}_b m_i/(m_1+m_2)$,
${\bf P}={\bf P}_b$ and $P_0\doteq P_{b0}$.

\subsubsection{Equal-time approximation}
\begin{figure}
\epsfig{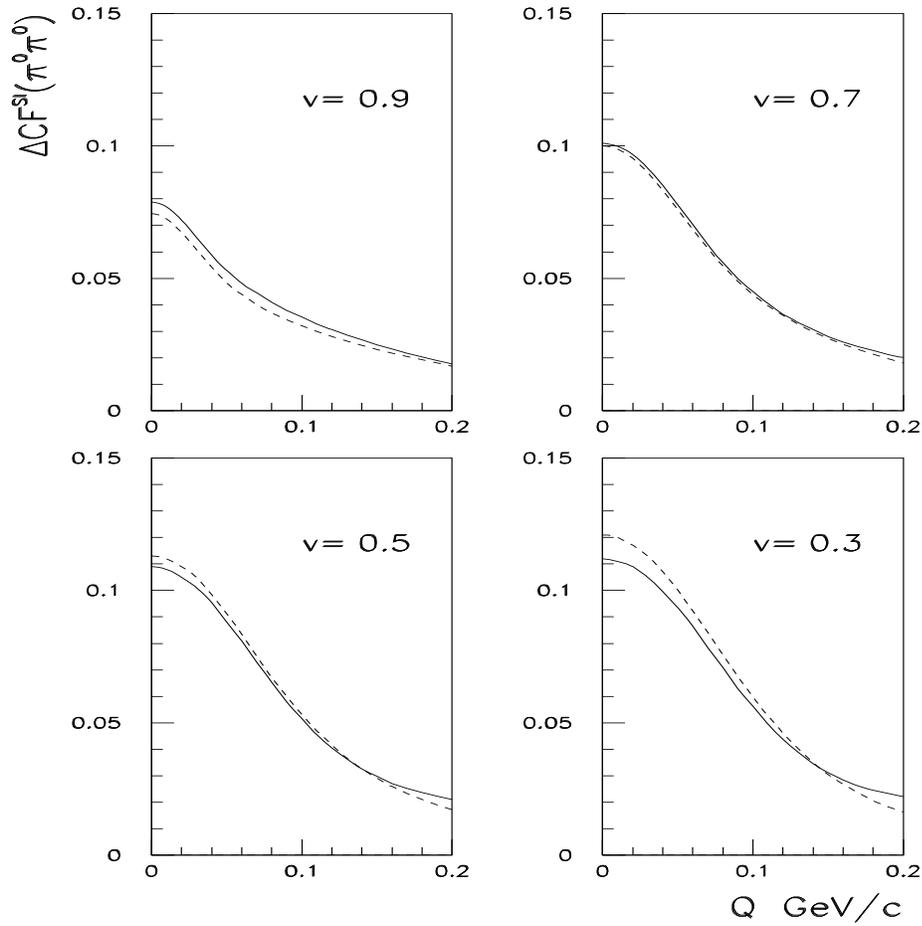}
\caption{The FSI contribution to the $\pi^0\pi^0$ correlation
function calculated for different values of the pair velocity
$v$ in a model of independent one-particle emitters distributed
according to a Gaussian law with the spatial and time width
parameters $r_0=2$ fm and $\tau_0=2$ fm/$c$. The exact results
(solid curves) are compared with those obtained in the
{\it equal-time} approximation (dash curves).}
\label{fig3}
\end{figure}

\begin{figure}
\epsfig{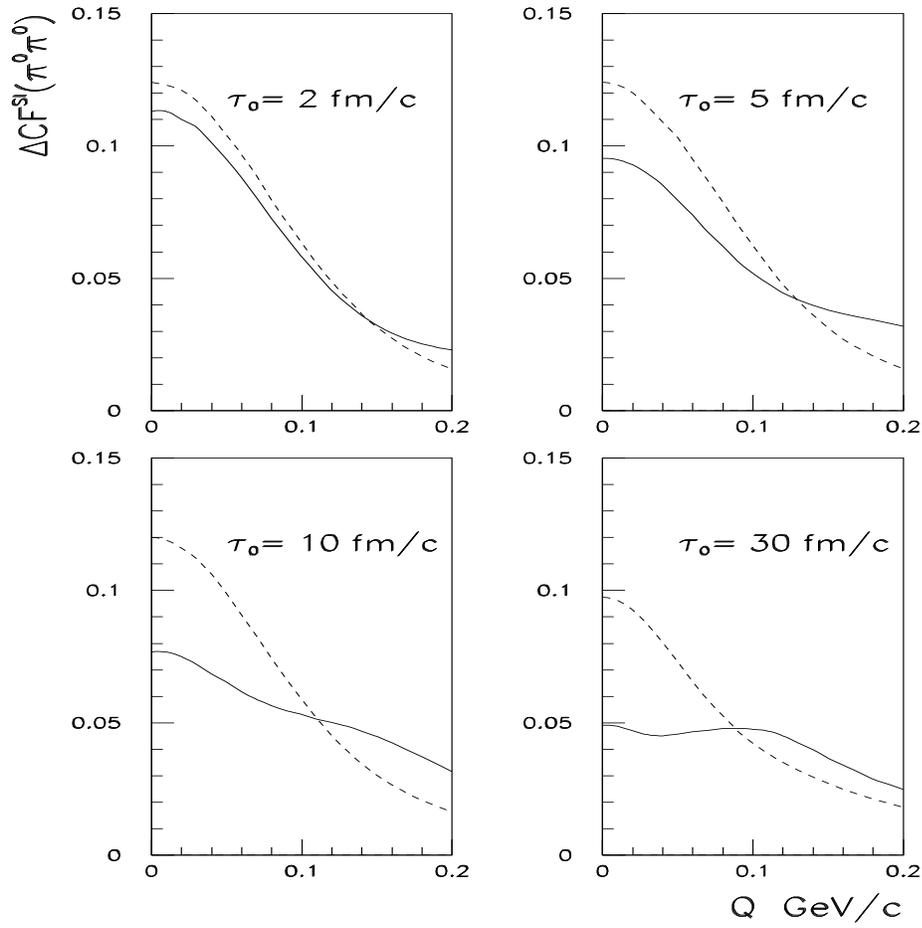}
\caption{The same as in figure~\ref{fig3} for the pair
velocity $v=0.1$,
the spatial width parameter $r_0=2$ fm and
different values of the time width parameter $\tau_0$.}
\label{fig4}
\end{figure}

For non-interacting particles, the non-symmetrized
Bethe--Salpeter amplitude
$\psi ^{(+)}_{\widetilde{q}}(x)=
{\rm e}^{-{\rm i}{\bf k}^*{\bf r}^*}$
is independent of the relative emission time $t^*$ in
the pair c.m. system.
On the contrary, the amplitude of two interacting
particles contains an explicit dependence on $t^*$ -- the
interaction effect vanishes at $|t^*|\rightarrow \infty$.
However, it can be shown \cite{ll82} (see Appendix A)
that the effect of non-equal times can be neglected on
condition
\begin{equation}
\label{10}
|t^{*}|\ll m(t^*){r}^{*2},
\end{equation}
where $m(t^{*}>0) = m_2$ and $m(t^*<0) = m_1$.
On this condition one can use the approximation of equal
emission times of the two particles in their c.m. system ($t^* = 0$)
and substitute the Bethe--Salpeter amplitude
by the usual non-relativistic two-particle wave function.
The applicability condition (\ref{10}) of the
{\it equal-time} approximation is usually satisfied
for heavy particles like kaons or nucleons.
But even for pions this approximation
merely leads to a slight overestimation (typically less than a
few percent) of the strong FSI contribution to the production
cross section \cite{ll82}. To demonstrate this, one can
use the simple static Gaussian model of independent one-particle
emitters described by the amplitude (\ref{188}).
The applicability condition (\ref{10}) of the
{\it equal-time}
approximation can then be written as \cite{ll82}
\begin{equation}
\label{10sta}
\tau_0\ll \mu\gamma r_0 (r_0{}^2+v^2\tau_0{}^2)^{1/2}.
\end{equation}
{
Recall, however,
that in high-energy collisions, the static model is relevant for a
limited rapidity region only. It means that the pair velocity
$v$ in the rest frame of the contributing emitters is
essentially determined by the distribution
of particle transverse momenta.
For pion pairs at $Q\to 0$
one then has $\langle v\rangle\approx 0.8$.
}
For $\tau_0 < \sim r_0$ condition (\ref{10sta})
requires sufficiently small Compton wave lengths of the
particles in the emitter rest frame:
$1/\omega_i \ll r_0$, while for large characteristic emission
times,
$\tau_0\gg r_0/v$, it requires small de Broglie wave lengths:
$1/{p}_i \ll r_0$. Clearly, this condition is not satisfied
for very slow particles emitted by the emitters of a long
lifetime. The increasing importance of the non-equal time
effect with the decreasing pair velocity and increasing
lifetime of the emitters is demonstrated in figures~\ref{fig3}
and \ref{fig4} for the FSI contribution in the $\pi^0\pi^0$
correlation function. For sufficiently large velocities $v>0.5$
and radii $r_0>1$ fm, we are interested in,
the effect is rather small, not exceeding 5$\%$ of the FSI
contribution in the low-${k}^*$ region, corresponding to the
effect of a few per mil in the correlation function.

As for the effect of non-equal times on the Coulomb FSI
it doesn't influence the leading zero-distance
($r^{*}=0$) part and, the effect of the subleading part
(expected on a similar percent level as in the case
of the strong FSI)
can be neglected when scaled by its contribution $\sim 2r^*/a$.
It concerns also the case of hadronic atoms since
the subleading part is the same as in the continuous spectrum
at $k^*=0$.

Adopting the {\it smoothness} and {\it equal-time}
approximations (with the accuracy of
a few per mil), we can rewrite (\ref{160}) and (\ref{16aa})
for the production cross sections of particles
$1$ and $2$ in continuous and discrete spectrum at low relative
or binding energies as follows:
\begin{equation}
\label{160app}
\gamma_1\gamma_2\frac{{\rm d}^6\sigma}{{\rm d}^3{\bf p}_1
{\rm d}^3{\bf p}_2}
\doteq
\gamma_1\gamma_2\frac{{\rm d}^6\sigma_0}{{\rm d}^3{\bf p}_1
{\rm d}^3{\bf p}_2}
\sum\limits_{S}
{\cal G}_S\left\langle \left|\psi ^{S}_{-{\bf k}^*}({\bf r}^*)
\right|^2
\right\rangle_{\widetilde{q}PS}
\end{equation}
\begin{equation}
\label{16aaapp}
\gamma_b\frac{{\rm d}^3\sigma_b^S}{{\rm d}^3{\bf P}_b} \doteq
(2\pi)^3\gamma_1\gamma_2\frac{{\rm d}^6\sigma_0}
{{\rm d}^3{\bf p}_1{\rm d}^3{\bf p}_2}
{\cal G}_S\left\langle \left|\psi ^{S}_{b}({r}^{*})
\right|^2\right\rangle_{0PS},
\end{equation}
where $b=\{n0\}$ and
${\bf p}_i = {\bf P}_b m_i/(m_1+m_2)$ in (\ref{16aaapp});
for equal-mass particles ${\bf p}_1={\bf p}_2={\bf P}_b/2$
and $\gamma_1=\gamma_2=\gamma_b$.
Particularly, for $\pi^+\pi^-$ production,
one can then rewrite the correction factors in (\ref{corcs}) and
(\ref{cords}) as
\begin{equation}
\label{corcs'}
1+\delta({\bf k}^*)\doteq
\left\langle \left|\psi_{-{\bf k}^*}({\bf r}^*)\right|^2
\right\rangle_{\widetilde{q}P}^{\mbox{\tiny SL}}
[A_c(\eta)]^{-1}
\end{equation}
\begin{equation}
\label{cords'}
1+\delta_n\doteq
\left\langle \left|\psi_{n0}({r}^{*}) \right|^2
\right\rangle_{0P}^{\mbox{\tiny SL}}
\left|\psi ^{\rm coul}_{n0}(0)\right|^{-2}.
\end{equation}

We will show that
the $r^*$-dependence of the wave functions
$\psi ^{S}_{-{\bf k}^*}$ and $\psi ^{S}_{b}$
for two oppositely charged particles in
continuous and discrete spectrum is practically the same at
separations $r^*$ in the pair c.m. system much smaller than the
Bohr radius $|a|$.
Therefore, the corrections to (\ref{160app}) and (\ref{16aaapp})
(arising due to the {\it smoothness} and {\it equal-time}
approximations used in their
respective derivation from (\ref{16}) and (\ref{16a}))
practically cancel out in the ratio of the numbers of pairs
produced in continuous and discrete spectrum provided
$\langle r^*\rangle^{\mbox{\tiny SL}}\ll |a|$.

\subsubsection{The effect of residual charge}
The formalism of section 2 assumes a free motion of a
given particle pair during the final stage of the collision.
Here we will estimate the FSI effect
of the residual charge which is known to substantially
influence particle spectra and, to a lesser extent,
also particle correlations in low energy
collisions involving nuclei \cite{3body}.
Since, at high energies,
this effect can be expected of minor importance,
we will estimate only its upper limit.

Generally, instead of the two-particle Bethe--Salpeter
amplitude
$\Psi^{(+)S}_{p_{1}p_{2}}(x_{1},x_{2})$,
the correlation function is determined by the amplitude
$\Psi^{(+)S\{\alpha\}}_{p_{1}p_{2}}(x_{1},x_{2})$
representing the solution
of a complicated multi-body problem, taking
into account interaction between the two particles and
also their interaction with the residual system described
by the quantum numbers $\{\alpha\}$.
For our purpose, it is sufficient to approximate these
quantum numbers by an effective
(comoving with a given pair)
point-like residual charge $Ze$ and consider a thermal
motion of the two particles with the temperature
$T\sim m_\pi$ in the rest frame of this charge.

Let us start with the hypothetical case of particles
that interact with the charge $Ze$ but their mutual
interaction is "switched off". In such a situation
we can treat the systems $(1,Z)$
and $(2,Z)$ independently. Then the interaction
with the Coulomb center just leads to the substitution
of the spatial parts of the plane waves
${\rm e}^{{\rm i}p_ix_i}$
by the usual Coulomb wave functions:
$
{\rm e}^{-{\rm i}{\bf p}_i{\bf r}_i} \rightarrow
{\rm e}^{-{\rm i}{\bf p}_i{\bf r}_i}
\Phi_{{\bf p}_i}^{z_iZ}({\bf r}_i)$, where
$
\Phi_{{\bf p}_i}^{z_iZ}({\bf r}_i)=
{\rm e}^{{\rm i}\delta_i}\sqrt{A_{c}(\eta_i)}
F\!\left(-{\rm i}\eta_i,1,{\rm i}\rho_i\right)$,
$\rho_i={\bf p}_i{\bf r}_i+
{\rm p}_ir_i$,
$\eta_i=({\rm p}_ia_i)^{-1}$,
$a_i=(\omega_iz_iZe^{2})^{-1}$ is the Bohr
radius of the system $(i,Z)$
(taking into account the sign of the interaction)
generalized to the relativistic case
by the substitution $m_i \rightarrow \omega_i$
of the particle masses by their energies,
$\delta_i$ is the Coulomb s-wave phase shift,
$A_{c}(\eta_i)$ is the Coulomb penetration factor
and $F(\alpha,1,z)$
is the confluent hypergeometrical function;
see (\ref{Ac}), (\ref{psicoul}) and (\ref{F}).
For the complete amplitude we have:
\begin{equation}
\label{eq:8}
\eqalign{
\widetilde{\Psi}_{p_{1}p_{2}}^{(+)Z}(x_{1},x_{2})=
{\rm e}^{{\rm i}p_{1}x_{1}+{\rm i}p_{2}x_{2}}
\Phi_{{\bf p}_{1}}^{z_{1}Z}({\bf r}_{1})
\Phi_{{\bf p}_{2}}^{z_{2}Z}({\bf r}_{2})
\cr
\qquad
\equiv
{\rm e}^{{\rm i}PX}{\rm e}^{-{\rm i}{\bf k}^{*}{\bf r}^{*}}
\Phi_{{\bf p}_{1}}^{z_{1}Z}({\bf r}_{1})
\Phi_{{\bf p}_{2}}^{z_{2}Z}({\bf r}_{2}).
}
\end{equation}
Note that a small contribution of spin-dependent
electro-magnetic forces is neglected here so that
$\widetilde{\Psi}^{(+)SZ}\equiv
\widetilde{\Psi}^{(+)Z}$
is independent of
the total spin $S$ of the particle pair.

\begin{figure}
\epsfig{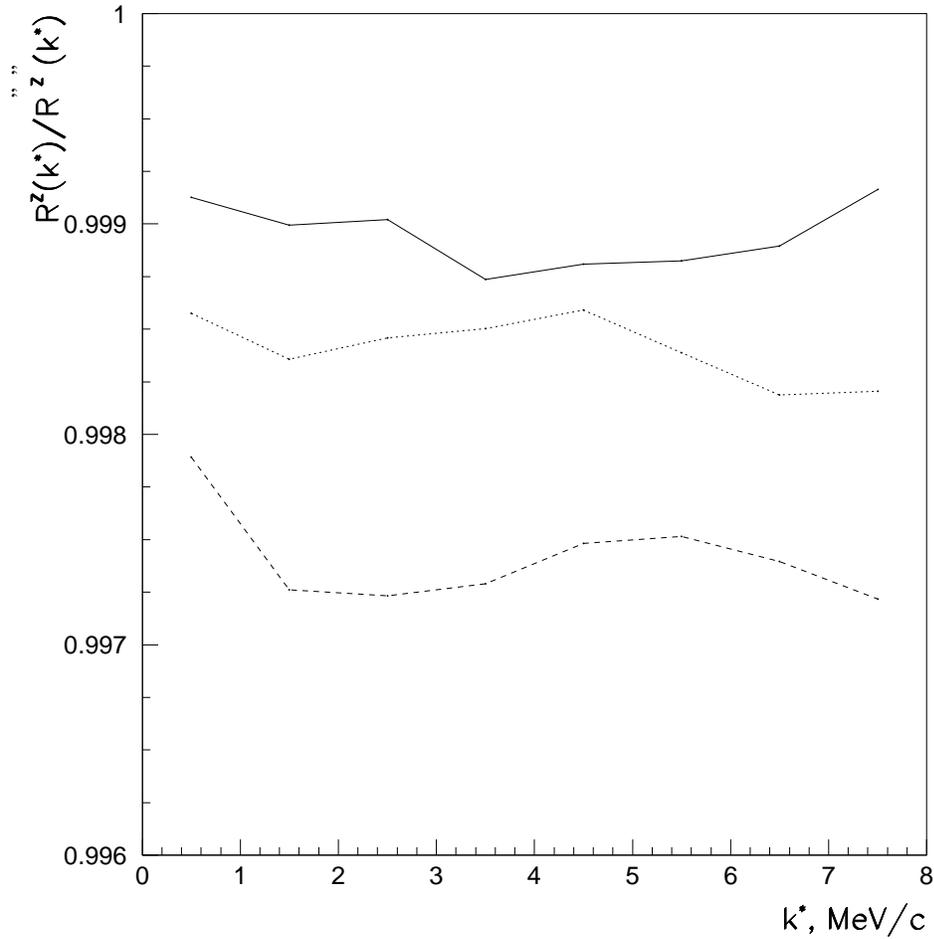}
\caption{The ratios of the $\pi^+\pi^-$ correlation functions
${\cal R}^{Z}$ and ${\cal R}^{"Z"}$. For the latter,
only one-particle spectra are influenced by the effective
comoving charge $Z$.
The pions are assumed to be emitted in the rest frame
of a point-like charge Z
according to the thermal law with a temperature of 140 MeV.
The distribution of the space-time coordinates of the particle
emitters is simulated as a product of Gauss functions
with the equal spatial and time width parameters
$r_{0}=c\tau_{0}$.
The full broken line corresponds to $Z=30, r_{0}=2$ fm,
the dash and dotted ones -- to $Z=60, r_{0}=2$ and 3 fm,
respectively.}
\label{figz}
\end{figure}

Let us now "switch on" the interaction between
particles $1$ and $2$. Since we consider the relative
motion of the two particles at characteristic distances
much slower compared with their motion with respect to
the Coulomb center, it is natural to assume that in such
a case the plane wave
${\rm e}^{-{\rm i}{\bf k}^{*}{\bf r}^{*}}$ in (\ref{eq:8})
will be basically substituted by the Bethe--Salpeter
amplitude $\psi_{\widetilde{q}}^{S}(x)$
describing the relative motion of isolated interacting
particles. After this substitution we get the amplitude
in so called adiabatic
(factorization) approximation \cite{3body}:
\begin{equation}
\label{eq:9}
\Psi_{p_{1}p_{2}}^{(+)SZ}(x_{1},x_{2})=
{\rm e}^{{\rm i}PX}
\psi_{\widetilde{q}}^S(x)
\Phi_{{\bf p}_{1}}^{z_{1}Z}({\bf r}_{1})
\Phi_{{\bf p}_{2}}^{z_{2}Z}({\bf r}_{2}).
\end{equation}
Instead of the six-dimensional correlation function
${\cal R}(p_{1},p_{2})$
we calculate the one-dimensional one,
${\cal R}^{Z}(k^{*})$, with the
numerator and denominator integrated over the simulated
particle spectra. In the {\it equal-time} approximation,
\begin{equation}
\label{eq:12}
{\cal R}^Z(k^{*}) =\frac
{\sum\limits_{i=1}^{N(k^*)}\sum_{S}\rho_{S}
|\psi_{-{\bf k}_i^{*}}^{S}({\bf r}_i^{*})
\Phi_{{\bf p}_{1_i}}^{z_{1}Z}({\bf r}_{1_i})
\Phi_{{\bf p}_{2_i}}^{z_{2}Z}({\bf r}_{2_i})|^{2}}
{\sum\limits_{i=1}^{N(k^*)}
|\Phi_{{\bf p}_{1_i}}^{z_{1}Z}({\bf r}_{1_i})
\Phi_{{\bf p}_{2_i}}^{z_{2}Z}({\bf r}_{2_i})|^{2}},
\end{equation}
where $N(k^*)$ is the number of generated particle pairs
in a given $k^*$ bin. To separate the pure effect of the
residual Coulomb field on particle correlations,
we compare the correlation function
${\cal R}^{Z}(k^{*})$
with the one, ${\cal R}^{"Z"}(k^{*})$, taking into account
for the latter the effect of the nucleus Coulomb field on
one-particle spectra but not on particle correlations
(i.e., simulating the argument ${\bf r}^{*}$
independently of the arguments ${\bf r}_{1}$
and ${\bf r}_{2}$).
Note that due to the velocity dependence of the correlation
function, ${\cal R}^{"Z"}={\cal R}^{Z=0}$ only at a fixed
pair velocity $v$.
In figure~\ref{figz}, we present
the ratios of the $\pi^+\pi^-$ correlation functions
${\cal R}^{Z}$ and ${\cal R}^{"Z"}$ assuming that
the pions are emitted in the rest frame
of the residual charge Z
according to the thermal law with a temperature of 140 MeV
at the space-time points distributed according to
a product of Gauss functions
with the equal spatial and time width parameters
$r_{0}=c\tau_{0}$.
One may see that even for the radius $r_0$ as low as
2 fm the effect of the residual comoving charge as large
as $Z=60$ is less than a few per mil. Taking into account
that the effective radius $r_0$ is larger than 2 fm even
for proton collisions with low-Z nuclei and that the
effective residual charge is only a fraction of the target
nucleus charge, one can conclude that the effect of the
residual charge is on a negligible level of a fraction of
per mil.

\subsubsection{Femtoscopy with non-identical particles}

The FSI effect allows one to access the space-time
characteristics of particle production also with the help
of correlations of non-identical particles.
One should be however careful when analyzing these correlations
in terms of simple models like those assuming the
Gaussian space-time parametrization of the source.
The simplified description of the $r^*$-separations
can lead to inconsistencies in the treatment of
QS and FSI effects.
While the QS and strong FSI effects are influenced by large
$r^*$-separations mainly through the correlation strength
parameter $\lambda$, the shape of the Coulomb FSI is sensitive
to the distances as large as the pair Bohr radius
(hundreds of fm for the pairs containing pions).

This problem can be at least partially overcome with the help
of imaging techniques \cite{bd97} or
transport simulations.
The former yield the ${\bf r}^*$-distribution inverting
the measured correlation function using the integral equation
(\ref{160app}) with the kernel given by the wave function squared.
The latter account for the dynamical
evolution of the emission process and provide the phase space
information required to calculate the QS and FSI effects on the
correlation function.

Thus, the transport RQMD v.2.3 code was used in a preliminary analysis
of the NA49 $\pi^+\pi^-$, $\pi^+p$ and $\pi^-p$ correlation data from
central Pb+Pb 158 $A$ GeV collisions \cite{led04}.
The model correlation functions ${\cal R}_{\rm RQMD}(Q;s_{r})$
have been calculated using the FSI code based on the formalism
developed in \cite{ll82},
weighting the simulated pairs by squares of the
corresponding wave functions.
The scale parameter $s_{r}$,
multiplying the simulated space-time coordinates of the
emitters, was introduced in the model correlation function
to account for a possible mismatch of the
$r^*$-distribution.
For this, a set of correlation functions
${\cal R}_{\rm RQMD}(Q;s_{r}^i)$ was calculated
at three chosen values $s_{r}^i$ of the scale parameter
and the quadratic interpolation was used
to calculate
${\cal R}_{\rm RQMD}(Q;s_{r})$ for arbitrary value of $s_{r}$:
\begin{equation}
\label{interp1}
\eqalign{
{\cal R}_{\rm RQMD}(Q;s_{r})= \sum_{i=1}^3
\frac{(s_{r}-s_{r}^j)(s_{r}-s_{r}^k)}
{(s_{r}^i-s_{r}^j)(s_{r}^i-s_{r}^k)}
{\cal R}_{\rm RQMD}(Q;s_{r}^1),
}
\end{equation}
where $\{i,j,k\}$ are permutations of the sequence $\{1,2,3\}$.
The NA49 correlation functions were then fitted by
\begin{equation}
{\cal R}(Q)= N \left[\lambda
{\cal R}_{\rm RQMD}(Q;s_{r})+ (1-\lambda)\right]
\label{Q1}
\end{equation}
with two additional parameters, the normalization $N$ and the
correlation strength $\lambda$.
The fitted values of the $\lambda$-parameter are in reasonable agreement
with the expected contamination of $\sim 15\%$
from strange particle decays and particle misidentification.
The fitted values of the scale parameter show that the RQMD
transport model overestimates the $r^*$-separations of the
pion and proton
emitters by 10-20$\%$ thus indicating an underestimation
of the collective flow in this model.

The shape of the correlation function is less influenced by large $r^*$-separations
in the case of two-particle systems with the absent Coulomb FSI,
e.g. in the case of $p\Lambda$ system.
The data on $p\Lambda$ correlations in heavy ion collisions show
a significant enhancement at low relative momentum, consistent with the known
singlet and triplet $p\Lambda$ $s$-wave scattering lengths. In fact, the fits using the
analytical expression for the correlation function \cite{ll82} yield
the Gaussian correlation radii of 3--4 fm in agreement with the radii obtained
from $pp$ correlations in the same experiments.
These radii are smaller than those obtained from two-pion and two-kaon
correlation functions at the same transverse momenta \cite{star06} and are in qualitative
agreement with the approximate $m_t$ scaling expected in the case of the collective
expansion, see (\ref{rz}) and (\ref{ry}).

The correlation function of non-identical particles, compared
with the identical ones, contains a principally new piece of information on the relative
space-time asymmetries in particle emission \cite{llen96}.
Since this information enters in the
two-particle FSI amplitude through the terms odd in
${\bf k}^*{\bf r}^* \equiv {\bf p}_1^*({\bf r}_1^*- {\bf r}_2^*)$,
it can be accessed studying the correlation functions ${\cal R}_{+i}$ and
${\cal R}_{-i}$ with positive and negative projection $k_i^*$ on a given direction
${\hat i}$ or, -- the ratio ${\cal R}_{+i}/{\cal R}_{-i}$. For example, ${\hat i}$ can be the
direction of the pair velocity or, any of the out (x), side (y), longitudinal (z)
directions. In LCMS, we have $r_i^* = r_i$, except for
$r_x^* \equiv \Delta x^* =  \gamma_t(\Delta x -v_t\Delta t)$,
where $\gamma_t$ and $v_t$ are the pair LCMS Lorentz factor and velocity.
One may see that the asymmetry in the out (x) direction depends on both space
and time asymmetries $\langle\Delta x\rangle^{\mbox{\tiny SL}}$
and $\langle\Delta t\rangle^{\mbox{\tiny SL}}$.
In case of a dominant Coulomb FSI, the intercept of the correlation function ratio
is directly related with the asymmetry
$\langle r_i^*\rangle^{\mbox{\tiny SL}}$ scaled by the pair Bohr radius $a$:
\begin{equation}
{\cal R}_{+i}/{\cal R}_{-i} \approx  1+2 \langle r_i^*\rangle^{\mbox{\tiny SL}} /a.
\end{equation}

It appears that the out correlation asymmetries
between pions, kaons and protons observed in heavy ion collisions at CERN
and BNL are in agreement with practically charge independent meson production
and, assuming $m_1 < m_2$, with a negative
$\langle\Delta x\rangle^{\mbox{\tiny SL}}=\langle x_1-x_2\rangle^{\mbox{\tiny SL}}$
and/or positive
$c\langle\Delta t\rangle^{\mbox{\tiny SL}}=c\langle t_1-t_2\rangle^{\mbox{\tiny SL}}$
on the level of several fm \cite{led04,star03}.
In fact they are in quantitative agreement with the
RQMD transport model as well as with the hydro-motivated blast wave parametrization,
both predicting the dominance of the spatial part of the asymmetries generated by
large transverse flows.

In the thermal approach, the mean thermal velocity is smaller
for heavier particle and thus washes out the positive spatial
shift due to the flow to a lesser extent. As a result,
$\langle x_\pi\rangle^{\mbox{\tiny SL}} < \langle x_K\rangle^{\mbox{\tiny SL}}
 < \langle x_p\rangle^{\mbox{\tiny SL}}$.
The observation of the correlation asymmetries in agreement with
the mass hierarchy of the shifts in the out direction
may thus be considered as one of the most direct signals of a universal
transversal collective flow \cite{led04}.
This is in contrast with the effect of
$m_t$ scaling of the correlation radii which
can be also explained by a large transverse
temperature gradient like in the Buda-Lund model \cite{ccbs04}.

\subsubsection{Correlation measurement of strong interaction}

One can also use the correlation measurements to improve knowledge of the strong
interaction for various two-particle systems.
In the collisions involving sufficiently heavy nuclei,
the effective radius $r_0$ of the emission region can be
considered much larger than the range of the
strong interaction potential.
The FSI contribution is then independent of the actual
potential form \cite{gkll86}.
At small $Q=2 k^*$ and a given total spin $S$,
it is determined by the s-wave scattering amplitude $f^S(k^*)$
\cite{ll82}.
In case of $|f^S|>r_0$, this contribution is of the order of
$|f^S/r_0|^2$ and dominates over the effect of QS.
In the opposite case,
the sensitivity of the correlation function to the scattering
amplitude is determined by the linear term $f^S/r_0$.

The possibility of the correlation measurement
of the scattering amplitudes has been demonstrated \cite{led04}
in a preliminary analysis of the
NA49 $\pi^+\pi^-$ correlation data within
the RQMD transport model.
For this, besides the $r^*$-scale $s_{r}$,
the strong interaction scale $s_f$
has been introduced in the RQMD correlation function
${\cal R}(Q;s_{r},s_f)$,
rescaling the original s-wave $\pi^+\pi^-$
scattering amplitude taken from \cite{nag79}:
$f(k^*) \rightarrow s_f f(k^*)$;
it approximately corresponds to the rescaling
of the original scattering length $f_0=$ 0.232 fm.
The fitted parameter $s_f=0.63\pm 0.08$ appears to be
significantly lower than unity. To a similar
but somewhat weaker rescaling ($\sim 0.8$)
point also the preliminary result of experiment DIRAC
on the pionium lifetime \cite{ade05}, the
BNL and CERN data on $K_{l4}$ \cite{pis01}
and $K^\pm\to \pi^\pm\pi^0\pi^0$ \cite{na48} decays
as well as the two-loop calculation
in the chiral perturbation theory with a standard value of the
quark condensate \cite{cgl01}.

Comparing the fit results with the theoretical predictions,
one should have in mind
that the latter are subject to the electro-magnetic corrections on the level
of several percent and that the correlation measurement may underestimate
$f(k^*)$ by a few percent due to the use of the equal-time
approximation.
A substantial systematic error can also arise from a simplified
fit of the strong FSI amplitude.
To avoid the latter, one can use the Roy equations and represent
the $\pi^+\pi^-$ strong interaction amplitude at low energies as
a unique function of the isoscalar and isotensor s-wave scattering
lengths $a_0^0$ and $a_0^2$, see Appendix D in \cite{ana00}.
The two-parameter dependence of the scattering amplitude can be
further reduced to a single-parameter one within the generalized chiral
perturbation theory predicting a strong correlation between the two
s-wave scattering lengths (see equation (13.2) in \cite{cgl01}).
The systematic error due to the uncertainty in the fitted
$r^*$-distribution (e.g., in the scale parameter $s_{r}$)
can be diminished in a simultaneous analysis of $\pi^+\pi^-$
and $\pi^\pm\pi^\pm$ correlation functions.
The high statistics DIRAC data
on two-pion correlations may thus allow one to determine
the s-wave scattering lengths $a_0^0$ and $a_0^2$ better than
to $10\%$ and serve as complementary to the pionium lifetime
measurement in the same experiment.

The correlation technique was also used to estimate the
singlet $\Lambda\Lambda$ $s$-wave scattering length
based on the fits of the $\Lambda\Lambda$ correlation data from Pb+Pb collisions
at 158 $A$ GeV \cite{led04}.
Though the fit results are not very restrictive, they likely exclude the possibility
of a large singlet scattering length comparable to that of
$\sim 20$ fm for the two-nucleon system.
Similarly, the fit of the $p\bar{\Lambda}$ and
$\bar{p}\Lambda$ correlation functions measured in
Au+Au collisions at $\sqrt{s}_{\rm NN}=200$ GeV allowed one to determine
the corresponding spin-averaged $s$-wave scattering length.
The fitted imaginary part of the scattering length of $\sim 1$ fm is in agreement with
the $\bar{p}p$ results (thus pointing to about the same $\bar{p}\Lambda$ and
$\bar{p}p$ annihilation cross sections) while the real part appears to be more
negative \cite{star06a}.

\section{One-channel wave functions}

\subsection{Continuous spectrum}

\subsubsection{Short-range FSI}
Let us start with the case when the two-particle
FSI is due to the short-range forces only.
In the considered region of small ${k}^{*}$ the short-range
particle interaction is dominated by s-waves. Since the radius
of the s-wave interaction is usually small compared with the
distance ${r}^{*}$ between the production points of
particles $1$ and $2$ in their c.m. system, the FSI effect is mainly
determined by the asymptotic behaviour of the scattered wave
outside the region of the strong interaction ${r}^* > d$:
\begin{equation}
\label{11}
\Delta\psi_{-{\bf k}^*}({\bf r}^*)=
f({k}^{*}){\rm e}^{{\rm i}{k}^{*}
{r}^{*}}/{r}^{*}.
\end{equation}
The s-wave amplitude $f$ depends
on the magnitude of the vector ${\bf k}^*$ only.
Assuming the absence of inelastic transitions,
it satisfies the
one-channel s-wave unitarity condition
$\Im f= k^*|f|^2$ or, equivalently $\Im f^{-1}= -k^*$,
and so can be represented as
\begin{equation}
\label{ampl0}
f= [\exp(2{\rm i}\delta_0)-1]/(2{\rm i}k^*)=
\left(K^{-1}-{\rm i}k^*\right)^{-1},
\end{equation}
where $\delta_0$ is the s-wave phase shift and
$K^{-1}=k^*\cot\delta_0$ is a real function of $k^*$.
Usually (for potentials vanishing with the distance
exponentially or faster), this function is real also for
negative kinetic energies $k^{*2}/(2\mu)$, so that its
expansion can contain only even powers of $k^*$ \cite{Lan77}.
Retaining near threshold only the first two terms
in the expansion, one can express
the function $K^{-1}$ or $K$ through the
corresponding two parameters: scattering length
$f_0$ and effective range $d_0$ or curvature $b_0$:
\begin{equation}
\label{K0}
K^{-1}\doteq f_0^{-1}+\frac12 d_0 k^{*2}\qquad
K\doteq f_0+b_0 k^{*2}\qquad b_0 \doteq -\frac12 d_0f_0{}^2.
\end{equation}
The expansion of $K^{-1}$ is superior for two-nucleon
systems (due to large scattering lengths, amounting to
about 20 fm in the singlet case) while for other systems,
the $K$-expansion is often preferred. To extend the latter
to a wider energy range, it is usually written in a
relativistic form and additional parameters are added.
For example \cite{cgl01}:
\begin{equation}
\label{K1}
K=\frac{2}{\sqrt{s}}\frac{s_{\rm th}-s_0}{s-s_0}
\sum_{j=0}^3 A_j x^{2j}\qquad x=2k^*/\sqrt{s_{\rm th}},
\end{equation}
where $s=(p_1+p_2)^2=(\omega_1^*+\omega_2^*)^2$,
$\omega_{1,2}^*=(m_{1,2}^2+k^{*2})^{1/2}$ and
$s_{\rm th}=(m_1+m_2)^2$.
The parameter $s_0$ takes into account the eventual
resonance, specifying the value
of the two-particle invariant mass
squared where the phase $\delta_0(k^*)$ passes
through $90^{o}$.
{The behaviour of the s-wave $K$-function in a wide
$k^*$-interval is however of minor importance
since we are interested in the near-threshold
region and, have already neglected the p-wave correction
$\Or(k^{*2}a_1/r^*)$
in (\ref{11}); here $a_1$ is a p-wave scattering length.
For $\pi^+\pi^-$ system, $a_1 \approx 0.1~\mbox{fm}^3$,
$f_0\approx 0.2~\mbox{fm}$, $d_0\approx -10~\mbox{fm}$ and
the relative p-wave contribution to
the $k^{*2}$ term due to the short-range FSI composes
in the production cross section
$\sim a_1/(a_1-d_0f_0{}^2/2-f_0{}^3/3)\sim 30\%$;
the relative contribution of the $k^{*2}$ term
$\sim (a_1-d_0f_0{}^2/2-f_0{}^3/3)k^{*2}/f_0$ being
less than $1\%$ of the total short-range FSI contribution
for $Q=2k^*< 30$ MeV/$c$.
}

Note that the extension of the asymptotic wave function in the
inner region leads to a relative shift in the production cross
section of the order
$ |f|^2\frac{{\rm d}}{{\rm d}k^{*2}}{\rm Re}(1/f)/
(\langle {r}^{*}\rangle^{\mbox{\tiny SL}})^3$
\cite{ll82,lll97}.
The leading part of this shift can be, in principle, corrected
for (see subsection 6.3). However, being quadratic in the amplitude
$f$, it is rather small for $\pi\pi$-, $\pi K$- or
$\pi p$-systems -- usually not exceeding several percent of the
short-range FSI contribution.

\subsubsection{Account of the Coulomb FSI}
\begin{figure}
\epsfig{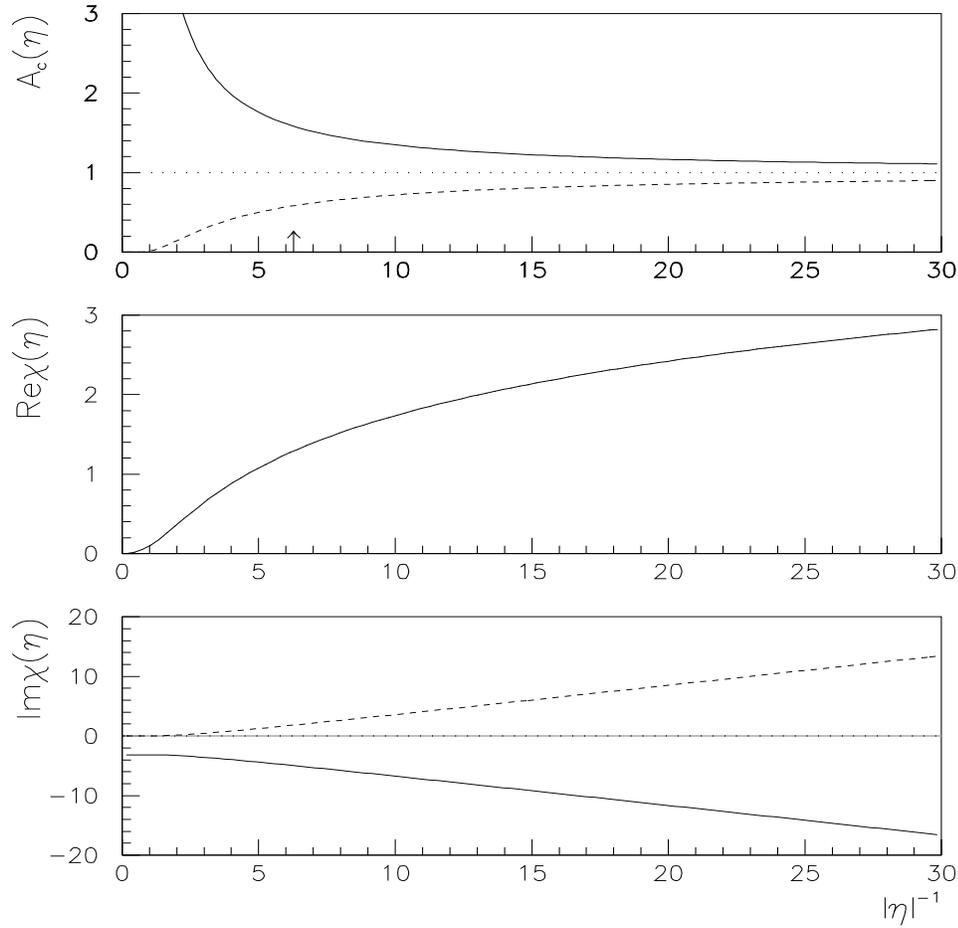}
\caption{The functions $A_c(\eta)$,
$\Re\chi(\eta)\equiv h(\eta)$  and
$\Im\chi(\eta)=A_c(\eta)/(2\eta)$
defined in (\ref{Ac}), (\ref{chi}) and (\ref{h}).
The solid and dash curves correspond to the
attraction ($\eta<0$) and repulsion ($\eta>0$) respectively.
For two-pion systems, the variable $|\eta|^{-1}\equiv |ak^*|$
approximately coincides with the relative three-momentum $Q=2k^*$
in MeV/$c$: $|\eta|^{-1}\doteq 0.98 Q/(\mathrm{MeV}/c)$.
The arrow in the first panel indicates the characteristic
width $|\eta|^{-1}=2\pi$ of the Coulomb effect.}
\label{figachi}
\end{figure}

Similar to the case of neutral particles,
we will approximate (with the same accuracy)
the wave function of two charged particles near threshold
by the asymptotic solution outside the
region of the strong interaction $r^*>d$.
It is well known that the long-range
Coulomb interaction modifies both the plane and spherical waves
\cite{Lan77}:
\begin{equation}
\label{psicoul}
\psi_{-{\bf k}^*}({\bf r}^*)={\rm e}^{{\rm i}\delta_c}
\sqrt{A_c(\eta)}\left[{\rm e}^{-{\rm i}{\bf k}^{*}{\bf r}^{*}}
F(-{\rm i}\eta,1,{\rm i}\xi)+f_c({k}^{*})
\frac{\widetilde{G}(\rho,\eta)}{{r}^{*}}
\right],
\end{equation}
where $\xi={\bf k}^{*}{\bf r}^{*}+{k}^{*}{r}^{*}\equiv
\rho(1+\cos\theta^*)$,
$\rho={k}^{*}{r}^{*}$, $\eta=({k}^{*}a)^{-1}$,
$a=(\mu z_1z_2e^2)^{-1}$ is the two-particle Bohr radius
including the sign of the interaction (see table \ref{brad}),
$\delta_c={\rm arg}\Gamma(1+{\rm i}\eta)$ is the Coulomb s-wave
phase shift,
\Table{\label{brad}
    The pair Bohr radius including the
    sign of the interaction, $a=(\mu z_1z_2e^2)^{-1}$, and
    the characteristic width
    of the Coulomb correlation effect,
    $Q_c\equiv 2k^*_c=4\pi/|a|$, corresponding
    to $|\eta|^{-1}=2\pi$ (see (\ref{Ac}) and the first
    panel in figure~\ref{figachi}).
    }
      \br
        Pair &
        $\pi^+\pi^{\pm}$&
        $\pi^+ K^{\pm}$&
        $\pi^{\pm} p$&
        $K^+K^{\pm}$ &
        $K^{\pm} p$ &
        $p p{\pm}$
        \\[0.1cm]
        \mr
$a,$ fm  & $\pm 387.5$ & $\pm 248.6$ & $\pm 222.5$ & $\pm 109.6$
& $\pm 83.6$ &$\pm 57.6$\\
$Q_c,$ MeV/$c$  & ~6.4 & 10.0 & 11.1 & 22.6 & 29.7 & 43.0\\
\br
\endTable
$A_c(\eta)$ is the Coulomb penetration factor defined in
(\ref{Ac}),
\begin{equation}
\label{F}
F(\alpha,1,z)=1+\alpha z/1!^2+\alpha(\alpha+1)z^2/2!^2+\cdots
\end{equation}
is the confluent hypergeometric function and
$\widetilde{G}=\sqrt{A_c}(G_0+{\rm i}F_0)$ is a combination of the
regular ($F_0$) and singular ($G_0$) s-wave Coulomb functions
(see, e.g., \cite{gkll86}):
\begin{eqnarray}
\label{GST}
\widetilde{G}(\rho,\eta)=P(\rho,\eta)+2\eta\rho\left[
\ln|2\eta\rho|+2C-1+\chi(\eta)\right]
B(\rho,\eta).
\end{eqnarray}
Here $C\doteq 0.5772$ is the Euler constant, the functions
\begin{eqnarray}
\label{BP0}
B(\rho,\eta)&=&\sum_{s=0}^{\infty}B_s\qquad B_0=1\qquad B_1=
\eta\rho\qquad
\dots
\nonumber \\
P(\rho,\eta)&=&\sum_{s=0}^{\infty}P_s\qquad P_0=1\qquad P_1=0\qquad
\dots
\end{eqnarray}
are given by the following recurrence relations:
\begin{eqnarray}
\label{BP}
(n+1)(n+2)B_{n+1}&=&2\eta\rho B_n-\rho^2B_{n-1}
\nonumber \\
n(n+1)P_{n+1}&=&2\eta\rho P_n-\rho^2P_{n-1}-
(2n+1)2\eta\rho B_n,
\end{eqnarray}
$B(\rho,\eta)\equiv F_0/(\rho \sqrt{A_c})
\rightarrow \sin(\rho)/\rho$ and
$P(\rho,\eta)\rightarrow \cos(\rho)$
in the limit $\eta\rho\equiv r^*/a\rightarrow 0$.
The function
\begin{equation}
\label{chi}
\chi(\eta)=h(\eta)+{\rm i}A_c(\eta)/(2\eta),
\end{equation}
where the function $h(\eta)$ is expressed through the
digamma function
$\psi(z)=\Gamma'(z)/\Gamma(z)$:
\begin{equation}
\label{h}
h(\eta)=\left[\psi(i\eta)+\psi(-{\rm i}\eta)-\ln(\eta^2)\right]/2.
\end{equation}
For $|\eta|<0.3$ the function $h(\eta)
\doteq 1.2\eta^2-\ln|\eta|-C$,
while at large $|\eta|$ this function can be represented
by a truncated series in inverse powers of $\eta^2$:
$h(\eta)=\eta^{-2}/12+ \eta^{-4}/120+\cdots$.
The amplitude
\begin{equation}
\label{fc1}
f_c({k}^{*})=f(k^*)/A_c(\eta),
\end{equation}
where
$f({k}^{*})$
is the amplitude of the low-energy s-wave elastic scattering
due to the short-range interaction renormalized by the
long-range Coulomb forces. Assuming again the absence of
inelastic transitions, the amplitude
$f({k}^{*})=({\rm e}^{2{\rm i}\delta_0}-1)/(2{\rm i}{k}^{*})$
and satisfies the one-channel s-wave unitarity condition.
Similar to the case of neutral particles, one then has
\cite{Lan77}:
\begin{equation}
\label{fc}
f_c({k}^{*})=
\left (K^{-1}-
\frac{2\chi(\eta)}{a}\right )^{-1},
\end{equation}
where the function $K$ can be parametrized according to (\ref{K0}) or
(\ref{K1}).

Note that $\delta_c\rightarrow 0$, $A_c\rightarrow 1$ for
$\eta\rightarrow 0$ (${k}^{*}\gg |a|^{-1}$) and
$\widetilde{G} \rightarrow {\rm e}^{{\rm i}\rho}$,
$F\rightarrow 1$
for $\eta\rho\equiv r^*/a \rightarrow 0$.
So, the  two-particle wave function in the absence of the
long-range Coulomb forces is recovered provided $r^*$, $f_0$
and $1/k^*$ are much smaller than the Bohr radius $|a|$.

In figure~\ref{figachi}, we plot $A_c(\eta)$ and $\chi(\eta)$
as functions of the variable $|\eta|^{-1}=|ak^*|$.
For the system of two charged pions, this variable
approximately corresponds to $Q=2k^*$ in MeV/$c$.
At $k^*\rightarrow 0$, the Coulomb
penetration factor $A_c(\eta)$ respectively tends to 0
and $\infty$ for like and unlike particle charges.
With the increasing $k^*$, this factor slowly approaches
unity:
$A_c(\eta)\approx 1-\pi\eta$ for $k^* > 2\pi/|a|$.
Note that the quadratic behaviour of
$\Re\chi(\eta)\equiv h(\eta)\approx \eta^{-2}/12$ at
$|\eta|^{-1}<1$ is changed by a steep quasi-linear rise
in the interval $1 < |\eta|^{-1} < 5$; the corresponding
slope being about 0.26.
As for $\Im\chi(\eta)\equiv A_c(\eta)/(2\eta)$, at $k^*=0$
it equals to 0 and $-\pi$
for like and unlike charges respectively, and,
for $k^* > 2\pi/|a|$, it
approaches the linear $k^*$-dependence:
$\Im\chi(\eta)\approx (\eta^{-1}-\pi)/2$.

\subsubsection{The small- and large-$r^*$ limits}
Since we are interested in the region of small relative
distances ${r}^{*}$ compared with the Bohr radius $|a|$
and small relative momenta $Q=2k^*$
compared with $1/r^*$, it is
useful to write the first terms in the expansion of the
hypergeometric functions
$F$ and $\widetilde{G}$ in ${r}^{*}/a$ and
$\rho\equiv {k}^{*}{r}^{*}$. We have ($x=\cos\theta^*$):
\begin{equation}
\label{FGexp}
\eqalign{
F(-{\rm i}\eta,1,{\rm i}\xi)=1+\frac{{r}^{*}}{a}(1+x)
\cr
\cdot
\left[1+\frac{{\rm i}\rho}{4}(1+x)-\frac{\rho^2}{18}(1+x)^2
+\Or(\rho^3)\right]+\Or
\left((\frac{{r}^{*}}{a})^2\right)
\cr
\widetilde{G}(\rho,\eta)=
1-\frac{\rho^2}{2}+2\frac{{r}^{*}}{a}
\cr
\cdot
\left[
\ln|2\frac{{r}^{*}}{a}|+2C-1+\chi(\eta)\right]
\left(1-\frac{\rho^2}{6}\right)+ \Or(\rho^4)+
\Or\!\left((\frac{{r}^{*}}{a})^2\right).
}
\end{equation}
For some systems of interest
($\pi\pi$, $\pi K$, $\pi p$),
$|f_0|^2< |f_0d_0|< \sim m_\pi{}^{-2} \ll r^{*2}$,
one can neglect the $Q$-dependence of the scattering
amplitude and, after the averaging over the uniform
$x$-distribution,
write the correlation function at a fixed separation $r^*$ as
\begin{equation}
\label{cfapp}
\eqalign{
{\cal R}(k^*;r^*)\equiv \langle|\psi_{-{\bf k}^*}({\bf r}^*)|^2\rangle
\cr
=A_c(\eta)\left\langle|F|^2+
2\Re\!\left({\rm e}^{{\rm i}{\bf k}^*{\bf r}^*}F^*\widetilde{G}
\frac{f_0}{r^*}\right)+
\Or\!\left((\frac{f_0}{{r}^{*}})^2\right)
\right\rangle
\cr
=A_c(\eta)\left\{1+2\frac{r^*}{a}+2\frac{f_0}{r^*}+
2\frac{f_0}{a}\left[1+2\left(\ln|\frac{2r^*}{a}|+
2C-1+h(\eta)\right)\right]\right.
\cr
\left.~~~~~~~
-\rho^2\left(\frac29\frac{r^*}{a}+
\frac43\frac{f_0}{r^*}\right)
+\Or\!\left((\frac{f_0}{{r}^{*}})^2\right)+
\Or\!\left((\frac{{r}^{*}}{a})^2\right)+
\Or(\rho^4)\right\}.
}
\end{equation}
Note that in the case of an anisotropic ${\bf r}^*$-distribution,
Eq. (\ref{cfapp}) implies the integration over the direction of the
vector ${\bf k}^*={\bf Q}/2$, distributed isotropically
for non-correlated particles at $Q\rightarrow 0$.
In the case of the cut $Q_T<Q_T^{\rm cut}$ on the component
of the vector ${\bf Q}$ transverse to the direction
of the pair three-velocity $\bf v$ and $Q>Q_T^{\rm cut}$,
Eq. (\ref{cfapp}) should be modified by the substitution
$2/9\to g_{\rm cut}2/9$,
\begin{equation}
\label{subst29}
g_{\rm cut}=
1+\frac{1}{2}\left\langle \frac{1}{2}(3\cos^2\theta_{r^*}-1)
\right\rangle (c_{\rm min}+c_{\rm min}^2)
\in \left(0.5,2\right),
\end{equation}
where $\theta_{r^*}$
is the angle between the vectors ${\bf r}^*$ and $\bf v$, and
$c_{\rm min}=[1-(Q_T^{\rm cut}/Q)^2]^{1/2}$ is the
minimal absolute value of the cosine of the angle
between the vectors ${\bf k}^*$ and ${\bf v}$.
For pion pairs containing pions from
resonance decays, one may expect
$\langle r_L^2\rangle>\langle r_T^2\rangle$ \cite{lp92}
(i.e. $\langle\cos^2\theta_{r^*}\rangle>1/3$)
and so $g_{\rm cut}>1$.

\begin{figure}
\epsfig{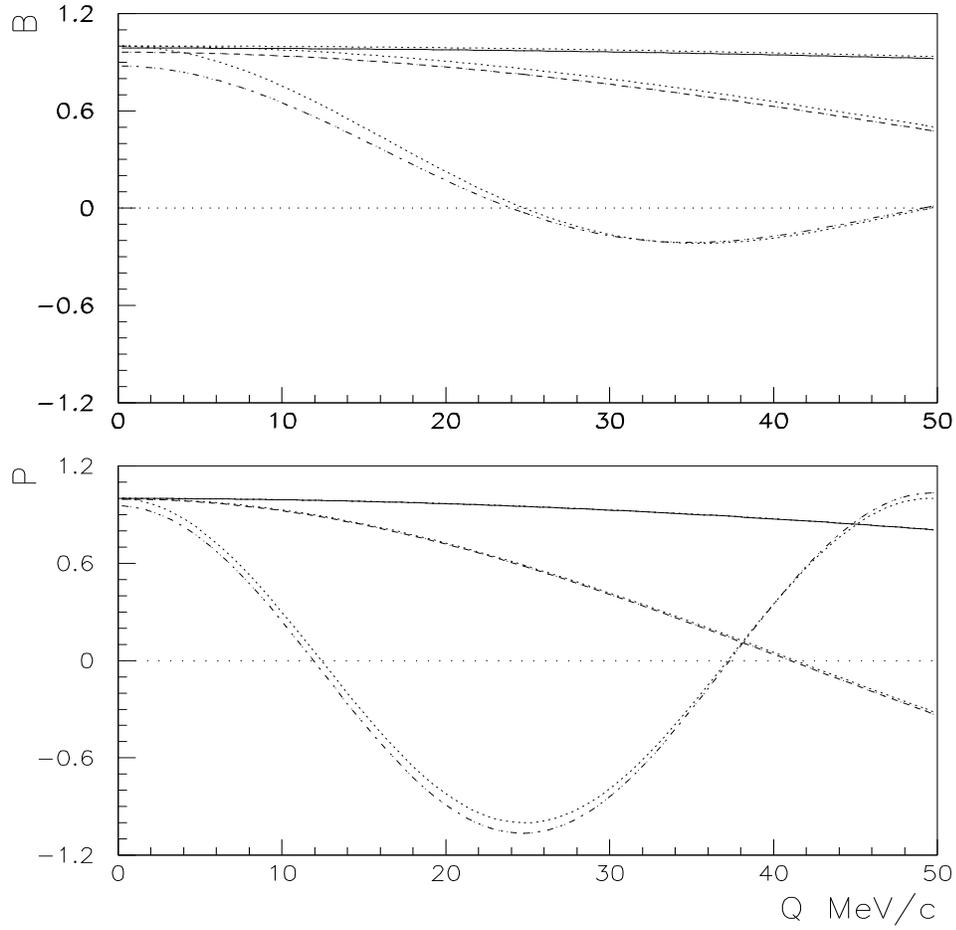}
\caption{The functions $B(\rho,\eta)$ and $P(\rho,\eta)$
defined in (\ref{BP0}), (\ref{BP}) and calculated for
the $\pi^+\pi^-$ system.
The solid, dash and dash-dotted curves correspond to
$r^*=$ 5, 15 and 50 fm respectively.
The dotted curves represent the functions
$B(\rho,0)=\sin\rho/\rho$ and $P(\rho,0)=\cos\rho$
corresponding to the case of neutral particles.}
\label{figbp}
\end{figure}

\begin{figure}
\epsfig{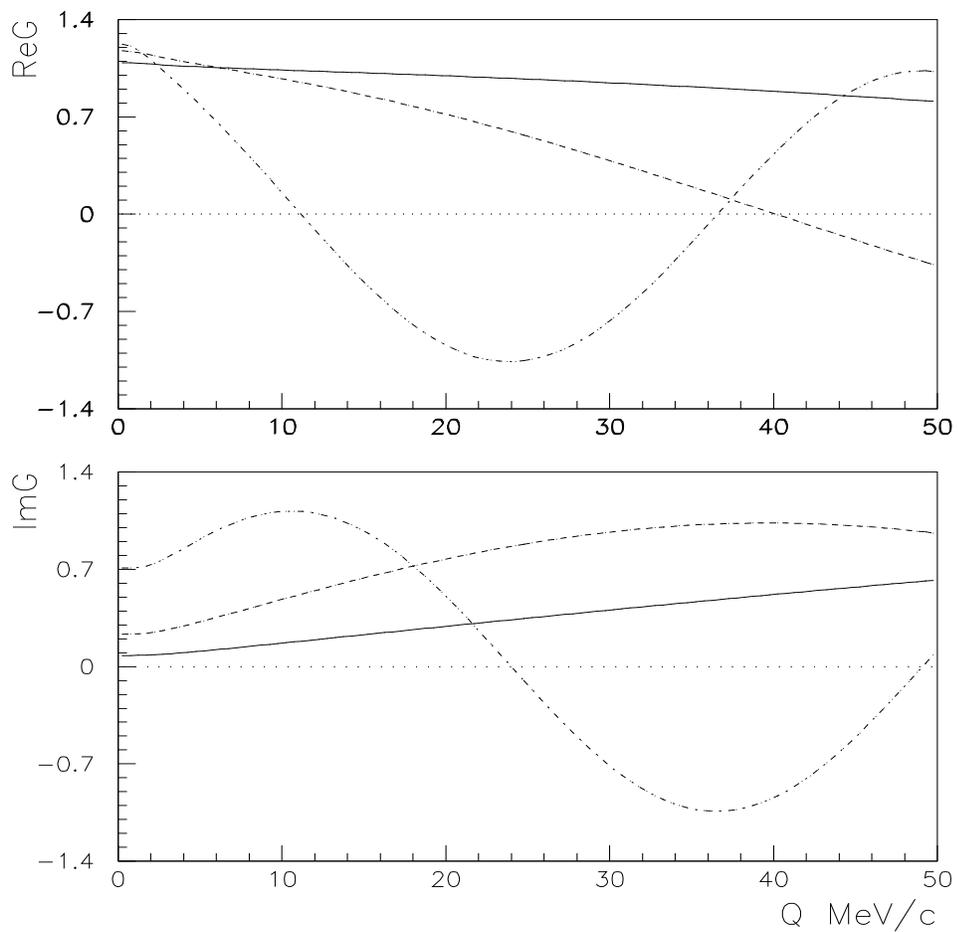}
\caption{The function $\widetilde{G}(\rho,\eta)$
defined in (\ref{GST}) and calculated for
the $\pi^+\pi^-$ system.
The solid, dash and dash-dotted curves correspond to
$r^*=$ 5, 15 and 50 fm respectively.}
\label{figgst}
\end{figure}

In figures~\ref{figbp}, \ref{figgst} and \ref{figgst1},
we show the $Q$-dependence of the functions
$B(\rho,\eta)$, $P(\rho,\eta)$,
$\widetilde{G}(\rho,\eta)$ and the reduced
correlation function ${\cal R}/A_c$
as well as
the corresponding main contributions due to the interference
term and the modulus squared of the hypergeometric function
for the $\pi^+\pi^-$ system at $r^*=$ 5, 15, 50 fm.
One may see that the almost universal quasilinear
decrease of ${\cal R}/A_c$ for $r^* < \sim 20$ fm
is due to the interference term, and that it is changed,
for higher $r^*$-values, by a steep rise due to the
$|F|^2$-term.
It appears that
the linear fit of ${\cal R}/A_c$ recovers the intercept better
than to 2 per mil for $r^* < \sim 20$ fm and -- better than
to 2 percent even for $r^*=50$ fm (see table~\ref{cffit}).

To clarify the origin of the quasilinear behaviour of the
reduced correlation function ${\cal R}/A_c$, one can use
(\ref{cfapp}) to estimate the slope at small $Q$:
\begin{eqnarray}
\label{cfder}
\left(\frac{{\cal R}}{A_c}\right)'\equiv\frac{{\rm d}}{{\rm d}Q}
\left(\frac{{\cal R}}{A_c}\right)\doteq
\pm 2f_0\frac{{\rm d}h}{{\rm d}|\eta|^{-1}}
-\left(\pm\frac19\frac{r^*}{|a|}+
\frac23\frac{f_0}{r^*}\right)r^{*2}Q,
\end{eqnarray}
where the sign $+$ ($-$) corresponds to the Coulomb repulsion
(attraction).
Using the fact that ${\rm d}h/{\rm d}|\eta|^{-1}\approx 0.26$ for
$1 < |\eta|^{-1} < 5$, one has
$({\cal R}/{A_c})'\approx -(0.6 + bQ)$ (GeV/$c$)$^{-1}$
for the $\pi^+\pi^-$-system
at $1 < Q < 5$ MeV/$c$ and $f_0=0.232$ fm, where
$b$ is small ($b< \sim 0.03$ (MeV/$c$)$^{-1}$) and positive
for $r^*< \sim 20$ fm and, for larger $r^*$-values, $b$ is
negative and its magnitude rapidly increases with $r^*$.
As a result, the slope
of the reduced $\pi^+\pi^-$ correlation function is negative
in this $Q$-interval
and nearly constant for small $r^*$-values,
while it becomes positive and rapidly increases with $Q$
for $r^*$-values of several tens of fm or larger.
For $Q > 5$ MeV/$c$,
the absolute value of the slope due to the $h$-function
decreases as $\sim 2.35|\eta|$. It appears that,
for the $\pi^+\pi^-$-system at $r^* < \sim 20$ fm,
this decrease is approximately compensated
by the $Q$-dependence of the functions $B$, $P$ and $F$,
(i.e., at $\rho\ll 1$, by the second term in (\ref{cfder})),
so that $({\cal R}/{A_c})'\approx -0.5$ (GeV/$c$)$^{-1}$ up to
$Q=50$ MeV/$c$.

Note that the $Q_T$-cut substantially influences the $Q$-dependence
of the reduced correlation function only for
sufficiently large values of $r^*/|a|$, leading to the substitution
${r^*}/{|a|}\to g_{\rm cut} {r^*}/{|a|}$
in (\ref{cfder}) at $Q>Q_T$ . Particularly, for $\pi^+\pi^-$ pairs
containing an $\omega$-decay pion, one may expect $g_{\rm cut}>1$
and so a more steep rise of the reduced correlation function
at $Q>Q_T$.

\Table{\label{cffit}
Results of the linear fits of the
reduced $\pi^+\pi^-$ correlation function:
${\cal R}/A_c=c_0+c_1 Q$
in different intervals $0<Q<Q_{\max}$.
The function ${\cal R}/A_c$ is calculated at
$r^*=$ 5, 15 and 50 fm
in the approximation of a constant
scattering amplitude $f_c(k^*)=f_0= 0.232$ fm and, assuming
the uniform distribution of the cosine
of the angle between the vectors ${\bf r}^*$ and
${\bf k}^*={\bf Q}/2$.
Also shown are the corresponding values of ${\cal R}/A_c$
at $Q=0$ (the intercepts).}
      \br
        $r^*$, fm&
        Intercept&
        $Q_{\max}$, MeV/$c$&
         10 & 20 & 30 & 40 & 50
        \\
        \mr
 5 &1.077 &$c_0$  & 1.077& 1.077& 1.077& 1.077& 1.078 \\
   &      &$c_1$, (GeV/$c$)$^{-1}$ &-0.55& -0.47& -0.48& -0.52& -0.57
\\
\mr
15 &0.961 &$c_0$  & 0.961& 0.961& 0.961& 0.960& 0.959 \\
   &      &$c_1$, (GeV/$c$)$^{-1}$ &-0.59& -0.56& -0.55& -0.51& -0.42
\\
\mr
50 &0.783 &$c_0$  & 0.778& 0.768& 0.766& 0.773& 0.783 \\
   &      &$c_1$, (GeV/$c$)$^{-1}$ &2.55& 4.61& 4.99& 4.38& 3.69
\\
\br
\endTable

To estimate the behaviour of the correlation function
at large $r^*$ or $k^*$, one can exploit the known asymptotic
expressions for hypergeometric functions.
Thus, at $\xi\gg 1+\eta^2$,
\begin{equation}
\label{Fexp2}
\sqrt{A_c(\eta)}F(-{\rm i}\eta,1,{\rm i}\xi) \rightarrow
\left(1-{\rm i}\frac{\eta^2}{\xi}\right)
{\rm e}^{{\rm i}(-\delta_c+\eta\ln\xi)}
+\frac{\eta}{\xi}{\rm e}^{{\rm i}(\delta_c+\xi-\eta\ln\xi)}
\end{equation}
and, at $\rho\gg 1+\eta^2$,
\begin{equation}
\label{Gexp2}
\widetilde{G}(\rho,\eta) \rightarrow \sqrt{A_c(\eta)}
{\rm e}^{{\rm i}(\delta_c+\rho-\eta\ln 2\rho)},
\end{equation}
so that both the effects of the Coulomb and strong FSI
vanish in the cross section as ${r}^{*-2}$.
In fact, the asymptotic expression for the $F$-function
in (\ref{Fexp2}) cannot be used in the case of nearly
opposite directions of the vectors ${\bf k}^*$ and
${\bf r}^*$ ($\cos\theta^*\approx -1$) when the variable
$\xi=\rho(1+\cos\theta^*)$
is suppressed even at large $\rho={k}^{*}{r}^{*}$.
This leads, after averaging over the angles,
to a slower vanishing of the Coulomb effect,
as ${r}^{*-1}$, in agreement with the classical
Jacobian factor
$[1-2/(a{r}^{*}{k}^{*2})]^{1/2}\approx
1-(a{r}^{*}{k}^{*2})^{-1}$.

\subsection{Discrete spectrum}
\subsubsection{General s-wave solution}
Since the Schr\"odinger equation at a small negative energy
$-\epsilon_b=-\kappa^2/(2\mu)$ practically coincides with
that in continuous spectrum at zero energy,
the ${r}^{*}$-dependence of the corresponding wave functions
at given orbital angular momentum $l$ and
${r}^{*}\ll \kappa^{-1}$ is the same.
{
This important conclusion was first stated by Migdal for the
$pn$-system \cite{mig55}.
}
In fact, both solutions (at positive and negative energies)
can be written in the same form for any ${r}^{*}$,
up to an energy dependent normalization factor ${\cal N}$.
Outside the region of the short-range interaction,
${r}^{*}>d$, we can write the s-wave solution as a combination
of the regular and singular Coulomb functions:
\begin{equation}
\label{psigen}
\psi_{l=0}({r}^{*})={\cal N}(\eta)\left[
\frac{F_0(\rho,\eta)}{\rho\sqrt{A_c(\eta)}}
+f_c(k^*)\frac{\widetilde{G}(\rho,\eta))}{{r}^{*}}\right].
\end{equation}
At $d<{r}^{*}\ll |a|$ and $|\rho|\ll 1$ it takes on the form:
\begin{equation}
\label{psigenA}
\eqalign{
\psi_{l=0}({r}^{*})
={\cal N}\left\{(1+\frac{{r}^{*}}{a})+
\Or\!\left((\frac{{r}^{*}}{a})^2\right)+\Or(\rho^2)
+\frac{f_c}{{r}^{*}}
~~~~~~~~~~~~~~~~~~~~ \right.
\cr
\left.
\cdot
\left[1+2\frac{{r}^{*}}{a}\left(
\ln|2\frac{{r}^{*}}{a}|+2C-1+{\chi}\right)
(1+\frac{{r}^{*}}{a})+
\Or\!\left((\frac{{r}^{*}}{a})^2\right)+\Or(\rho^2)
\right]\right\}.
}
\end{equation}
For positive energies, ${\cal N}={\rm e}^{{\rm i}\delta_c}
\sqrt{A_c(\eta)}$
and, at ${k}^{*}\rightarrow 0$,
$f_c=f_0/[1-2f_0\chi(\pm\infty)/a]$, $\chi(+\infty)=0$
($a>0$) or
$\chi(-\infty)=-{\rm i}\pi$ ($a<0$).
In the case of opposite charges ($a<0$),
(\ref{psigenA}) yields:
\begin{equation}
\label{psi+}
\eqalign{
\psi_{{k}^{*},l=0}={\rm e}^{{\rm i}\delta_c}\sqrt{A_c}
\left\{\left(1-\frac{{r}^{*}}{|a|}\right)
\left[1-2\frac{f_0}{|a|}\left(\ln\left
|\frac{2{r}^{*}}{a}\right|+2C-1\right) \right]
+\frac{f_0}{{r}^{*}}+
\right.
\cr
\left.
+\left(1+\frac{f_0}{r^*}\right)
\left[
2{\rm i}\pi\frac{f_0}{|a|}
+ \Or\!\left((\frac{{r}^{*}}{a})^2\right)
+ \Or\!\left(\rho^2\right)
\right]
\right\}.
}
\end{equation}
For the discrete levels at negative energies, the
substitution ${k}^{*}\rightarrow i\kappa_n$ has
to be done, particularly yielding \cite{Lan77,lll97}:
\begin{equation}
\label{chi0}
\chi(\eta_n)=
\frac{\pi}{2}\cot(\frac{\pi}{\kappa_n|a|})+\frac12
\left[2\ln(\kappa_n|a|)+
\psi\!\left(\frac{1}{\kappa_n|a|}\right)+
\psi\!\left(-\frac{1}{\kappa_n|a|}\right)\right],
\end{equation}
where $\eta_n=(i\kappa_n a)^{-1}$.
A more compact form of
(\ref{chi0}) follows from the relation
$\psi(-x)=\psi(x)+\pi\cot(\pi x)+x^{-1}$:
\begin{equation}
\label{chi0'}
\chi(\eta_n)=
\pi\cot(\pi x_n)-(2x_n)^{-1}[\phi(x_n)-3]\qquad
x_n=(\kappa_n |a|)^{-1}
\end{equation}
\begin{equation}
\label{phi}
\phi(x)= 2+2x[\ln{x}-\psi(x)].
\end{equation}

\subsubsection{Energy levels}
For a pure Coulombic
atom ($a<0$, $f_0=0$), only
the solution $F_0/\rho$, regular at $r^*\rightarrow 0$,
contributes and the requirement of its exponential
damping at large distances fixes the energy levels.
The corresponding $\kappa$-values at a given principle
quantum number $n$ are equal to $\kappa_n^{c}=(n|a|)^{-1}$.
The wave functions $\psi_{nl}^{\rm coul}({r}^{*})$
can then be expressed in terms of the Laguerre polynomials
$L_{n+l-1}^{2l+1}(z)$. For $l=0$,
\begin{equation}
\label{psicapp1}
\psi_{n0}^{\rm coul}({r}^{*})=\psi_{n0}^{\rm coul}(0)
\exp\!\left(-\frac{{r}^{*}}{n|a|}\right)
L_{n-1}^1\!\left(\frac{{2r}^{*}}{n|a|}\right)
(n\cdot n!)^{-1}.
\end{equation}
The square of the wave function $\psi_{n0}^{\rm coul}(0)$
at zero separation
is given in (\ref{wfds0}) and the Laguerre polynomials
are defined by the following recurrence relations:
\begin{equation}
\label{Lag}
\eqalign{
L_{n-1}^1(z)=(n\cdot n!)\sum_{s=0}^{n-1}l_{n-1}^s(z)
\cr
l_{n-1}^0(z)=1\qquad
l_{n-1}^s(z)=-\frac{z(n-s)}{s(s+1)}l_{n-1}^{s-1}(z).
}
\end{equation}
At ${r}^{*}\ll n|a|$,
\begin{equation}
\label{psicapp1A}
\psi_{n0}^{\rm coul}({r}^{*})= \psi_{n0}^{\rm coul}(0)
\left[1-\frac{{r}^{*}}{|a|}+
\Or\!\left((\frac{{r}^{*}}{na})^2\right)\right].
\end{equation}

The strong interaction slightly shifts the Coulombic energy
levels thus making the regular part of the general solution
(\ref{psigen}) divergent at large distances. Therefore, the
amplitude $f_c$ has to have a pole at $k^*=i\kappa_n$,
and so, according to (\ref{fc}),
\begin{equation}
\label{chi1}
\chi(\eta_n)
=-\frac{|a|}{2K(i\kappa_n)}
=-\frac{|a|}{2f_0}\left[1+
\Or\!\left(\frac{f_0d_0}{(na)^2}\right)\right].
\end{equation}
Using (\ref{chi0'}) and (\ref{chi1}),
one can fix the energy levels
$E_n=-\kappa_n{}^2/(2\mu)$ in discrete
spectrum with the relative error of
$\Or\!\left(a^{-3}\right)$:
\begin{equation}
\label{kappa}
\eqalign{
\kappa_n= \kappa_n^{\rm c}\left\{1+2f_0\kappa_n^{\rm c}
\left[1+f_0\kappa_n^{\rm c}[\phi(n)-1]
-\frac{4\pi^2}{3}\Or\!\left(\frac{f_0{}^2}{a^2}\right)
\right.\right.
\cr
\qquad
\left.\left.
+\Or\!\left(\frac{f_0d_0}{n^2a^2}\right)
\right]\right\}\qquad \kappa_n^{\rm c}=(n|a|)^{-1}.
}
\end{equation}
To show this, one can put $\kappa_n=\kappa_n^c(1+\epsilon)$,
$x_n=|\kappa_n a|^{-1}=n/(1+\epsilon)$ and use the equality
$\tan(\pi x_n)=-\tan(\pi x_n\epsilon)=-(\pi x_n\epsilon)[1+
(\pi^2 n^2/3)\Or(\epsilon^2)]$ and the inequality
$|\phi(x_n)-\phi(n)|<\Or(\epsilon)$, the latter
following from the fact that $\phi'(n)$ vanishes faster than
$n^{-1}$. Equation (\ref{kappa}) is in agreement with
the result of \cite{ras82} for the relative energy shift
$\epsilon(n,0)\equiv (2+\epsilon)\epsilon
\doteq\epsilon_0(n,0)[1+\epsilon_0(n,0)p_1(n,0)]$,
where $\epsilon_0(n,0)=4f_0\kappa_n^c$ and
$p_1(n,0)=\phi(n)/4$.
The function $\phi(n)$ is defined in (\ref{phi})
with the digamma function for the integer values of the
argument given by the recurrence relation:
\begin{equation}
\label{psin}
\psi(n+1)=\psi(n)+1/n\qquad
\psi(1)=-C\doteq -0.5772.
\end{equation}
Note that $\phi(n)\approx 3$ is nearly constant:
$\phi(1)=2+2C\doteq 3.15443$, $\phi(2)\doteq 3.08145$,
$\phi(3)\doteq 3.05497$, $\dots$,
$\phi(\infty)=3$.

\subsubsection{Normalization}
Since $N(\eta_n)=0$ (to compensate for the pole of the
amplitude $f_c$ at $k^*=i\kappa_n$), the s-wave solutions
in discrete spectrum are now given
(for $r^*>d$) by the second term in (\ref{psigen}),
exponentially vanishing at large distances:
\begin{equation}
\label{psigenC}
\eqalign{
\psi_{n0}({r}^{*})={\cal N}'(n)
K(i\kappa_n)\frac{{\widetilde{G}(\rho_n,\eta_n)}}{r^*}
\cr
\qquad = {\cal N}'(n)
f_0\frac{{\widetilde{G}(\rho_n,\eta_n)}}{r^*}
\left[1+\Or\!\left(\frac{f_0d_0}{n^2a^2}\right)\right].
}
\end{equation}
The arguments $\rho_n$ and $\eta_n$ are taken at
${k}^{*}=i\kappa_n$ and the normalization factor
\begin{equation}
\label{norm'}
{\cal N}'(n)={\cal N}(\eta_n)f_c(i\kappa_n)/K(i\kappa_n)
\end{equation}
is set by the requirement
\begin{equation}
\label{norm}
\int |\psi_{n0}({r}^{*})|^2 {\rm d}^3{\bf r}^* =1.
\end{equation}
Note that the extension in the integral (\ref{norm})
of the asymptotic wave function (\ref{psigenC})
into the inner region $r^*<d$
leads to negligible relative errors
$\Or(f_0 d^2/(na)^3)$, $\Or(f_0^2d/(na)^3)$
in the normalization factor ${\cal N}'$.
Using the expansion of the $\widetilde{G}$-function in the square
brackets in (\ref{psigenA}) and the expression
for $\chi(\eta_n)$ in (\ref{chi1}),
one can write for distances $d<r^*\ll |a|$:
\begin{equation}
\label{psi-}
\eqalign{
\psi_{n0}({r}^{*})={\cal N}'(n)
\left\{\left(1-\frac{{r}^{*}}{|a|}\right)
\left[1-2\frac{f_0}{|a|}\left(\ln\!\left
|\frac{2{r}^{*}}{a}\right|+2C-1\right) \right]
+\frac{f_0}{{r}^{*}}
\right.
\cr
\left.
\qquad \qquad
+ \Or\!\left(\frac{f_0d_0}{n^2 a^2}\right)
+ \Or\!\left(\frac{f_0{r}^{*}}{a^2}\right)\right\}.
}
\end{equation}
Comparing (\ref{psi-})
with the low-$r^*$ expansion (\ref{psicapp1A})
of the pure Coulombic wave function and,
also taking into account
the exponential damping at large distances,
one can approximate the wave function (\ref{psigenC})
at $r^* \ll |a^2/f_0|$ by the expression:
\begin{equation}
\label{psiapp}
\psi_{n0}^{\rm app}({r}^{*})=
\frac{{\cal N}'(n)}{\psi_{n0}^{\rm coul}(0)}
\psi_{n0}^{\rm coul}({r}^{*})
\left[1-2\frac{f_0}{|a|}\left(\ln\!\left|\frac{2{r}^{*}}{a}
\right|+2C-\frac32\right) +\frac{f_0}{{r}^{*}}\right].
\end{equation}
From the results of calculations for the s-wave
$\pi^+\pi^-$ atoms, presented in upper panel of
figure~\ref{figdel}, one can see that the squares of the
approximate and exact expressions (\ref{psiapp}) and
(\ref{psigenC}) practically coincide for the distances
up to several tens fm and that the agreement is better
than percent even at $r^*\sim|a|$.

It follows from (\ref{psiapp}) that the relative
difference of the normalization factors ${\cal N}'(n)$ and
$\psi_{n0}^{\rm coul}(0)$ scales as $\Or(f_0/a)$.
In fact, this difference can be fixed when extending the
theory to a multichannel case and requiring the equality
of the total width $\Gamma_n=-2\Im E_n$ and the sum of the
partial widths (see (\ref{rate}) and (\ref{width''}) or,
(\ref{Aaa2a}), (\ref{Aaa2}) and (\ref{width'})).
As a result:
\begin{equation}
\label{normapp}
|{\cal N}'(n)/\psi_{n0}^{\rm coul}(0)|^2-1
\doteq \phi(n)\frac{2f_0}{n|a|}.
\end{equation}
We have checked (\ref{normapp}),
calculating ${\cal N}'$ from the integral (\ref{norm}) for
various values of the scattering length $f_0$, Bohr radius
$|a|$ and the principle quantum number $n$.

\subsection{Universality}
\begin{figure}
\epsfig{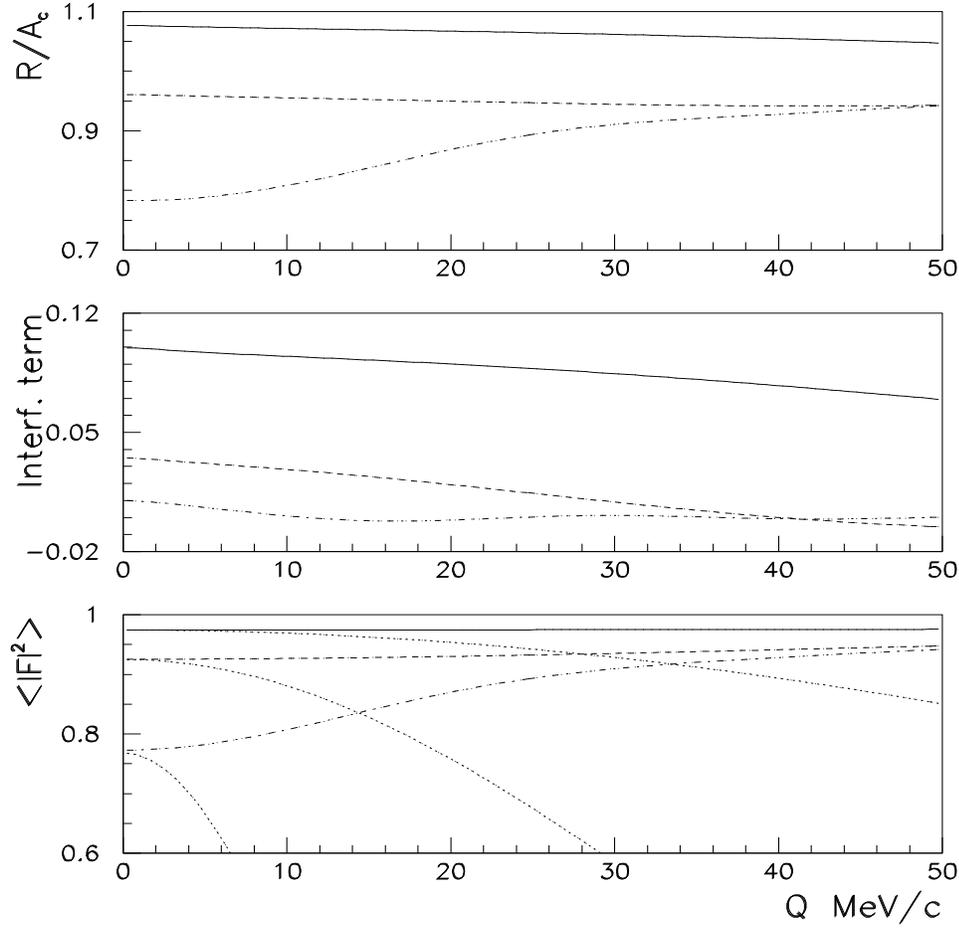}
\caption{The $\pi^+\pi^-$ correlation function
at a fixed separation $r^*$ divided by
the Coulomb penetration factor: ${\cal R}/A_c=
\langle|{\rm e}^{{-{\rm i}{\bf k}^*}{\bf r}^*}F+
f_c\widetilde{G}/r^*|^2\rangle$,
and the corresponding main contributions due to the interference
term and the modulus squared of the hypergeometric function
(see (\ref{cfapp})).
The solid, dash and dash-dotted curves correspond to
$r^*=$ 5, 15 and 50 fm respectively.
The calculation is done in the approximation of a constant
scattering amplitude $f_c(k^*)=f_0= 0.232$ fm, the
averaging assumes the uniform distribution of the cosine
of the angle between the vectors ${\bf r}^*$ and
${\bf k}^*={\bf Q}/2$.
The dotted curves in the lower panel represent
the s-wave Coulomb contribution $B^2(\rho,\eta)$ to the
quadratic term.}
\label{figgst1}
\end{figure}

\begin{figure}
\epsfig{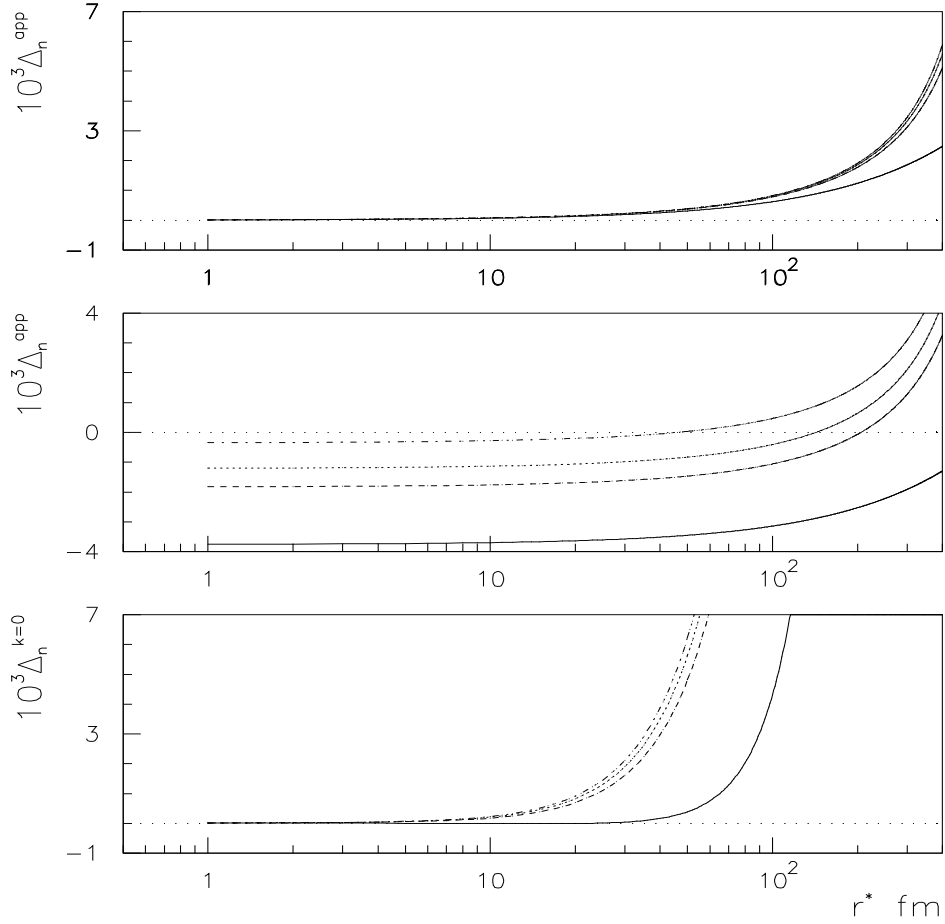}
\caption{Comparison of the approximate $\pi^+\pi^-$ atomic
wave function $\psi_{n0}^{\rm app}(r^*)$
and the $\pi^+\pi^-$ wave function in continuous spectrum
$\psi_{-{\bf k}^*}({\bf r}^*)$ at $k^*\rightarrow 0$,
respectively defined in (\ref{psiapp}) and
(\ref{psicoul}) ($f_0=0.232$ fm), with the exact s-wave
solution outside the range of the strong interaction
$\psi_{n0}(r^*)$ given in (\ref{psigenC}):
$\Delta_n^{\rm app}(r^*)=
[\psi_{n0}^{\rm app}(r^*)/\psi_{n0}(r^*)]^2-1$
and
$\Delta_n^{k^*=0}(r^*)= \langle
|[\psi_{-{\bf k}^*}({\bf r}^*)/
\psi_{-{\bf k}^*}^{\rm coul}(0)]/
[\psi_{n0}(r^*)/{\cal N}'(n)]|^2\rangle-1$,
$k^*\rightarrow 0$;
the averaging in the latter expression is done
over the uniform distribution of the cosine
of the angle between the vectors ${\bf r}^*$ and
${\bf k}^*={\bf Q}/2$.
The central panel shows $\Delta_n^{\rm app}(r^*)$
assuming ${\cal N}'(n)=\psi_{n0}^{\rm coul}(0)$
in (\ref{psiapp}) in
correspondence with the ansatz (\ref{nemansatz}).
The curves in the increasing order correspond to
$n= 1, 2, 3, 10$.
}
\label{figdel}
\end{figure}

Comparing (\ref{psi+}) and (\ref{psi-}),
valid for the distances $d<{r}^{*}\ll |a|$, one confirms
the important conclusion,
already stated at the beginning of subsection 5.2,
about the universality of the
$r^*$-behaviour of the moduli squared of the
s-wave solutions in continuous ($k^*\rightarrow 0$)
and discrete spectrum,
up to corrections vanishing as inverse squares of the
Bohr radius $|a|$. Assuming $f_0 <\sim d$, one has:
\begin{equation}
\label{delta_k0}
\eqalign{
\Delta_{n0}^{k^*}(r^*)\equiv
\left|\frac{\psi_{{k}^*0}({\bf r}^*)/
\psi_{{k}^*0}^{\rm coul}(0)}
{\psi_{n0}(r^*)/{\cal N}'(n)}\right|^2 -1
\cr
\qquad
=
4\pi^2\Or\!\left(\frac{f_0{}^2}{a^2}\right)+
\Or\!\left(\frac{f_0d_0}{n^2 a^2}\right)+
\Or\!\left(\frac{r^{*2}}{a^2}\right)+\Or(\rho^2).
}
\end{equation}
The universality holds with the same accuracy
also if the s-wave solution
in continuous spectrum were substituted by the
complete wave function
(recall that $\psi_{-{\bf k}^*}^{\rm coul}(0)=
\psi_{k^*0}^{\rm coul}(0)\equiv A_c^{1/2}$),
provided the averaging over the angle
between the vectors ${\bf r}^*$ and ${\bf k}^*$:
\begin{equation}
\label{delta_k}
\eqalign{
\Delta_{n}^{k^*}(r^*)\equiv
\left\langle\left|\frac{
\psi_{-{\bf k}^*}({\bf r}^*)
/\psi_{-{\bf k}^*}^{\rm coul}(0)}
{\psi_{n0}(r^*)/{\cal N}'(n)}\right|^2\right\rangle -1
\cr
\qquad
=
4\pi^2\Or\!\left(\frac{f_0{}^2}{a^2}\right)+
\Or\!\left(\frac{f_0d_0}{n^2 a^2}\right)+
\Or\!\left(\frac{r^{*2}}{a^2}\right)+\Or(\rho^2).
}
\end{equation}
This result follows from the fact that,
at $k^*\rightarrow 0$ and typical distances $r^*\ll |a|$,
the total wave function in continuous spectrum almost
coincides with the s-wave amplitude $\psi_{k^*0}(r^*)$
(see the lower panel in figure~\ref{figgst1}):
\begin{equation}
\label{psi++}
\psi_{-{\bf k}^{*}}({\bf r}^{*})=
\psi_{{k}^{*}0}({r}^{*})+{\rm e}^{{\rm i}\delta_c}\sqrt{A_c}
\frac{{\bf k}^*{\bf r}^{*}}{{k}^{*}a}
+ \Or\!\left(\frac{{r}^{*2}}{a^2}\right)+
\Or(\rho^2)
\end{equation}
and, that the relatively significant correction term
$\Or(r^*/a)$ in the square of the wave function
$\psi_{-{\bf k}^{*}}({\bf r}^{*})$ vanishes
after averaging over the direction of the relative
three-momentum ${\bf Q}=2{\bf k}^*$ or, -- after
suppressing the signs of the components $Q_i$ of the
vector ${\bf Q}$
{
(assuming a symmetric detector acceptance with respect
to the reflection $Q_i \to -Q_i$).
}
From the lower panel of figure~\ref{figdel}, one can see
that for the
$\pi^+\pi^-$ system, the universality holds to better
than percent for  $r^*< \sim 50$ fm.
{
Note that $\Delta_{n0}^0$ (not shown in figure~\ref{figdel})
is negative and, contrary to $\Delta_{n}^0$, it shows the
strongest deviation from zero for $n=1$, achieving a
per mil level already at $r^* \approx 20$ fm.
}

Comparing (\ref{160app}) and (\ref{16aaapp}),
one can see that the number $N_A$ of produced
$\pi^+\pi^-$ atoms is determined by the number of non-atomic
$\pi^+\pi^-$ pairs in the region of small $k^*$.
So, $N_A$ is actually proportional to the ratio
$\langle |\psi_{n0}({r}^{*}) |^2\rangle_{0P}/
\langle |\psi_{-{\bf k}^*}({\bf r}^*)|^2
\rangle_{\widetilde{q}P}$
in which the effects of
the $r^*$-dependence as well as the corrections
due to non-equal emission times ($t^*\ne 0$)
and {\it smoothness} assumption are to a large
extent compensated for, being practically the same
for the wave functions in continuous spectrum at
$k^*\rightarrow 0$ and discrete spectrum at
$r^* \ll |a|$.
In fact, according to (\ref{normapp}-\ref{delta_k}),
one can write the ratio of the finite-size correction
factors at small relative momenta ($k^*\ll
\langle 1/r^*\rangle^{\mbox{\tiny SL}}$)
and moderate distances between the particle emitters
($\langle r^*\rangle^{\mbox{\tiny SL}} \ll |a|$) as
\begin{equation}
\label{nemansatz}
\eqalign{
\frac{1+\delta_n}{1+\delta(k^*)}\equiv
\frac{\langle |\psi_{n0}({r}^{*})/
\psi_{n0}^{\rm coul}(0)|^2
\rangle_{0P}^{\mbox{\tiny SL}}}
{\langle |\psi_{-{\bf k}^*}({\bf r}^*)/
\psi_{k^*0}^{\rm coul}(0)|^2
\rangle_{\widetilde{q}P}^{\mbox{\tiny SL}}}
\equiv \left| \frac{N'(n)}
{\psi_{n0}^{\rm coul}(0)}\right|^2
\frac{1+\delta_n'}{1+\delta(k^*)}
\cr
\qquad
=
\left[1+\phi(n)\frac{2f_0}{n|a|}
\right]
\left\{1+
\Or\left(\frac{\langle r^{*2}
\rangle^{\mbox{\tiny SL}}}{a^2}\right)
+\Or\left(k^{*2}\langle r^{*2}
\rangle^{\mbox{\tiny SL}}\right)\right\}.
}
\end{equation}
thus leading to (\ref{migdal_mod}) up to a small
correction due to the transition
$\pi^0\pi^0 \rightarrow \pi^+\pi^-$
(see (\ref{nemansatz1})).
Recall that though the $k^*$-dependence of the correction
factor in braces is quadratic at very low
values of $k^*$, in fact, in a wider $k^*$-interval
and for sufficiently small values
$\langle r^*\rangle^{\mbox{\tiny SL}} < \sim 10$ fm,
it shows a quasi-linear and almost universal behaviour
(see figure~\ref{figgst1} and \cite{ade95}).

\subsection{The $n$-dependence}
Neglecting the production of the $\pi^+\pi^-$ atoms with the
orbital
angular momentum $l>0$, suppressed by powers of the
$\pi^+\pi^-$ Bohr radius $|a|$,
the pionium production probability is given in
(\ref{16anem''}) and
depends on the main quantum number $n$ as
\begin{equation}
\label{wn}
w_n\propto(1+\delta_n)|\psi^{\rm coul}_{n0}(0)|^2
\propto (1+\delta_n)/n^3.
\end{equation}
The correction factor $(1+\delta_n)$ slightly
modifies the $n^{-3}$ law of simple
ansatz in (\ref{16anem}).
It follows from (\ref{nemansatz}) that
the $n$-dependence of the short-distance
part of the correction $\delta_n$ is dominated by the
renormalization effect of the strong FSI on
the two-pion atomic wave function and that
the renormalized correction (see also (\ref{migdal_mod}))
\begin{equation}
\label{corfac}
\delta_n'\doteq \delta(0) +
\Or(\langle r^{*2}/a^2\rangle^{\mbox{\tiny SL}}).
\end{equation}
The most right panel in figure~\ref{figroa1b1}
confirms that the short-distance part of the correction
$\delta_n'$ is practically
independent of $n$ and equal to $\delta(0)$.
The renormalization correction
$2\phi(n)f_0/(n|a|)\approx 6f_0/(n|a|)$
is the largest for low values of $n$. For example,
for pionium at $n=1$ it composes $\sim 0.3\%$
(see also the central panel of figure~\ref{figdel}).
As for the $\omega$ and $\eta'$ contributions to
$\delta_n'$, their $n$-dependence is not negligible
and the shifts from $\delta(0)$ compose up to
$\sim -0.004$ and $\sim -0.1$, respectively.

In \cite{ami99,grw01}, the effect of the strong
interaction on the
$n$-dependence of the pionium wave function has been
studied numerically, solving the corresponding
Schr\"odinger equations. Thus, in \cite{ami99},
the ratio $R_n=\psi_{n0}/\psi_{n0}^{\rm coul}$
and the difference $\Delta R_n=R_1-R_n$ have been
calculated for $n=1-3$ using an exponential form of the
short-range potential.
According to (\ref{psicapp1A}), (\ref{psi-}) and
(\ref{normapp}), one has, up to
corrections $\Or(f_0/a)$ and $\Or(r^{*2}/a^2)$:
\begin{equation}
\label{Rn}
\eqalign{
R_n\equiv \frac{\psi_{n0}(r^*)}{\psi_{n0}^{\rm coul}(r^*)}
\doteq 1+\frac{f_0}{r^*}
\cr
\Delta R_n\equiv R_1-R_n\doteq
\frac{f_0}{|a|}\left\{\phi(1)-
\frac1n\phi(n)\right\}
\left(1+\frac{f_0}{r^*}\right).
}
\end{equation}
From figure~1 of \cite{ami99}, one can deduce a value
of $\sim 0.15$ fm for the scattering length $f_0$ to
achieve an agreement with the prediction of (\ref{Rn})
for the ratio $R_n$ at $d<r^*\ll |a|$.
The differences $\Delta R_n$, presented in figure~1 of
\cite{ami99} for $n=$ 2 and 3, are however by a factor
of 1.6 higher than the corresponding predictions of
(\ref{Rn}). For example, for $10^3\Delta R_n$ at
$r^*=8$ fm, $n=2$ and 3, one can read from this figure
the values\footnote
{
One should correct the figure by interchanging the curves.
The author is grateful to O.~Voskresenskaya for pointing out
this misprint.
}
1.0 and 1.3 while, (\ref{Rn}) respectively predicts 0.6
and 0.8. This discrepancy may indicate that the calculation
error, declared in \cite{ami99} to be better than
$10^{-4}$, was underestimated by a factor of 5.

In \cite{grw01}, a more refined numerical study
of the $n$-dependence has been done
accounting for the second channel ($\pi^0\pi^0$) and extended
charges. The hadronic $\pi\pi$ potentials have been chosen to
reproduce the phase shifts given by two-loop chiral
perturbation theory. The quantity
$d_n=n^{3/2}\psi_{n0}/\psi_{10}-1$ has been calculated
for $n=1-4$. Similar to (\ref{Rn}), one has
for $d<r^*\ll |a|$
\begin{equation}
\label{dn}
d_n\equiv n^{3/2}\frac{\psi_{n0}(r^*)}{\psi_{10}(r^*)}-1
\doteq -\frac{f_0}{|a|}\left\{\phi(1)-
\frac1n\phi(n)\right\},
\end{equation}
up to corrections
${\cal O}(f_0r^*/a^2)$ and $\Or(r^{*2}/a^2)$.
The results of numerical calculations presented in figure~2
of \cite{grw01} are in qualitative agreement with
(\ref{dn}), $d_n$ being almost constant
(except for the region of very small
$r^*$) and showing the right $n$-dependence:
$d_n\sim -(1-1/n)$. Similar to \cite{ami99},
the numerical results for $|d_n|$ are however higher,
now by a factor of 2.5, than the predictions
of (\ref{dn}) calculated with $f_0=0.2$ fm which
should correspond within $\sim 10\%$ to the choice of the
potentials in \cite{grw01}. Since the presence of the
second channel leads to a negligible modification of
(\ref{dn}) ($f_0\to\Re A^{\alpha\alpha}\approx f_0$;
see next chapter)
and the correction due to
the extended charges is also expected to be negligible
($\sim -\frac16 \langle r^2\rangle_\pi/a^2$),
the discrepancy in the size of the correction $d_n$ has
to be attributed to the insufficient calculation accuracy
or, to the incorrect matching of the scattering length.

\section{Two-channel wave functions}
\subsection{Continuous spectrum in both channels}
It was implied until now that a long-time FSI takes place
and can be separated in the Bethe--Salpeter amplitudes
in the near-threshold final-state elastic transitions
$1 + 2\rightarrow 1 + 2$ only.
In principle, however, it can be separated
also in the inelastic transitions,
$1 + 2\rightarrow 3 + 4$, characterized by a slow
relative motion in both entrance and exit channels.
The necessary condition for such a
separation is an approximate equality of the sums of particle
masses in the intermediate ($m_3+m_4$) and final ($m_1+m_2$)
states. Some examples are the transitions
$\pi^+\pi^-\leftrightarrow \pi^0\pi^0$,
$\pi^-K^+\leftrightarrow \pi^0 K^0$ or
$\pi^- p\leftrightarrow \pi^0 n$.
For such processes only the second
term in the upper diagram in figure~\ref{fig1} contributes,
now with the particles $3, 4$ in the intermediate
state. In the {\it equal-time} approximation, the corresponding
amplitudes reduce to the wave functions describing a
two-channel scattering of the particles $1, 2$ with the
inverse direction of the relative three-momentum:
${\bf k}^*\rightarrow -{\bf k}^*$
(the scattering is viewed in the diagram from right
to left so that the final-state particles $1, 2$ are in the
entrance scattering channel).
We will denote the channels as $\alpha=\{1 + 2\}$ and
$\beta=\{3 + 4\}$, and the corresponding wave functions
describing the scattering
$\alpha \rightarrow \alpha$ and $\alpha \rightarrow \beta$ -
as $\psi^{\alpha}$ and $\psi^{\beta}$ respectively.
Outside the range of the strong FSI, ${r^*}>d$,
they can be written, for the $\alpha$- and
$\beta$-channel continuous spectrum,
as \cite{lll97}:
\begin{equation}
\label{psiab}
\eqalign{
\psi ^{\alpha}_{-{\bf k}^*}({\bf r}^*)=
{\cal N}(\eta_{\alpha})
\left[{\rm e}^{-{\rm i}{\bf k}^{*}{\bf r}^{*}}
F(-{\rm i}\eta_{\alpha},1,{\rm i}\xi_{\alpha})+
f_c^{\alpha\alpha}({k}^{*})
\frac{\widetilde{G}(\rho_{\alpha},\eta_{\alpha})}
{{r}^{*}}\right]
\cr
\psi ^{\beta}_{-{\bf k}^*}({\bf r}^*)=
{\cal N}(\eta_{\alpha})f_c^{\beta\alpha}({k}^{*})
\sqrt{\frac{\mu_{\beta}}{\mu_{\alpha}}}
\frac{\widetilde{G}(\rho_{\beta},\eta_{\beta})}
{{r}^{*}}
\cr
\qquad ~~\rightarrow ~~
{\cal N}(\eta_{\alpha})f_c^{\beta\alpha}({k}^{*})
\sqrt{\frac{\mu_{\beta}}{\mu_{\alpha}}}
\frac{\exp(i\rho_{\beta})}
{{r}^{*}},~~~~~~
}
\end{equation}
where ${\cal N}(\eta_{\alpha})=
{\rm e}^{{\rm i}\delta_c(\eta_{\alpha})}
\sqrt{A_c(\eta_{\alpha})}$,
${\bf k}^*\equiv {\bf k}^*_{\alpha}$ and
\begin{equation}
\label{kab}
\eqalign{
{k}^{*}_{\beta}{}^2=
\frac{[m_4{}^2-m_3{}^2+(\omega_1^*+\omega_2^*)^2]^2}
{4(\omega_1^*+\omega_2^*)^2}-m_4{}^2
\cr
\qquad
\doteq\frac{\mu_{\beta}}
{\mu_{\alpha}}{k}^{*}_{\alpha}{}^2
+2\mu_{\beta}(m_1+m_2-m_3-m_4).
}
\end{equation}
The approximate equality in (\ref{kab}) corresponds
to the non-relativistic expressions for the energies:
$\omega_j^*=m_j+k_\alpha^{*2}/(2m_j), j=1,2$.
We consider here the systems with the Coulomb interaction
absent in the channel $\beta$,
so $a_\beta=\infty$, $\eta_\beta=0$, $A_c(\eta_{\beta})=1$,
$\widetilde{G}(\rho_\beta,\eta_\beta)=\exp(i\rho_\beta)$
and $\chi(\eta_\beta)/a_\beta = {\rm i} k^*_\beta$;
the amplitude $\psi^\beta$ in (\ref{psiab}) then
reduces to the expression indicated by the arrow.
The $\beta$-channel momenta at the $\alpha$-channel
thresholds
(${k}^{*}_{\alpha}=0$) for $\pi\pi$-, $\pi K$-,
$\pi N$-, $K K$-, $K N$- and $\bar{N} N$-systems
are given in table~\ref{k2}. This table also demonstrates
that even close to the $\alpha$-channel threshold,
the use of the non-relativistic approximation can lead
to noticeable shifts in ${k}^{*}_{\beta}$.

    \Table{\label{k2}
    The $\beta$-channel momenta ${k}^{*}_{\beta}$
calculated at the $\alpha$-channel thresholds
${k}^{*}_{\alpha}=0$.
Also shown are the relative shifts
$\Delta {k}^{*}_{\beta}/{k}^{*}_{\beta}$ arising from the
non-relativistic approximation in the second formula in
(\ref{kab}).
    }
      \br
        $\alpha$ &
        $\pi^+\pi^-$&
        $\pi^-K^+$&
        $\pi^- p$&
        $K^+K^-$&
        $K^-p$&
        $\bar{p}p$\\
        $\rightarrow\beta$ &
        $\rightarrow \pi^0\pi^0$&
        $\rightarrow \pi^0 K^0$&
        $\rightarrow \pi^0 n$&
        $\rightarrow K^0\bar{K}^0$&
        $\rightarrow \bar{K}^0 n$&
        $\rightarrow \bar{n} n$
        \\
        \mr
$k_\beta^*,$ MeV/$c$  & 35.5 & 11.3 & 28.0
& i~62.9& i~58.6 & i~49.3 \\
$k_\beta^{*-1},$ fm  & 5.6 & 17.5 & 7.0
&-i~3.1 &-i~3.4 &-i~4.0 \\
$\Delta k_\beta^*/k_\beta^*,$~~~$\%$& -0.84 & -0.07 &-0.46
&0.20 &0.13 &0.03
\\
\br
\endTable

Similar to equation (\ref{fc1}) in the single-channel case,
the amplitudes
\begin{equation}
\label{fc-def}
f_c^{\lambda\lambda'}=f^{\lambda\lambda'}
[A_c(\eta_{\lambda})A_c(\eta_{\lambda'})]^{-1/2},
\end{equation}
where $f^{\lambda\lambda'}$
are the amplitudes of the low-energy s-wave scattering
due to the short-range interaction renormalized by the
long-range Coulomb forces,
$\lambda,\lambda'=\alpha, \beta$.
The time-reversal invariance requires
$f^{\lambda\lambda'}=f^{\lambda'\lambda}$.
It is convenient to consider the amplitudes
$f_c^{\lambda\lambda'}$ and
$f^{\lambda\lambda'}$ as the elements of the symmetric
matrices $\hat{f}_c$ and $\hat{f}$ related by the
matrix equation
\begin{equation}
\label{fcf}
\hat{f}({k}^{*})=[A_c(\hat{\eta})]^{1/2}
\hat{f}_c({k}^{*})[A_c(\hat{\eta})]^{1/2}.
\end{equation}
The single-channel expression (\ref{fc}) for the
amplitude $f_c$ can then be rewritten in a matrix form:
\begin{equation}
\label{fc2}
\hat{f}_c({k}^{*})=
\left (\hat{K}^{-1}-
\frac{2\chi(\hat{\eta})}{\hat{a}}\right )^{-1},
\end{equation}
where
$\hat{a}$, $\hat{\eta}$, $\chi(\hat{\eta})$ and
$A_c(\hat{\eta})$
are diagonal matrices in the $(\alpha,\beta)$-channel
representation, for example,
$[A_c(\hat{\eta})]_{\lambda\lambda'}=A_c({\eta}_\lambda)
\delta_{\lambda\lambda'}$.
The symmetric matrix $\hat{K}$ has to be real for the
energies above both thresholds due to the
two-channel s-wave unitarity condition \cite{Lan77}
\begin{equation}
\label{two-channal_unitarity}
\Im{\hat{f}}= \hat{f}^+\Re\hat{k}~\hat{f},
\end{equation}
where the diagonal matrix $k_{\lambda\lambda'}=
k^*_\lambda\delta_{\lambda\lambda'}$.
Usually, the $\hat{K}$-matrix is real also for negative
kinetic energies (provided sufficiently fast vanishing of
the short-range potential with the distance), and so it
can be expanded in even powers of $k^*\equiv k^*_\alpha$,
similar to (\ref{K0}) or (\ref{K1}) with the
parameters substituted by the corresponding matrices
(e.g., $f_0\rightarrow \hat{f}_0$).

Since, in the cases of practical interest, the particles
(pions, kaons, nucleons) in the channels $\alpha$ and
$\beta$ are members of the corresponding isotopic multiplets,
one can assume the parameter matrices diagonal in the
representation of the total isospin \cite{lll97}.
The elements of the parameter or
$\hat{K}$ ($\hat{K}^{-1}$) matrices in the channel
representation are then given by the corresponding
isospin projections. Particularly, for
$\alpha=\{\pi^+\pi^-\}$, $\beta=\{\pi^0\pi^0\}$,
one has:
\begin{equation}
\label{proj1}
\eqalign{
f_0^{\alpha\alpha}=\frac23f_0^{(0)}+\frac13f_0^{(2)}\qquad
f_0^{\alpha\beta}=f_0^{\beta\alpha}=
-\frac{\sqrt{2}}{3}(f_0^{(0)}-f_0^{(2)})
\cr
f_0^{\beta\beta}=\frac13f_0^{(0)}+\frac23f_0^{(2)}.
}
\end{equation}
Analogous relations, with the substitutions
$(0)\rightarrow (1/2)$ and
$(2)\rightarrow (3/2)$, take place for the channels
$\alpha=\{\pi^-p,\pi^-K^+,\pi^+K^-\}$,
$\beta=\{\pi^0n,\pi^0K^0,\pi^0\bar{K}^0\}$.
For the channels $\alpha=\{K^+K^-,K^-p,\bar{p}p\}$,
$\beta=\{K^0\bar{K}^0,\bar{K}^0n,\bar{n}n\}$, one has
\begin{equation}
\label{proj2}
f_0^{\alpha\alpha}=f_0^{\beta\beta}=
\frac12(f_0^{(0)}+f_0^{(1)})\qquad
f_0^{\alpha\beta}=f_0^{\beta\alpha}=
-\frac12(f_0^{(0)}-f_0^{(1)}),
\end{equation}
where the parameters $f_0^{(0)}$ and $f_0^{(1)}$
have now positive imaginary parts due to the
effective inclusion of the additional channels
opened at the energies of the elastic thresholds
($k^*_\alpha=0$) in the reactions
$K\bar{K}\rightarrow \pi\pi, \pi\eta,~
\bar{K}N\rightarrow \pi's\Lambda, \pi's\Sigma,~
\bar{N}N\rightarrow$ mesons.

Note that the use of the isospin relations (\ref{proj1})
and (\ref{proj2}) implies that the violation of isotopic
invariance is solely associated with the Coulomb factors
$A_c({\eta_j})$ (strongly deviating from unity at
$k^*_j < 2\pi/|a_j|$) and the mass differences between
the members of the same multiplets
($k^*_\alpha \ne k^*_\beta$).
These relations however neglect the direct violation
of isotopic invariance of order $\Or(f_0/a)$
due to the renormalization effect
of the Coulomb interaction on the scattering lengths,
usually leading to the shifts on the level
of several percent.
Within this uncertainty, one can also use
(\ref{proj1}) or (\ref{proj2})
directly for the elements of the matrices
$\hat{K}^{-1}$ or $\hat{K}$.

The difference between the
channel momenta can be neglected sufficiently far from the
threshold. Then, one can apply
(\ref{proj1}) or (\ref{proj2}) to the
amplitudes $\widetilde{f}_{jj'}$ in the absence of the Coulomb
interaction
and, switch on this interaction in a similar way as in the
single-channel case \cite{bj50}:
\begin{equation}
\label{fjj'}
\hat{f}(k^*)=[A_c(\hat{\eta})]^{1/2}
\left\{\hat{\widetilde{f}}{}^{-1}(k^*)+{\rm i}\hat{k}-
\frac{2\chi(\hat{\eta})}{\hat{a}}
\right\}^{-1}[A_c(\hat{\eta})]^{1/2}.
\end{equation}
{
One may note that
(\ref{proj1}), (\ref{proj2}) correspond to the
two-dimensional unitary transformation
$\hat{f}_0=\hat{U}^{-1}\hat{f}_0'\hat{U}$,
$U_{11}=U_{22}=\cos{\varphi},~U_{12}=-U_{21}=\sin{\varphi}$.
Since it applies also to the $\hat{d}_0$-matrix, one immediately
arrives at the same transformation of the complete amplitude $\hat{f}$
in the case of absent Coulomb interaction and $\hat{k}=k^*\hat{1}$.
}

\subsection{Discrete spectrum in the $\alpha$-channel}
One can repeat the same arguments as for the single-channel case,
starting from the general solution in (\ref{psigen})
with the substitution $f_c\rightarrow f_c^{\alpha\alpha}$.
For a discrete energy level $E_n=-\kappa_n{}^2/(2\mu)$,
the amplitude $f_c$ has to have a pole or, equivalently,
$\det \hat{f}_c^{-1}(i\kappa_n)=0$.
Following \cite{ras82} and introducing the matrix
\begin{equation}
\label{A-def}
(\hat{A}^{-1})^{\lambda\lambda'}=(\hat{K}^{-1})^{\lambda\lambda'}
-{\rm i}\delta_{\lambda\lambda'}\delta_{\lambda\beta}k_\beta^*,
\end{equation}
one can rewrite this requirement in a form of
equation (\ref{chi1})
modified by the substitution
$K(i\kappa_n)\rightarrow A^{\alpha\alpha}(i\kappa_n)$
and thus, fix the discrete energy levels similar to
(\ref{kappa}):
\begin{equation}
\label{kappa2}
\kappa_n= \kappa_n^{\rm c}\left\{1+2A^{\alpha\alpha}
\kappa_n^{\rm c}
\left[1+A^{\alpha\alpha}\kappa_n^{\rm c}[\phi(n)-1]+\Or(a^{-2})
\right]\right\}
\end{equation}
\begin{equation}
\label{Aaa}
A^{\alpha\alpha}=\frac{K^{\alpha\alpha}-{\rm i}k_\beta^*
\det\hat{K}}
{1-{\rm i}k_\beta^* K^{\beta\beta}}=K^{\alpha\alpha}+
\frac{{\rm i}k_\beta^* (K^{\beta\alpha})^2}
{1-{\rm i}k_\beta^* K^{\beta\beta}},
\end{equation}
where $a=a_\alpha$.
Since
$\hat{K}(i\kappa_n)=\hat{K}(0)
[1+{\rm Tr}\Or(\hat{f}_0\hat{d}_0(na)^{-2})]$
and
$k_\beta^*(i\kappa_n)=k_\beta^*(0)
[1+\Or((nak_\beta^*)^{-2})]$,
one can safely make the substitutions
$\hat{K}(i\kappa_n)\rightarrow \hat{f}_0\equiv \hat{K}(0)$ and
$k_\beta^*(i\kappa_n)\rightarrow k_\beta^*(0)$
and write, with the relative errors $\Or(a^{-2})$
less than a fraction of per mil,
\begin{equation}
\label{A-matrix}
A^{\lambda\lambda'} =
\frac{K^{\lambda\lambda'}-
ik_\beta^* \det\hat{K}
\delta_{\lambda\lambda'}\delta_{\lambda\alpha}}
{1-{\rm i}k_\beta^* K^{\beta\beta}} \doteq
\frac{f_0^{\lambda\lambda'}-
ik_\beta^* \det\hat{f}_0
\delta_{\lambda\lambda'}\delta_{\lambda\alpha}}
{1-{\rm i}k_\beta^* f_0^{\beta\beta}},
\end{equation}
particularly,
\begin{equation}
\label{Aaa1}
\eqalign{
\Re A^{\alpha\alpha}=
K^{\alpha\alpha}-K^{\beta\beta}
\frac{(k_\beta^*K^{\beta\alpha})^2}
{1+(k_\beta^* K^{\beta\beta})^2}
\doteq f_0^{\alpha\alpha}-f_0^{\beta\beta}
\frac{(k_\beta^*f_0^{\beta\alpha})^2}
{1+(k_\beta^* f_0^{\beta\beta})^2}
\cr
\Im A^{\alpha\alpha} =
\frac{k_\beta^*(K^{\beta\alpha})^2}
{1+(k_\beta^* K^{\beta\beta})^2}
\doteq \frac{k_\beta^*(f_0^{\beta\alpha})^2}
{1+(k_\beta^* f_0^{\beta\beta})^2}.
}
\end{equation}
In (\ref{A-matrix}) and (\ref{Aaa1}), $k_\beta^*$ simply
denotes $k_\beta^*(0)$ or $k_\beta^*(i\kappa_n)$.
It can be seen from table~\ref{k2} that $k_\beta^{*-1}$ represents
a scale which is intermediate between the Bohr radius $|a|$ and
the elements of the matrix $\hat{f}_0$.
As a result, the terms like
$\Or(k_\beta^*(f_0^{\lambda\lambda'})^2/a)$ or
$\Or((ak_\beta^*)^{-2})$ contribute less than
a fraction of per mil and can be omitted. As for the terms
$\Or((k_\beta^* f_0^{\lambda\lambda'})^2)$,
their contribution is on a per mil level and is retained.

The s-wave solutions corresponding to the $\alpha$-channel
discrete spectrum are again given by the second term in
(\ref{psigen})
(${\cal N}(\eta_n)=0$)
with the finite normalization
${\cal N}'={\cal N}f_c^{\alpha\alpha}/A^{\alpha\alpha}$
introduced in the same way as in
(\ref{norm'}) modified by the substitution
$K\rightarrow A^{\alpha\alpha}$.
As for the corresponding $\beta$-channel s-wave solutions
$\psi_{n0}^{\beta}({r}^{*})$,
they are given by the second of equations (\ref{psiab}) with
\begin{equation}
\label{N''}
{\cal N}f_c^{\beta\alpha}={\cal N}'A^{\alpha\alpha}
\frac{f_c^{\beta\alpha}}{f_c^{\alpha\alpha}}=
{\cal N}'\frac{K^{\beta\alpha}}
{1-{\rm i}k_\beta^*K^{\beta\beta}}
\equiv {\cal N}'A^{\beta\alpha},
\end{equation}
the second equality following from (\ref{Aaa}) and the
explicit inversion of the symmetric matrix
$\hat{f}_c^{-1}$:
\begin{equation}
\label{f-1inver}
\eqalign{
D f_c^{\alpha\alpha}=K^{\alpha\alpha}-{\rm i}k_\beta^*
\det\hat{K}\qquad
D f_c^{\beta\alpha}=K^{\beta\alpha}\qquad
\cr
D f_c^{\beta\beta}=K^{\beta\beta}+\frac{2\chi}{|a|}
\det\hat{K}\qquad
\det\hat{K}=
K^{\alpha\alpha}K^{\beta\beta}-(K^{\beta\alpha})^2
\cr
D=\det\hat{f}_c^{-1}\det\hat{K}=
1-{\rm i}k_\beta^*K^{\beta\beta}
+\frac{2\chi}{|a|}(K^{\alpha\alpha}-{\rm i}k_\beta^*\det\hat{K}),
}
\end{equation}
where $\chi$ denotes here $\chi(\eta_\alpha)$; recall that
$\chi(\eta_\beta)/a_\beta=i k_\beta^*$ due to
the absent Coulomb interaction in the channel $\beta$.
{Note that the product $D f_c^{\lambda\lambda'}$
is finite since the amplitude pole for a bound state
is compensated by the corresponding zero of
the factor $D \propto \det\hat{f}_c^{-1}$.
For the continuous spectrum at the $\alpha$-channel threshold,
$\hat{K}=\hat{f}_0$ and $D=1-{\rm i}k_\beta^*f_0^{\beta\beta}-
(2{\rm i}\pi/|a|)(f_0^{\alpha\alpha}-{\rm i}k_\beta^*\det\hat{f}_0)$.
}

As a result,
\begin{equation}
\label{psicapp'}
\eqalign{
\psi_{n0}^{\alpha}({r}^{*})= {\cal N}'(n)
A^{\alpha\alpha}
\frac{\widetilde{G}}{{r}^{*}}
\cr
\qquad =
{\cal N}'(n)
\frac{f_0^{\alpha\alpha}-{\rm i}k_\beta^*\det\hat{f}_0}
{1-{\rm i}k_\beta^*f_0^{\beta\beta}}
\frac{\widetilde{G}}{{r}^{*}}
\left[1+{\rm Tr}\Or\!\left(\frac{\hat{f}_0\hat{d}_0}
{n^2a^2}\right)\right]
\cr
\psi_{n0}^{\beta}({r}^{*})=
{\cal N}'(n)A^{\beta\alpha}
\sqrt{\frac{\mu_{\beta}}{\mu_{\alpha}}}
\frac{{\rm e}^{{\rm i}k^*_{\beta}r^*}}{{r}^{*}}
\cr
\qquad =
\frac{{\cal N}'(n)f_0^{\beta\alpha}}
{1-{\rm i}k_\beta^*f_0^{\beta\beta}}
\sqrt{\frac{\mu_{\beta}}{\mu_{\alpha}}}
\frac{{\rm e}^{{\rm i}k^*_{\beta}r^*}}{{r}^{*}}
\left[1+{\rm Tr}\Or\!\left(\frac{\hat{f}_0\hat{d}_0}
{n^2a^2}\right)\right],
}
\end{equation}
where $\widetilde{G}=\widetilde{G}(\rho_n,\eta_n)$ with
the arguments $\rho_n$ and $\eta_n$ taken at
$k^*_{\alpha}=i\kappa_n$ ($\kappa_n$ is expressed through
$\hat{f}_0$ in (\ref{kappa2}) and (\ref{Aaa1})), and
${\cal N}'(n)=\psi_{n0}^{\rm coul}(0)
[1+\Or(f_0^{\alpha\alpha}/a)]$
is fixed by the normalization integral (\ref{norm}) for the
wave function $\psi_{n0}^{\alpha\alpha}$.
It can be calculated also analytically using (\ref{normapp})
with the substitution $f_0\rightarrow \Re A^{\alpha\alpha}
\approx f_0^{\alpha\alpha}$ (see Appendix B):
\begin{equation}
\label{normapp1}
|{\cal N}'(n)/\psi_{n0}^{\rm coul}(0)|^2-1
= \phi(n)\frac{2\Re A^{\alpha\alpha}}{n|a|}
-4\pi^2\Or\!\left(\left(\frac{\Re A^{\alpha\alpha}}{a}
\right)^2\right).
\end{equation}

Using (\ref{f-1inver}), one can express the
amplitudes $f_c^{\lambda\lambda'}(k^*)$ at
$k_\alpha^*=0$ ($\chi=-{\rm i}\pi, \hat{K}=\hat{f}_0$)
through the elements of the A-matrix
(related to the scattering lengths $f_0^{\lambda\lambda'}$
in (\ref{A-matrix}))
with the relative error
$\Or(a^{-2})$
less than a fraction of per mil:
\begin{equation}
\label{fA}
f_c^{\lambda\lambda'}(0)= A^{\lambda\lambda'}
\left[1+\frac{2{\rm i}\pi}{|a|}A^{\alpha\alpha}+
\Or(a^{-2})\right]-
\frac{2{\rm i}\pi}{|a|}\frac{\det\hat{f}_0}
{1-{\rm i}k_\beta^*f_0^{\beta\beta}}
\delta_{\lambda\lambda'}\delta_{\lambda\beta}.
\end{equation}

\subsection{Universality}

Comparing (\ref{psiab}) and (\ref{psicapp'}),
one may see that the universal
$r^*$-behaviour of the s-wave amplitudes $\psi^{\lambda}$
in continuous ($k^*\rightarrow 0$)
and discrete spectrum takes place with similar accuracy
as in the single-channel case. Thus,
using the expansions (\ref{psi+}) and (\ref{psi-})
for the amplitudes $\psi^{\alpha}$,
modified by the substitutions
$f_0\rightarrow f_c^{\alpha\alpha}(0)$ and
$f_0\rightarrow A^{\alpha\alpha}$ respectively, one has
for the measures of the
universality violation defined as in (\ref{delta_k0}),
\begin{equation}
\label{delta_k0_2}
\eqalign{
\Delta_{n0}^{\alpha,k^*}(r^*)=
4\pi\Or\!\left(\frac{k^*_\beta(f_0^{\beta\alpha})^2}{a}\right)+
4\pi^2\Or\!\left(\frac{(f_0^{\alpha\alpha})^2}{a^2}\right)
\cr
\qquad \qquad +
{\rm Tr}\Or\!\left(\frac{\hat{f}_0\hat{d}_0}{n^2 a^2}\right)+
\Or\!\left(\frac{r^{*2}}{a^2}\right)+\Or(\rho^2)
\cr
\Delta_{n0}^{\beta,k^*}(r^*)=
{\rm Tr}\Or\!\left(\frac{\hat{f}_0\hat{d}_0}{n^2 a^2}\right)+
\Or\!\left(\frac{r^{*2}}{a^2}\right)+\Or(\rho^2).
}
\end{equation}
The presence of the second channel manifests itself
through a new scale
$k_\beta^*$ (see table~\ref{k2}), basically leading to the
additional correction of
$4\pi\Or(k_\beta^*(f_0^{\beta\alpha})^2/a))$
which is still on the negligible level less than a fraction
of per mil.

For the production cross sections,
instead of (\ref{160app})
and (\ref{16aaapp}), we now have:
\begin{eqnarray}
\label{160app1}
\gamma_1\gamma_2\frac{{\rm d}^6\sigma}{{\rm d}^3{\bf p}_1
{\rm d}^3{\bf p}_2} \doteq
\gamma_1\gamma_2\frac{{\rm d}^6\sigma_0^{\alpha}}
{{\rm d}^3{\bf p}_1{\rm d}^3{\bf p}_2}
\sum\limits_{S}
{\cal G}_{S,\alpha}\left\langle
\left|\psi ^{S,\alpha}_{-{\bf k}^*}({\bf
r}^*)\right|^2\right\rangle_{\widetilde{q}PS}
\nonumber\\
\qquad \qquad +
\gamma_3\gamma_4\frac{{\rm d}^6\sigma_0^{\beta}}
{{\rm d}^3{\bf p}_3{\rm d}^3{\bf p}_4}
\sum\limits_{S}
{\cal G}_{S,\beta}\left\langle
\left|\psi ^{S,\beta}_{-{\bf k}^*}({\bf r}^*)
\right|^2\right\rangle_{\widetilde{q}PS}
\end{eqnarray}
\begin{eqnarray}
\label{16aaapp1}
\gamma_b\frac{{\rm d}^3\sigma_b^S}{{\rm d}^3{\bf P}_b} \doteq
(2\pi)^3\gamma_1\gamma_2\frac{{\rm d}^6\sigma_0^{\alpha}}
{{\rm d}^3{\bf p}_1
{\rm d}^3{\bf p}_2}{\cal G}_{S,\alpha}\left\langle
\left|\psi ^{S,\alpha}_{b}({r}^{*})
\right|^2\right\rangle_{0PS}
\nonumber\\
\qquad \qquad +
(2\pi)^3\gamma_3\gamma_4\frac{{\rm d}^6\sigma_0^{\beta}}
{{\rm d}^3{\bf p}_3
{\rm d}^3{\bf p}_4}{\cal G}_{S,\beta}
\left\langle \left|\psi ^{S,\beta}_{b}({r}^{*})
\right|^2\right\rangle_{0PS},
\end{eqnarray}
where ${\bf p}_i = {\bf P}_b m_i/(m_1+m_2)$ in (\ref{16aaapp1})
and $b=\{n0\}$.
Since the particles $1,3$ and $2,4$ are usually the members
of the same isospin multiplets, we can take
$\gamma_1\gamma_2 {\rm d}^6\sigma_0^{\alpha}\doteq
\gamma_3\gamma_4{\rm d}^6\sigma_0^{\beta}$ as a common factor in
(\ref{160app1}) and (\ref{16aaapp1}) and also put
${\cal G}_{S,\alpha}\doteq {\cal G}_{S,\beta}$.

The two-channel effects
in the production cross section, being quadratic in the amplitude
$f_0^{\beta\alpha}$, usually represent less than several percent
of the strong FSI contribution (a fraction of percent in the cross
section).
Thus, for a near-threshold two-pion system produced according to
a Gaussian $r^*$-distribution (\ref{gs-r})
with the characteristic radius $r_{\cal G}=3$ fm and,
taking the two-pion s-wave amplitudes from
\cite{cgl01} ($f_0^{\alpha\alpha}=0.186$ fm,
$f_0^{\beta\alpha}=-0.176$ fm),
the contributions of the FSI transitions
$\pi^+\pi^-\leftrightarrow \pi^+\pi^-$ and
$\pi^+\pi^-\leftrightarrow \pi^0\pi^0$ to the $\pi^+\pi^-$
production cross section respectively compose $7.72\%$ and
$0.16\%$, ; these contributions
are somewhat higher, $9.66\%$ and
$0.20\%$, for the amplitudes from
\cite{nag79} ($f_0^{\alpha\alpha}=0.232$ fm,
$f_0^{\beta\alpha}=-0.192$ fm). At large $r_{\cal G}$, the elastic
and inelastic contributions vanish
as $f_0^{\alpha\alpha}/r_{\cal G}$ and
$|f_0^{\beta\alpha}/r_{\cal G}|^2$
respectively. One should also account for the correction
due to the deviation of the solutions in (\ref{psiab}) and
(\ref{psicapp'}) from the exact ones in the inner region $r^*<d$.
Though this correction vanishes as $r_{\cal G}^{-3}$,
at $r_{\cal G}=3$ fm
it is still comparable to the contribution of the inelastic
two-pion transition, composing $0.25\%$ and $0.20\%$ for the
amplitudes from \cite{cgl01} and \cite{nag79}, respectively.

Note that assuming $\gamma_1\gamma_2{\rm d}^6\sigma_0^{\alpha}
{\cal G}_{S,\alpha}\doteq \gamma_3\gamma_4{\rm d}^6\sigma_0^{\beta}
{\cal G}_{S,\beta}$, the correction to the correlation function
at a given total spin $S$, total four-momentum $P$ and a small
generalized relative four-momentum
$\widetilde{q}=\{0,2{\bf k}^*\}\rightarrow 0$
can be written as
\begin{eqnarray}
\label{corr1}
\Delta{\cal R}\doteq \int {\rm d}^3{\bf r}^* W_P({\bf r}^*)
\left\{\left[\left|\psi_{-{\bf k}^*}^\alpha({\bf r}^*)\right|^2+
\left|\psi_{-{\bf k}^*}^\beta({\bf r}^*)\right|^2\right]
\right.
\nonumber\\
\left.
\qquad \qquad-
\left[\left|\widetilde{\psi}_{-{\bf k}^*}^\alpha({\bf r}^*)\right|^2+
\left|\widetilde{\psi}_{-{\bf k}^*}^\beta({\bf r}^*)
\right|^2\right]\right\},
\end{eqnarray}
where $W_P({\bf r}^*)=\int {\rm d}t^* g_P(t^*,{\bf r}^*;0)/
\int d^4 x g_P(x;0)$
is the normalized distribution of the vector ${\bf r}^*$
of the relative distances between the emission points in the
pair c.m. system
and $\widetilde{\psi}$ denotes the solutions in
(\ref{psiab}) extended to the inner region $r^*<d$.
In the case of only two open channels $\alpha$ and $\beta$,
the leading part of the correction scaled by $W_P(0)$ is
expressed through bilinear products of the amplitudes
$f^{\lambda\lambda'}$ in equation (44) of \cite{lll97}.
After a straightforward though lengthy algebra,
it can be written in a more explicit form:
\begin{eqnarray}
\label{corr2}
\Delta{\cal R}\approx -4\pi W_P(0)A_c(\eta_\alpha)
\left[|f_c^{\alpha\alpha}|^2\frac{{\rm d}}{{\rm d}k^{*2}}
(\hat{K}^{-1})^{\alpha\alpha}
\right.
\nonumber\\
\left.
\qquad +
|f_c^{\beta\alpha}|^2\frac{{\rm d}}{{\rm d}k^{*2}}
(\hat{K}^{-1})^{\beta\beta}+
2\Re (f_c^{\alpha\alpha}f_c^{\beta\alpha *})
\frac{{\rm d}}{{\rm d}k^{*2}}(\hat{K}^{-1})^{\beta\alpha} \right];
\end{eqnarray}
at $k^*=0$, twice the derivatives of the inverse $\hat{K}$-matrix
elements coincide with the effective radii
$d_0^{\lambda\lambda'}$.
Similarly, in the case of discrete spectrum, the leading
correction to $\langle |\widetilde{\psi}_{n0}^\alpha({\bf r}^*)|^2+
|\widetilde{\psi}_{n0}^\beta({\bf r}^*)|^2\rangle$
is also given by (\ref{corr2}) with the substitutions
$A_c(\eta_\alpha)\rightarrow N'(n)$ and
$f_c^{\lambda\lambda'}\rightarrow A^{\lambda\lambda'}$.
For the Gaussian $r^*$-distribution, (\ref{corr2})
is valid up to subleading contributions
$\Or(k^{*2}a_1/r_{\cal G})$ (see a discussion after
(\ref{K1})) and
$\Or(f_0^{\alpha\alpha}d^{4}/r_{\cal G}^5)$.

It is important that the presence of the second channel
does not practically modify the ratio (\ref{nemansatz})
of the finite-size correction factors in discrete and continuous
($k^*\rightarrow 0$) spectrum at moderate distances $r^*\ll |a|$.
The only modifications are
the substitution $f_0\rightarrow \Re A^{\alpha\alpha}
\approx f_0^{\alpha\alpha}$ and
the appearance of the negligible
correction $4\pi\Or(k_\beta^*(f^{\beta\alpha}_0)^2/|a|)$:
\begin{equation}
\label{nemansatz1}
\eqalign{
\frac{1+\delta_n}{1+\delta(k^*)}\equiv
\frac{[{\cal G}_{\alpha}\langle |\psi_{n0}^\alpha({r}^{*})|^2
\rangle_{0P}^{\mbox{\tiny SL}}+
{\cal G}_{\beta}\langle |\psi_{n0}^\beta({r}^{*}) |^2
\rangle_{0P}^{\mbox{\tiny SL}}]
|\psi_{n0}^{\rm coul}(0)|^{-2}}
{[{\cal G}_{\alpha}\langle |\psi_{-{\bf k}^*}^\alpha({\bf r}^*)|^2
\rangle_{\widetilde{q}P}^{\mbox{\tiny SL}}+
{\cal G}_{\beta}\langle |\psi_{-{\bf k}^*}^\beta({\bf r}^*)|^2
\rangle_{\widetilde{q}P}^{\mbox{\tiny SL}}]
|\psi_{k^*0}^{\rm coul}(0)|^{-2}}
\cr
\qquad
\equiv \left| \frac{N'(n)}
{\psi_{n0}^{\rm coul}(0)}\right|^2
\frac{1+\delta_n'}{1+\delta(k^*)}
=\left[1+\phi(n)\frac{2\Re A^{\alpha\alpha}}{n|a|}\right]
\cr
\qquad
\cdot
\left\{1+
\Or\!\left(\frac{\langle r^{*2}
\rangle^{\mbox{\tiny SL}}}{a^2}\right)
+\Or(k^{*2}\langle r^{*2}\rangle^{\mbox{\tiny SL}})+
4\pi\Or\!\left(\frac{k_\beta^*(f_0^{\beta\alpha})^2}{|a|}\right)
\right\}.
}
\end{equation}

\section{Finite-size effect in the experiment DIRAC}
\subsection{$\pi^+\pi^-$ system}
We will use the results of the UrQMD transport code simulations
of the pion production in pNi interactions at 24 GeV in the
conditions of the DIRAC experiment at CERN \cite{smol}.
Since we are interested in the region of very small relative
momenta $Q=2k^* <$ 20 MeV/$c$, where the angular distribution
of the vector ${\bf Q}$ is isotropic for non-correlated pions
and, for $Q <$ 10 MeV/$c$,
the detector acceptance is practically independent of
the direction of the vector ${\bf Q}$,
one can simplify the analysis integrating over this direction.
The finite-size effect is then determined by the distribution
of the relative distance $r^*$ between the pion production points
in the pair c.m. system, irrespective of the angular distribution
of the vector ${\bf r}^*$.
In fact, due to the applied cut $Q_T<Q_T^{\rm cut}=4$ MeV/$c$ \cite{ade05},
this is true for $Q<Q_T^{\rm cut}$ only,
see the discussion after (\ref{cfapp}) and (\ref{cfder})).
For larger $Q$-values or, in the case of a two-dimensional
$(Q_T,Q_L)$-analysis, one needs
two-dimensional $(r^*_T,r^*_L)$- or
$(r^*,\cos\theta_{r^*})$-distributions.
We will neglect this complication here.

The simulated $r^*$-distribution is shown in figure~\ref{fig_rst}.
The tail of this distribution ($r^* > 50 fm$)
is dominated by pion pairs containing a pion from
the decays of $\omega$ and $\eta'$ resonances,
except for the pairs with both pions from one and
the same decay.
The respective decay lengths in the rest frame
of the decay pion are about 30 and 900 fm;
the decay length
$l \approx \tau \langle p_{\rm dec}\rangle/m_\pi$
is determined by the resonance lifetime $\tau$ and the
four-velocity $p_{\rm dec}/m_\pi$ of the decay pion.
As a consequence of the exponential decay law, the form of
the corresponding $r^*$-distributions is nearly exponential,
except for the region of small $r^*$ dominated by the phase space
suppression factor $\propto r^{*2}$.
The exponential form is also distorted due to the averaging over the
continuous spectrum of the decay momenta and over
the emission points of the second pion.
For $r^*$ less than 2000-3000 fm,
the simulated $\eta'$ contribution
($\sim 1\%$ of pion pairs at $Q < 50$ MeV/$c$) can be
sufficiently well parametrized by an exponential-like formula
interpolating between the phase space and exponential behaviour:
\begin{equation}
\label{etapr-r}
\sum_i \frac{d N(\pi_{\eta'}\pi_i)}{d r^*}\doteq
n_{\eta'} {\cal F}(r^*;r_{\eta'},l_{\eta'})
\end{equation}
\begin{equation}
\label{exp-r}
\eqalign{
{\cal F}(r^*;r_{\eta'},l_{\eta'})
=
\frac{x^2}{2.2}\left\{1-
\exp\!\left[-\frac{2.2}{x^2}
\left(1+0.2 x^2 \frac{1+0.15 x^2 y}{1+x^5/125}
\right)\right]\right\}
{\rm e}^{-y}
\cr
x=\frac{r^*}{r_{\eta'}}\qquad y=\frac{r^*}{l_{\eta'}},
}
\end{equation}
where $r_{\eta'} = 2$ fm, $l_{\eta'} = 790$ fm.
At the same time, a good description of the
$\omega$ contribution
($\sim 19\%$ of low-$Q$ pion pairs) requires a
superposition of two exponential-like expressions:
\begin{equation}
\label{omega-r}
\sum_{i\ne \eta'} \frac{d N(\pi_{\omega}\pi_i)}{d r^*}
\doteq
n_{1\omega} {\cal F}(r^*;r_{1\omega},l_{1\omega})+
n_{2\omega} {\cal F}(r^*;r_{2\omega},l_{2\omega}).
\end{equation}
The parameters
$r_{1\omega} = 1.07$ fm, $l_{1\omega} = 43.0$ fm,
$r_{2\omega} = 2.65$ fm, $l_{2\omega} = 25.5$ fm,
$n_{1\omega}/n_{2\omega}=0.991$
in the interval 2-200 fm
and
$r_{1\omega} = 1.00$ fm, $l_{1\omega} = 44.0$ fm,
$r_{2\omega} = 2.55$ fm, $l_{2\omega} = 25.8$ fm
$n_{1\omega}/n_{2\omega}=0.845$
in the interval 2-350 fm.
We will use the former
parameter set, but we have checked that the use of the
latter one leads to a negligible change ($<0.1\%$)
of the breakup probability.
The rest of the $r^*$-distribution due to the pions
produced directly in the collision, in the rescatterings
or in the decays of resonances with the decay lengths
shorter than $l_\omega$
is peaked at $\sim 3$ fm and its main part
($\sim 60\%$ of low-$Q$ pion pairs) including the tail
for $r^* = 10-100$ fm
can be effectively described by a power-like expression:
\begin{equation}
\label{ml-r}
{\cal M}(r^*;r_{\cal M},\alpha,\beta)
= r^{*2}\left[1+\left(
\frac{r^*}{r_{\cal M}}\right)^{2\alpha}\right]^{-2\beta},
\end{equation}
where $r_{\cal M}= 9.20$ fm, $\alpha= 0.656$, $\beta= 2.86$;
note that the tail vanishes as $(r^*)^{-5.5}$, i.e.
much faster than the Lorentzian ($\alpha=\beta=1$).
The remaining short-distance part of the
$r^*$-distribution
($\sim 20\%$ of low-$Q$ pion pairs) is strongly shifted
towards the origin because the UrQMD code assumes the
point-like regions of the decays and rescatterings;
particularly,
$r^*=0$ for $\sim 8\%$ of low-$Q$ $\pi^+\pi^-$ pairs.
Therefore, we will represent this part by a Gaussian
distribution:
\begin{equation}
\label{gs-r}
{\cal G}(r^*;r_{\cal G})
= r^{*2}\exp\!\left(-
\frac{r^{*2}}{4r_{\cal G}^2}\right),
\end{equation}
where the Gaussian radius $r_{\cal G}\approx 1-2$ fm.
As a result,
\begin{equation}
\label{sd-r}
\sum_{i,j \ne \omega,\eta'}
\frac{d N(\pi_i\pi_j)}{d r^*} \doteq
n_{\cal M} {\cal M}(r^*;r_{\cal M},\alpha,\beta)+
n_{\cal G} {\cal G}(r^*;r_{\cal G}).
\end{equation}
We will also represent the short-distance part
of the $r^*$-distribution
by the Gaussian contribution alone,
i.e. put $n_{\cal M}=0$ and $r_{\cal G}= 3$ and 2 fm
in (\ref{sd-r}).

\begin{figure}
\epsfig{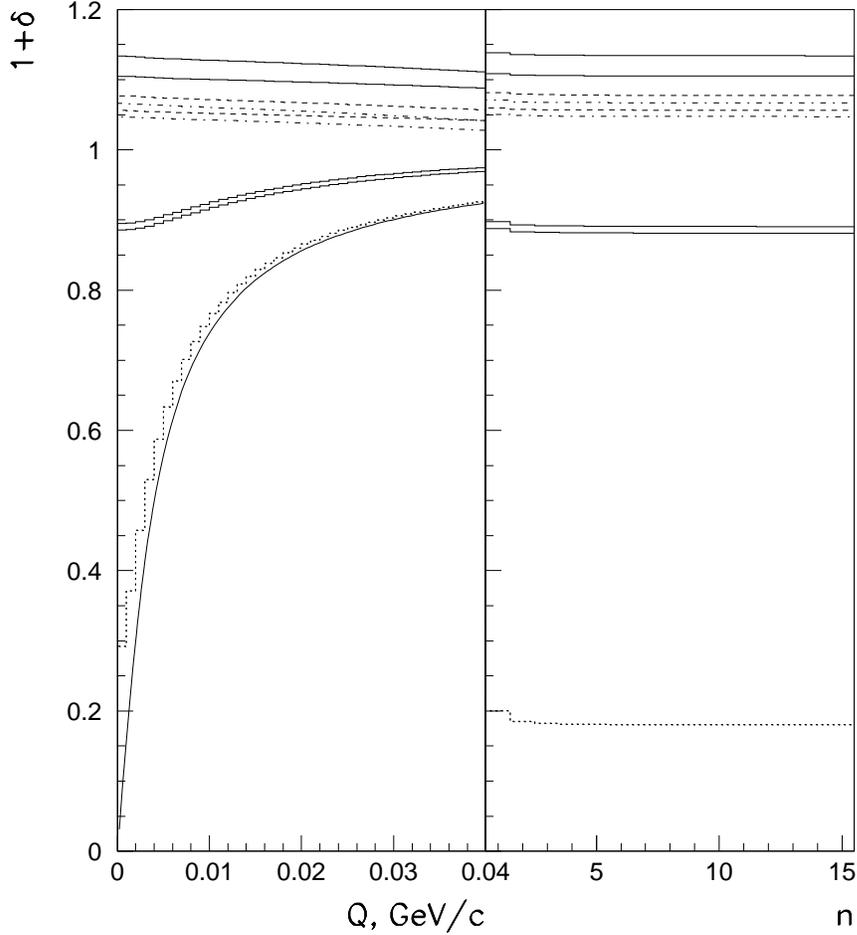}
\caption{The correction factors $1+\delta(k^*)$
(left)
and $1+\delta_{n}$ (right) as functions of the
relative momentum $Q=2k^*$ and the main atomic quantum number $n$
respectively. They are required to calculate the $\pi^+\pi^-$
production cross sections in the continuous and
discrete spectrum according to (\ref{16nem''}) and
(\ref{16anem''}).
The two sets of histograms
denoted by the same lines (dotted, full, dash-dotted, dashed
and full) correspond to the two-pion scattering amplitudes from
\cite{cgl01} (lower) and \cite{nag79} (upper).
In increasing order, they correspond to the $r^*$-distributions
$\eta'$, $\omega$, ${\cal G}(r^*;3 {\rm fm})$,
${\cal M}(r^*;9.20 {\rm fm},0.656,2.86)$ and
${\cal G}(r^*;2 {\rm fm})$ defined in
(\ref{etapr-r})-(\ref{gs-r}).
The calculation was done according to the two-channel expressions
given in (\ref{corcspi}) and (\ref{cordspi}),
taking into account the correction in (\ref{corr2}).
Note that the infinite-size correction factors
$1+\delta^\infty(k^*)=1/A_c(\eta)$ (the curve)
and $1+\delta^\infty_{n}=0$.
}
\label{figroa1_dif}
\end{figure}

The correction factors $1+\delta(k^*)$ and $1+\delta_n$
corresponding to the $r^*$-distributions
$\eta', \omega, {\cal M}, {\cal G}$,
required to calculate the $\pi^+\pi^-$ production cross
section in the continuous and discrete spectrum,
are shown in figure~\ref{figroa1_dif}. The two sets of
histograms denoted by the same lines (dotted, full,
dash-dotted, dashed and full) correspond to the two-pion
scattering amplitudes from \cite{cgl01} (lower)
and \cite{nag79} (upper).
In increasing order, they correspond to the
$r^*$-distributions
$\eta'$, $\omega$, ${\cal G}(r^*;3 {\rm fm})$,
${\cal M}(r^*;9.20 {\rm fm},0.656,2.86)$ and
${\cal G}(r^*;2 {\rm fm})$.
One may see that the correction factors corresponding to
the $\eta'$ contribution are practically independent of
the two-pion scattering amplitudes and, noticeably
deviate from the infinite-size correction factors
$1+\delta^\infty(k^*)=1/A_c(\eta)$ (the curve)
and $1+\delta^\infty_{n}=0$.
We thus do not include the $\eta'$-meson in the class
of LL emitters, unlike the $\eta$-meson with the
decay length of $\sim 10^5$ fm.

The calculation of the correction factors was done
according to the two-channel expressions
given in the numerator and denominator of the first
equality in (\ref{nemansatz1}):
\begin{equation}
\label{corcspi}
1+\delta({\bf k}^*)\doteq
\left\langle \left|\psi ^{\alpha}_{-{\bf k}^*}({\bf r}^*)\right|^2+
\left|\psi ^{\beta}_{-{\bf k}^*}({\bf r}^*)\right|^2
\right\rangle_{\widetilde{q}P}^{\mbox{\tiny SL}}
[A_c(\eta)]^{-1},
\end{equation}
\begin{equation}
\label{cordspi}
1+\delta_n\doteq
\left\langle \left|\psi ^{\alpha}_{n0}(r^*)\right|^2+
\left|\psi ^{\beta}_{n0}(r^*)\right|^2
\right\rangle_{0P}^{\mbox{\tiny SL}}
\left|\psi ^{\rm coul}_{n0}(0)\right|^{-2},
\end{equation}
where $\alpha$ and $\beta$ respectively denote the channels
$\pi^+\pi^-$ and $\pi^0\pi^0$.
However, the account of the coupled $\pi^0\pi^0$ channel
and of the leading correction due to the
approximate treatment of the wave function inside the
range of the strong interaction, does not practically
influence the results corresponding to the $\eta'$ and
$\omega$ contributions and only slightly ($<1\%$)
shifts up the correction factors corresponding
to the short-distance ${\cal M}$ and ${\cal G}$ ones.
A shift of the correction factors can
arise also from the uncertainty in the s-wave
elastic $\pi^+\pi^-$ scattering length $f_0$.
The shift due to $\sim 20\%$ difference of
the two-pion scattering amplitudes from \cite{cgl01}
($f_0=0.186$ fm) and \cite{nag79} ($f_0=0.232$ fm) is
$\sim 2-3\%$ for the short-distance ${\cal M}$ and ${\cal
G}$ contributions and $\sim 1\%$ for the $\omega$ one.
The global shifts are however not important since they
can be absorbed in the product
$\lambda g=\Lambda$ in (\ref{16nem''}) and
(\ref{16anem''}).

\begin{figure}
\epsfig{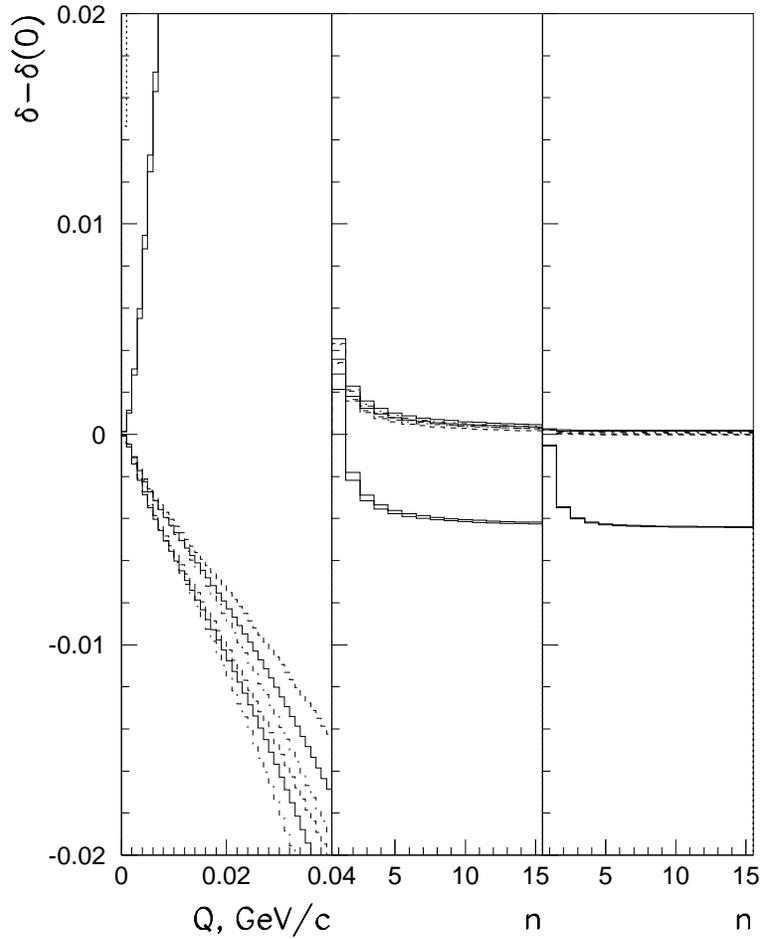}
\caption{
The differences $\delta(k^*)-\delta(0)$ (left panel),
$\delta_n-\delta(0)$ (middle panel)
and $\delta_n'-\delta(0)$ (right panel)
calculated from the $\pi^+\pi^-$ correction factors given in
figure~\ref{figroa1_dif}. The latter difference is corrected for the
effect of the strong interaction on the normalization of the
pionium wave function according to (\ref{norm_corr}).
The differences
corresponding to the $\eta'$ contribution (dotted histograms)
are not seen except for the first bin in the left panel;
in the middle and right panels they compose $\sim -0.1$.
}
\label{figroa1b1}
\end{figure}

In accordance with the results in
table \ref{cffit} and figure \ref{figgst1},
one may see in figure~\ref{figroa1_dif} the nearly universal
slope of the factors $1+\delta(k^*)$ corresponding to the
short-distance ${\cal M}$ and ${\cal G}$ contributions.
In accordance with (\ref{cfder}), the slope scales with
$f_0$ and is $\sim 20\%$ steeper when using the two-pion
amplitudes from \cite{nag79}
instead of those from \cite{cgl01}.
This is clearly seen in figure~\ref{figroa1b1},
where we plot the same correction factors as in
figure~\ref{figroa1_dif} in a larger scale and with
the subtracted intercepts $1+\delta(0)$. At $Q> 20$ MeV/$c$,
there is also seen $\sim 5-10\%$ variation of the slope
corresponding to different short-distance distributions.

Figures~\ref{figroa1_dif} and \ref{figroa1b1} also
demonstrate the violation of the universality
relation $\delta_n\doteq \delta(0)$ up to $\sim 0.4\%$
for the short-distance and $\omega$ contributions
and up to $\sim 9\%$ for the $\eta'$ one.
The most right panel in figure~\ref{figroa1b1} shows that,
in the case of the short-distance contribution,
this violation is mainly related to the effect of the
strong interaction on the normalization of the pionium
wave function.
Indeed,
the difference $\delta_n' - \delta(0)$,
corrected for this effect according to (\ref{norm_corr}),
practically vanishes.

\begin{figure}
\epsfig{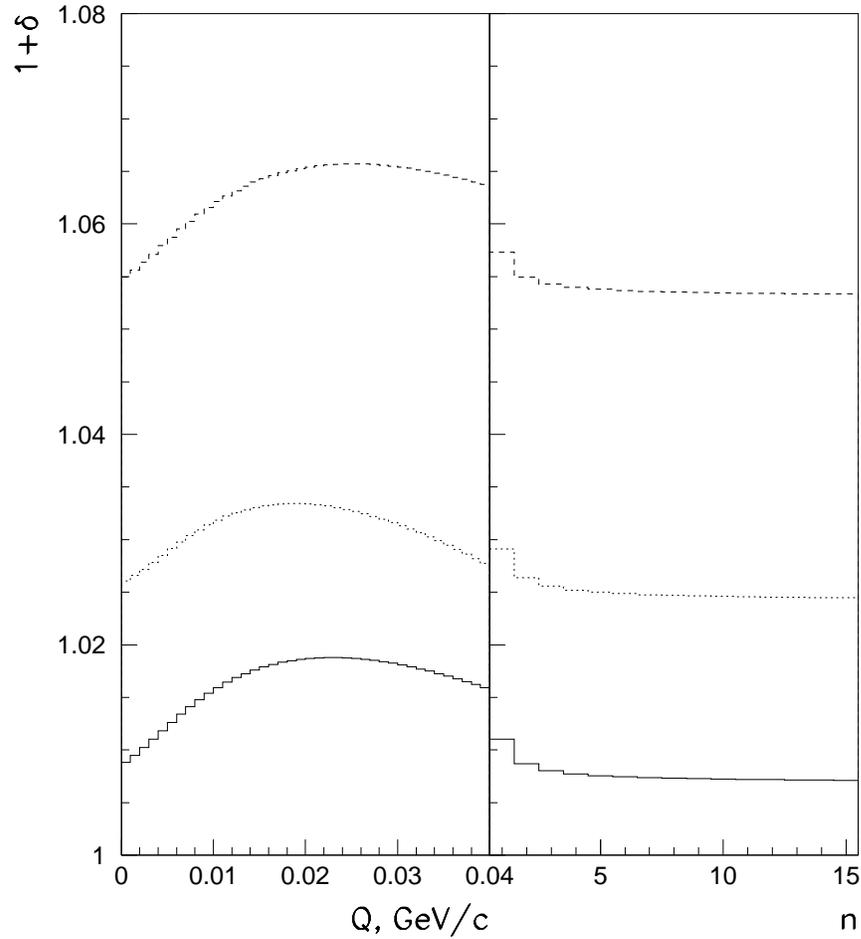}
\caption{
The $\pi^+\pi^-$ correction factors $1+\delta(k^*)$
(left) and $1+\delta_{n}$ (right)
calculated in the same way as in figure~\ref{figroa1_dif}
assuming
the mixtures of $1\%$ $\eta'$, $19\%$ $\omega$ and $80\%$
${\cal G}$
contributions with $r_{\cal G}=3$ fm (lower and middle) and
$r_{\cal G}=2$ fm (upper).
The lower and upper histograms correspond to
the two-pion scattering amplitudes from \cite{cgl01},
the middle one -- to those from \cite{nag79}.
}
\label{figroa2_dif}
\end{figure}

In figure~\ref{figroa2_dif} we plot the correction factors
corresponding to the mixture of $1\%$ $\eta'$,
$19\%$ $\omega$ and $80\%$ short-distance contributions,
as expected from the UrQMD simulation of low-$Q$ pairs
of charged pions in conditions of the experiment DIRAC.
We neglect here the dependence of the contributions
on $Q$, as well as the
dependence on the pion charges. In fact,
within the analysis region of $Q<15$ MeV/$c$, the simulated
$\omega$ contribution increases with decreasing $Q$
by $\sim 0.01$ and
its average value for $\pi^+\pi^-$ pairs composes
$\sim 0.15$ \cite{smol}.

\begin{figure}
\epsfig{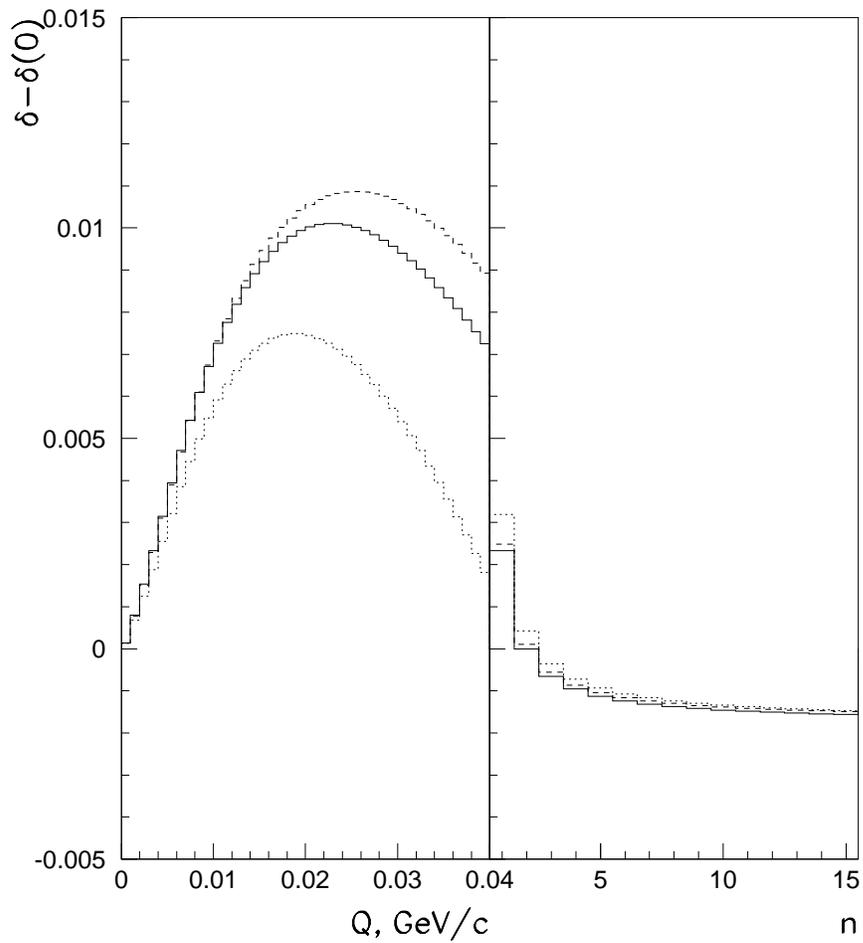}
\caption{
The differences $\delta(k^*)-\delta(0)$ (left)
and $\delta_n-\delta(0)$ (right)
corresponding to the correction factors in
figure~\ref{figroa2_dif}.
}
\label{figroa2a_dif}
\end{figure}

To show the effect of a possible
uncertainty in the short-distance part,
we describe it by the Gaussians with different
characteristic radii $r_{\cal G}= 3$ and 2 fm.
To account for the uncertainty in the two-pion
scattering amplitudes, we have used those from
\cite{cgl01} ($f_0=0.186$ fm) and
\cite{nag79} ($f_0=0.232$ fm).
One may see that the corresponding global
variations of the correction
factors compose $\sim 5\%$ and $\sim 2\%$, respectively.
In figure~\ref{figroa2a_dif}, we plot the same factors with the
subtracted values of the intercept $1+\delta(0)$.
One may see that after the subtraction, the correction factors
calculated for the same two-pion
scattering amplitude but at different values of $r_{\cal G}$
practically coincide for any $n$ in discrete spectrum
and for $Q<20$ MeV/$c$ in continuous spectrum.
Since the subtraction can be included in the overall
normalization factor, one may
conclude that the uncertainty in the
short-distance part of the $r^*$-distribution is of minor
importance for the relative momenta $Q<20$ MeV/$c$.
As for the effect of $\sim 20\%$ increase of the
s-wave elastic $\pi^+\pi^-$ scattering length,
it leads to $\sim 20\%$ increase of $\delta_n-\delta_\infty$
and to $\sim 20\%$ decrease of $\delta(k^*)-\delta(0)$
at $Q=12$ MeV/$c$.

\begin{figure}
\epsfig{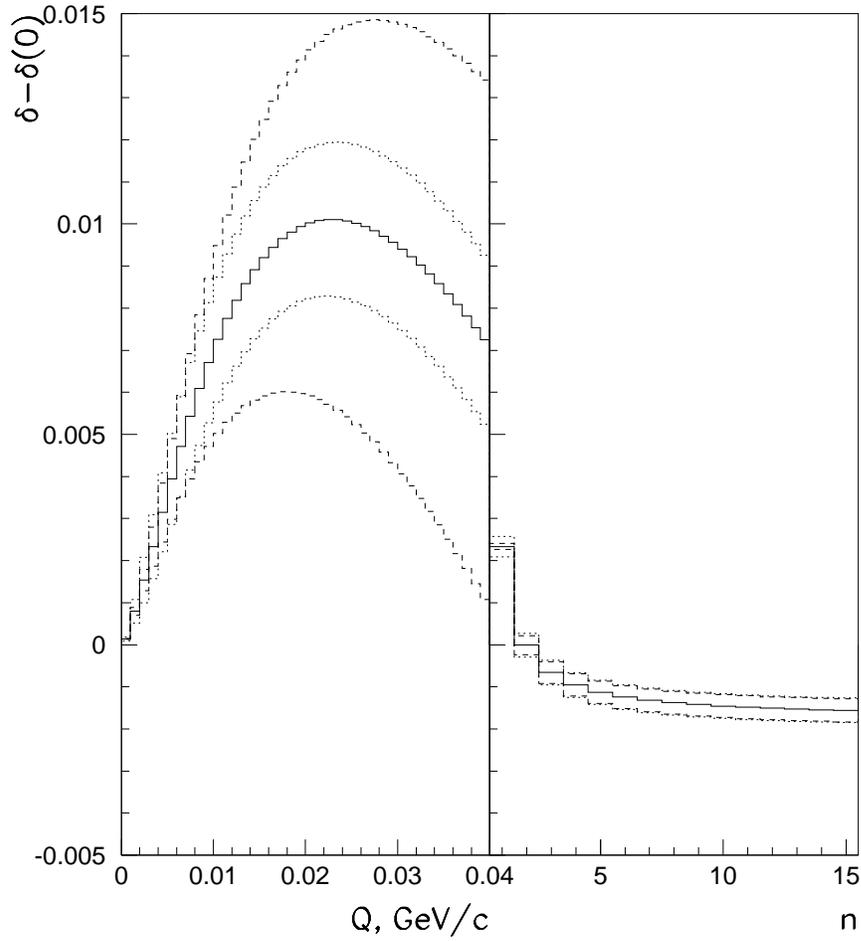}
\caption{
The differences $\delta(k^*)-\delta(0)$ (left)
and $\delta_n-\delta(0)$ (right).
The full histogram coincides with that in
figure~\ref{figroa2a_dif}.
The dashed ones correspond to the $0.19\pm 0.06$
$\omega$ contributions and the dotted ones -- to the
$0.010\pm 0.003$ $\eta'$ contributions.
}
\label{figroa2ab_col}
\end{figure}

To estimate the effect of the uncertainties in the
$\omega$ and $\eta'$ contributions, we plot in
figure~\ref{figroa2ab_col} the differences $\delta-\delta(0)$,
varying these contributions by $\sim 30\%$.
One may see that the corresponding variations of the
differences respectively compose
$\sim 30\%$ and $\sim 20\%$ for
$\delta(k^*)-\delta(0)$ at $Q=12$ MeV/$c$
and, they are quite small ($<0.0003$) for
$\delta_n-\delta(0)$.
It should be noted that
the $\eta'$ contribution to the correction factor at $Q>Q_c$
is quite close to the infinite-size contribution $1/A_c$.
The latter is included in the fit of the non-atomic
$\pi^+\pi^-$ correlation function thus
essentially reducing the corresponding uncertainty in
the breakup probability.
This is demonstrated in table \ref{fits}, where
the contributions $-\Delta N_A/N_A$ and
$\Delta N_A^{\rm br}/N_A^{\rm br}$ to the relative shifts
$\Delta P_{\rm br}/P_{\rm br}=-\Delta N_A/N_A +
\Delta N_A^{\rm br}/N_A^{\rm br}$
of the breakup probability due to the neglect
of finite-size corrections,
corresponding to different mixtures of the $\eta'$,
$\omega$ and ${\cal G}$ contributions
and different fit and signal intervals,
are presented
(see (\ref{DPbr})-(\ref{DNAbr})).
One may see that the $30\%$ uncertainty
in the $\eta'$ contribution leads to
negligible variations in the relative shifts
$\Delta N_A/N_A ~(<0.03\%)$ and
$\Delta N_A^{\rm br}/N_A^{\rm br} ~(<0.2\%)$.
  \Table{\label{fits}
  The contributions $-\Delta N_A/N_A$ and
  $\Delta N_A^{\rm br}/N_A^{\rm br}$ to the relative shift
  $\Delta P_{\rm br}/P_{\rm br}=-\Delta N_A/N_A +
  \Delta N_A^{\rm br}/N_A^{\rm br}$
  of the breakup probability (\ref{pbr}) due to the neglect
  of finite-size corrections.
  The non-atomic
  $\pi^+\pi^-$  correlation functions,
  calculated according to (\ref{Rcna}) for
  different mixtures of the
  $\eta'$, $\omega$ and ${\cal G}(r_{\cal G})$ contributions,
  were fitted by (\ref{Rapp}) (fits $i=1-7$)
  and $\Delta N_A/N_A $ and
  $\Delta N_A^{\rm br}/N_A^{\rm br}$ were calculated
  according to (\ref{DNA}) and (\ref{DNAbr}).
  In approximate correspondence with \cite{ade05},
  a uniform population of non-correlated pion pairs in $Q$
  was assumed in the considered fit intervals $(Q_1,Q_2)$
  and the ratios $N^{\rm cna}_{\pi^+\pi^-}/N_A^{\rm br}$ in the
  signal intervals $(0,Q_{\rm cut})$ were
  set equal to 16, 9 and 4
  for $Q_{\rm cut}=$ 4, 3 and 2 MeV/$c$, respectively.
    }
     \br
        FIT $i$ &
        1&
        2&
        3&
        4&
        5&
        6&
        7&
        fit&
        signal\\
        $r_{\cal G}$~fm&
        3&
        3&
        3&
        3&
        3&
        3&
        2&
        region&
        cut
        \\
        $\omega~\%$ &
        19&
        25&
        13&
        19&
        19&
        19&
        19&
        $Q_1,Q_2$&
        $Q_{\rm cut}$
        \\
        $\eta'~\%$ &
        1.0&
        1.0&
        1.0&
        1.3&
        0.7&
        1.0&
        1.0&
        MeV/$c$&
        MeV/$c$
        \\
        phase shifts&
        \cite{cgl01}&
        \cite{cgl01}&
        \cite{cgl01}&
        \cite{cgl01}&
        \cite{cgl01}&
        \cite{nag79}&
        \cite{cgl01}& &
        \\
        \mr
$-\Delta N_A/N_A~\%$ &1.09&1.61&0.58&1.11&1.06&0.89&1.14&4, 20&-\\
$\Delta N_A^{\rm br}/N_A^{\rm br}~\%$ &4.06&6.77&1.30&4.22&3.86&2.69&4.48& &4\\
                              &3.34&5.46&1.18&3.47&3.18&2.26&3.63& &3\\
                              &2.08&3.37&0.80&2.17&1.98&1.44&2.24& &2
       \\
       \mr
$-\Delta N_A/N_A~\%$ &1.02&1.43&0.61&1.04&0.99&0.90&1.03&4, 15&-\\
$\Delta N_A^{\rm br}/N_A^{\rm br}~\%$ &3.62&5.57&1.50&3.70&3.42&2.66&3.82& &4\\
                              &3.00&4.59&1.32&3.09&2.85&2.25&3.14& &3\\
                              &1.89&2.88&0.88&1.96&1.80&1.44&1.96& &2
\\
\mr
$-\Delta N_A/N_A~\%$ &0.83&1.11&0.55&0.85&0.81&0.78&0.83&4, 10&-\\
$\Delta N_A^{\rm br}/N_A^{\rm br}~\%$ &2.43&3.62&1.16&2.53&2.35&1.92&2.51& &4\\
                              &2.12&3.14&1.06&2.21&2.04&1.70&2.17& &3\\
                              &1.39&2.05&0.72&1.46&1.33&1.12&1.41& &2
       \\
       \br
\endTable

As for the uncertainty of the short-distance part of the
$r^*$-distribution, introduced by $30\%$ decrease of the
Gaussian radius from 3 to 2 fm, it also leads to
negligible changes of the relative shifts
$\Delta N_A/N_A ~(<0.05\%)$ and
$\Delta N_A^{\rm br}/N_A^{\rm br} ~(<0.4\%)$
that rapidly decrease with decreasing upper boundaries
of the fit and signal intervals.

One can also neglect the present $\sim 5\%$ uncertainty
in the $\pi^+\pi^-$ scattering length $f_0$.
Thus even the variation of $f_0$ by $20\%$ leads to rather
small variations of the relative shifts
$\Delta N_A/N_A$ and $\Delta N_A^{\rm br}/N_A^{\rm br}$;
e.g., for $Q_{\rm cut}=4$ MeV/$c$
and the fit interval $(4,15)$ MeV/$c$ they compose only
$0.12\%$ and $0.96\%$, respectively.

\begin{figure}
\epsfig{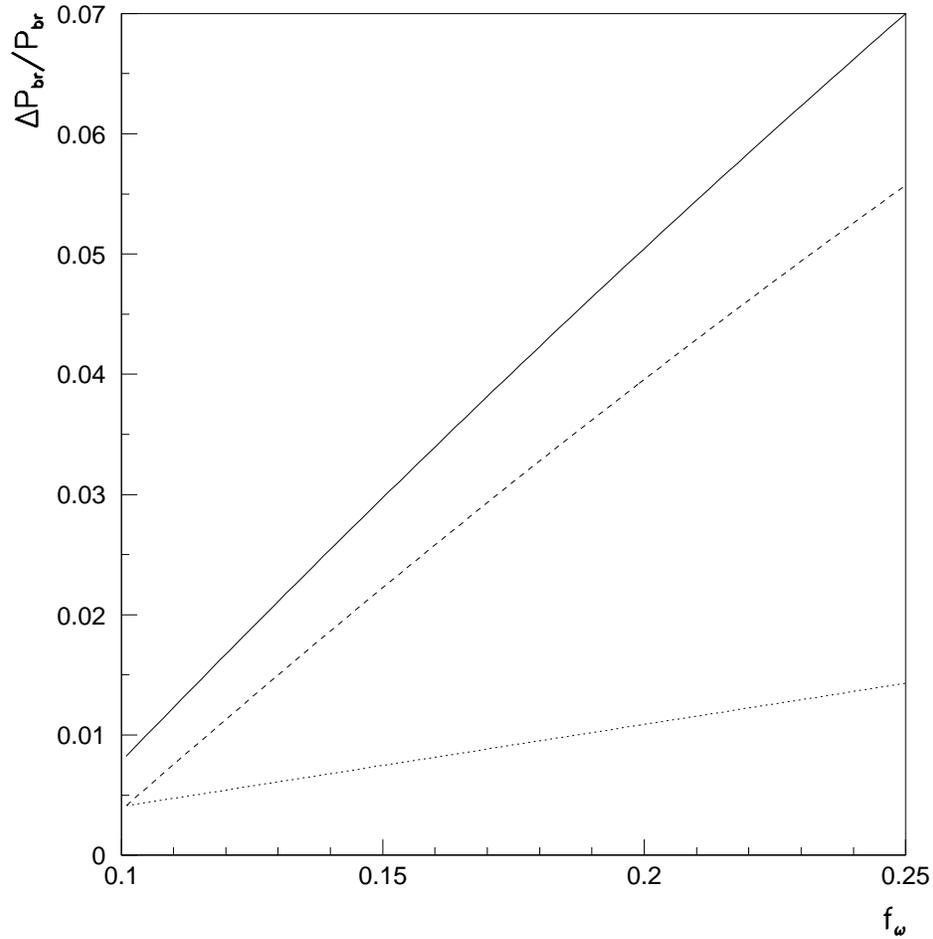}
\caption{Relative shift of the breakup probability
$\Delta P_{\rm br}/P_{\rm br}=-\Delta N_A/N_A +
\Delta N^{\rm br}_A/N^{\rm br}_A$
(full curve) due to the neglect
of the finite-size effect as a function of the fraction
$f_\omega$ of $\pi^+\pi^-$ pairs containing a pion from
$\omega$ decay and the other pion from any short-lived
source, except for pion pairs from one and the same $\omega$ decay.
The fit and signal intervals are $(4,15)$ and $(0,4)$
MeV/$c$ respectively (see table \ref{fits}).
Also shown are contributions of the
relative shifts $-\Delta N_A/N_A$ (dotted curve)
and $\Delta N^{\rm br}_A/N^{\rm br}_A$ (dashed curve)
in the calculated numbers of produced atoms
and breakup atoms respectively.
}
\label{figdpbr}
\end{figure}

The dominant uncertainty in the finite-size correction
to the breakup probability arises from the uncertainty
in the $\omega$ contribution.
One may see from figure \ref{figdpbr} that the
correction $\Delta P_{\rm br}$ almost linearly increases with the fraction
$f_\omega$ of $\pi^+\pi^-$ pairs containing
a pion from $\omega$ decay and the other pion from any short-lived
source, except for pion pairs from one and the same $\omega$ decay;
$\Delta P_{\rm br}/P_{\rm br}\approx -0.032+0.41 f_\omega$
for $Q_{\rm cut}=4$ MeV/$c$
and the fit interval $(4,15)$ MeV/$c$.
Thus, taking $f_\omega =0.19$, a $30\%$ ($\pm 0.06$) variation in $f_\omega$
leads to $\sim 50\%$ ($\pm 0.024$) variation
in $\Delta P_{\rm br}/P_{\rm br}$.

The correction rapidly decreases with decreasing
upper boundaries of the fit and signal intervals.
A decrease of the boundaries is however limited due to
the increase of statistical errors.
Also, the decrease of $Q_{\rm cut}$ below 3 MeV/$c$
introduces a systematic shift of $\sim 5\%$
in the breakup probability due to possibly
insufficiently accurate description of the shape
of the $Q$-spectrum of the atomic breakup
$\pi^+\pi^-$ pairs \cite{ade05}.
The optimal choice seems to be $Q_{\rm cut}=4$ MeV/$c$
and the fit interval $(4,15)$ MeV/$c$.

Taking
$f_\omega= 0.15$,
the overestimation of the breakup probability
in these fit and signal intervals composes
$3\%$
and corresponds to
$\sim 7.5\%$
overestimation of the pionium lifetime.
Correcting for this overestimation and
assuming rather conservative $30\%$
uncertainty in the $\omega$ contribution, the
uncertainty in the breakup probability composes
$2\%$,
corresponding to
$\sim 5\%$
uncertainty in the extracted pionium lifetime.

One may expect an increase of the finite-size correction
and its uncertainty when taking into account the increased slope of
$\delta(k^*)$ at $Q > Q_T^{\rm cut}= 4$ MeV/$c$.
On the other hand, one may expect a reduction of the correction
when extending the fit region down to $Q=0$ and performing a more
constrained 2-dimensional fit of the $(Q_T,Q_L)$-distribution
taking into account also the shape of the spectrum of atomic pion
pairs.

\subsection{$\pi^-\pi^-$ and $\pi^+\pi^+$ systems}
As a by-product, the experiment DIRAC provides
data on the correlation functions of identical
charged pions which contains the information on the
space-time characteristics of pion production and can be
used to check the results of the UrQMD simulations.
Thus, $\sim 5\cdot 10^5$ $\pi^-\pi^-$ pairs
have been collected in pNi interactions in 2001,
representing $\sim 40\%$ of the available statistics
\cite{smol}. The $Q$-distribution of these pairs is
peaked at $\sim 60$ MeV/$c$ and drops essentially
outside the interval $(20,120)$ MeV/$c$ due to
a decrease of the phase space and detector acceptance
at small and large $Q$ respectively.

Contrary to the case of the $\pi^+\pi^-$ system, the
correlation effect in the system of identical pions extends
and is measured up to the relative momenta $Q\sim 200$ MeV/$c$,
so neither the distribution of the vector ${\bf Q}$ nor
the detector acceptance can be considered independent of
the direction of this vector.
Since further the angular distribution of the vector
${\bf r}^*$ is not isotropic (particularly,
the characteristic width of the out component of the
${\bf r}^*$-distribution increases with the transverse
momentum while those of the side and longitudinal
ones decrease), the required space-time information
does not reduce to the distribution
of the relative distance $r^*$ between the pion production
points in the pair c.m. system; generally, the 3-dimensional
distribution of the vector ${\bf r}^*$ is required.
Here we however neglect this complication and calculate the
1-dimensional correlation function of two identical charged
pions in the same way as for the previously considered case
of the near-threshold $\pi^+\pi^-$ system, i.e. assuming
the uniform distribution of the cosine of the angle between
the vectors ${\bf Q}$ and ${\bf r}^*$ for the non-correlated
pions.

\begin{figure}
\epsfig{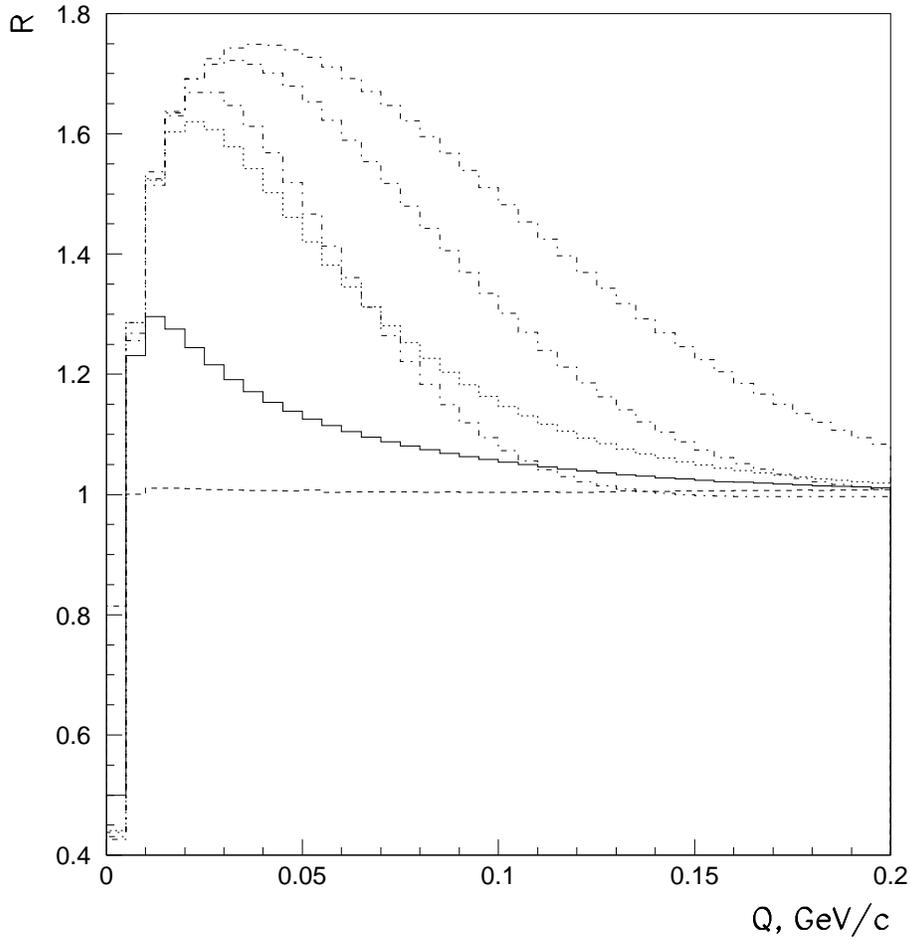}
\caption{The $\pi^-\pi^-$ correlation functions.
The histograms in the increasing order of the peak
values correspond to the $r^*$-distributions
$\eta', \omega, {\cal M}(r^*;9.20 {\rm fm},0.656,2.86)$,
${\cal G}(r^*;3\mbox{fm})$, ${\cal G}(r^*;2\mbox{fm})$
and ${\cal G}(r^*;1.5\mbox{fm})$ respectively.
}
\label{figcfmm1}
\end{figure}

\begin{figure}
\epsfig{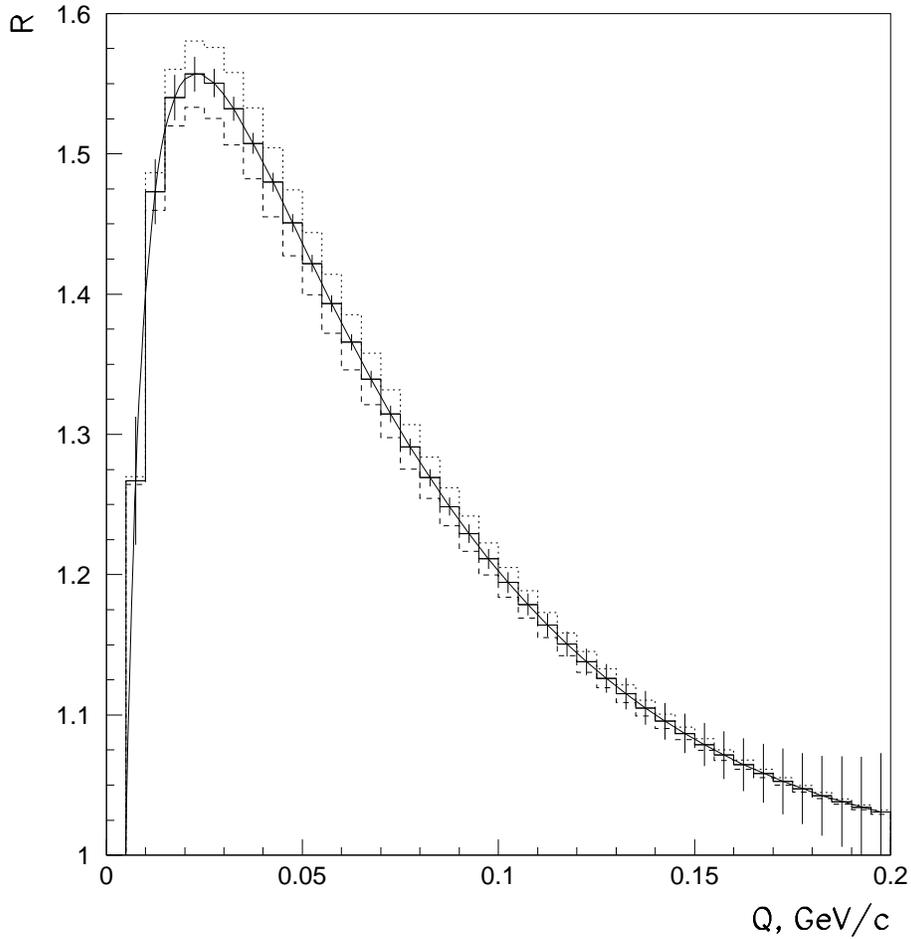}
\caption{The $\pi^-\pi^-$ correlation functions.
The middle histogram and the fit curve
correspond to $1\%$ $\eta'$, $19\%$ $\omega$, $60\%$
${\cal M}(r^*;9.20 {\rm fm},0.656,2.86)$
and $20\%$ ${\cal G}(r^*;1.5 {\rm fm})$ contributions.
The errors are taken from the DIRAC pNi 2001
data \cite{smol}.
The upper and lower histograms correspond to the
$1\%$ $\eta'$, $19\%\pm 6\%$ $\omega$ contributions
and the unchanged ratio 3:1 of the
${\cal M}$ and ${\cal G}$ contributions
(unchanged form of the short-distance contribution).
}
\label{figcfmm2}
\end{figure}

The calculated $\pi^-\pi^-$ correlation functions
${\cal R}^{--}_{\eta'}$, ${\cal R}^{--}_{\omega}$,
${\cal R}^{--}_{\cal M}$ and ${\cal R}^{--}_{\cal G}$
corresponding to the $r^*$-distributions
$\eta', \omega, {\cal M}(r^*;9.20 \mbox{fm},0.656,2.86)$,
and ${\cal G}(r^*;r_{\cal G})$, $r_{\cal G}=3,2,1.5~\mbox{fm}$,
are shown in figure~\ref{figcfmm1}.
In figure~\ref{figcfmm2}, we
show the correlation function corresponding to
$1\%$ $\eta'$, $19\%$ $\omega$, $60\%$
${\cal M}(r^*;9.20 \mbox{fm},0.656,2.86)$
and $20\%$ ${\cal G}(r^*;1.5\mbox{fm})$ contributions,
as expected for the pairs of charged pions from the
UrQMD simulations; the errors are taken from the
DIRAC pNi 2001 data.
To demonstrate the sensitivity to the
relative $\omega$ contribution,
$f_\omega$, we show in this figure also the correlation functions
calculated with $f_\omega$ varied by $\sim 30\%$.
One can conclude, that the different shape of the
$\omega$ contribution as compared with
the shapes of the short-distance ones
(${\cal M}$ and ${\cal G}$) allows, in principle,
to determine $f_\omega$ -- the most critical parameter
required to calculate the finite-size $\pi^+\pi^-$
correction factors.
To estimate the statistics required to determine $f_\omega$
better than to $30\%$, we have fitted the correlation function
in figure~\ref{figcfmm2} by
\begin{equation}
\label{Rmm}
\eqalign{
{\cal R}(Q)=N \left\{
\lambda\left[
f_{\eta'}{\cal R}^{--}_{\eta'}(Q)+
f_\omega{\cal R}^{--}_{\omega}(Q)+
f_{\cal G}{\cal R}^{--}_{\cal G}(Q;r_{\cal G})
\right.\right.
\cr
\left.\left.
+
(1-f_{\eta'}-f_\omega-f_{\cal G})
{\cal R}^{--}_{\cal M}(Q;r_{\cal M},\alpha,\beta)
\right]+(1-\lambda)
\right\}
(1+b Q).
}
\end{equation}
The parameter $N$ cares for possible normalization mismatch,
the correlation strength parameter $\lambda$
takes into account the contribution of LL
emitters, particle misidentification as well as
coherence effects and
the slope parameter $b$ cares for a possible mismatch
in the $Q$-dependence of the reference sample in the
denominator of the correlation function.
Equation (\ref{Rmm}) does not take into account
a decrease of the fractions
$f_{\eta'}$ and $f_{\omega}$ with increasing $Q$
which can amount to $\sim 20\%$ in the interval
0-200 MeV/$c$ \cite{smol}.

The dependence of the correlation function on the
parameters $r_{\cal G}$, $r_{\cal M}$, $\alpha$,
$\beta$ is calculated with the help of quadratic
interpolation. For example, to calculate
the $r_{\cal G}$-dependence, three values of this parameter
are chosen and the corresponding correlation functions
${\cal R}^{--}_{\cal G}(Q;r_{\cal G}^i)$, $i=1,2,3$,
are used to interpolate to the correlation function
at a given value of $r_{\cal G}$ according to
(\ref{interp1}) with the substitutions
${\cal R}_{\rm RQMD}\to {\cal R}^{--}_{\cal G}$ and
$s_{r}\to r_{\cal G}$.
Similarly, the quadratic interpolation in the three-parameter
space $r_{\cal M}$, $\alpha$, $\beta$ requires the calculation
of $3^3$ correlation functions
${\cal R}^{--}_{\cal M}(Q;r_{\cal M}^i,\alpha^j,\beta^k)$,
$i,j,k=1,2,3$.

The fit recovers the input parameters with rather small
parabolic errors, particularly, $f_\omega=0.189\pm 0.048$.
It appears, however, that the real error, corresponding to
the increase of the $\chi^2$ by one unit, is one order
of magnitude larger. Thus, to achieve the error in $f_\omega$
smaller than $30\%$, one has to
collect the statistics of $\sim 5\cdot 10^7$ $\pi^-\pi^-$ pairs.
The fit results corresponding to such a statistics are shown in
table \ref{fitmm}, fits 1-6.
  \Table{\label{fitmm}
  The results of the fits according to (\ref{Rmm}) of the
  $\pi^-\pi^-$  correlation function corresponding to
  the mixture of the $1\%$
  $\eta'$, $19\%$ $\omega$, $60\%$
  ${\cal M}(r_{\cal M},\alpha,\beta)$ and
  $20\%$ ${\cal G}(r_{\cal G})$ contributions
  with $r_{\cal G}=1.5$ fm, $r_{\cal M}=9.2$ fm,
  $\alpha=0.656$, $\beta=2.863$. The errors correspond
  to the statistics of $5\cdot 10^7$ $\pi^-\pi^-$
  pairs from pNi interactions in the conditions of the
  experiment DIRAC.
  Only the parabolic errors are shown,
  absent error means a fixed parameter.
    }
     \br
        FIT $i$ &~~1&~~2&~~3&~~4&~~5&~~6\\
        \mr
        $N$&~~1.000&~~0.999 &~~1.000&~~1.002&~~0.998 &~~1.017 \\
        &$\pm 0.001$&$\pm 0.001$ &$\pm 0.001$&$\pm 0.001$&$\pm 0.001$ &$\pm 0.001$  \\
        $\lambda$&~~1.001&~~0.993 & ~~0.999& ~~0.973&~~1.024 &~~0.984  \\
        &$\pm 0.004$&$\pm 0.003$ &$\pm 0.004$&$\pm 0.003$&$\pm 0.003$ &$\pm 0.003$& \\
        $r_{\cal M}$~fm&~~9.253&~~9.239 &~~9.161& ~~7.381&~10.844 &~~6.322   \\
        &$\pm 0.048$&$\pm 0.049$ &$\pm 0.048$&$\pm 0.045$&$\pm 0.065$ &$\pm 0.032$ \\
        $\alpha$&~~0.657&~~0.651 &~~0.655& ~~0.611&~~0.746 &~~0.561 & \\
        &$\pm 0.003$&$\pm 0.003$ &$\pm 0.003$&$\pm 0.002$& $\pm 0.004$&$\pm 0.001$ \\
        $\beta$&~~2.864&~~2.893 &~~2.853& ~~2.717&~~2.813 &~~3.001 & \\
        &$\pm 0.010$& $\pm 0.010$&$\pm 0.009$&$\pm 0.008$&$\pm 0.013$ &$\pm 0.006$ \\
        $r_{\cal G}$~fm&~~1.495&~~1.492 &~~1.496& ~~1.505&~~1.554 &~~1.5 \\
        &$\pm 0.016$& $\pm 0.016$&$\pm 0.015$&$\pm 0.022$&$\pm 0.013$ & \\
        $f_{\cal G}$&~~0.202&~~0.199 &~~0.200& ~~0.147&~~0.261 &~~0. \\
        &$\pm 0.003$&$\pm 0.003$ &$\pm 0.003$&$\pm 0.004$&$\pm 0.003$ &~~ \\
        $f_\omega$&~~0.187&~~0.190 &~~0.184& ~~0.131&~~0.238 &~~0.234 \\
        &$\pm 0.004$&$\pm 0.005$ &$\pm 0.005$&~~&~~ &$\pm 0.004$ \\
        $f_{\eta'}$&~~0.010&~~0.&~~0.01&~~0.01&~~0.01&~~0.01 \\
        &$\pm 0.004$&~~ &~~&~~&~~ &~~ \\
        $b~(\mathrm{GeV}/c)^{-1}$&~~0.001&~~0.000 &~~0.000&~-0.033&~~0.036 &~-0.162 \\
        &$\pm 0.008$&$\pm 0.008$ &$\pm 0.007$&$\pm 0.008$&$\pm 0.008$ &$\pm 0.006$ \\
        $\chi^2$&~~0.01&~~0.02 &~~0.01&~~1.01&~~1.01 &~~33.9
        \\
       \br
\endTable
Comparing fits 1-3, one may see that they are practically
insensitive to the nearly flat $\eta'$ contribution
to the correlation function,
except for $\sim 20\%$ drop in the first 5 MeV/$c$ bin.
Since the drop contribution of $0.2\cdot 0.01$ is much
smaller than the correlation function error of 0.02 in
the first bin,
the 0.01 shift in $f_{\eta'}$ in fit 2 ($f_{\eta'}=0$)
leaves the parameters and $\chi^2$
practically unchanged, except for the compensating
$\sim 0.01$ shift in the
correlation strength parameter $\lambda$.
Comparing further fits 3-5, one can conclude that the true error in
$f_\omega$, corresponding to the increase of $\chi^2$ by one unit,
composes $\sim 25\%$ ($\sim 0.05$).
Finally, fit 6 ($f_{\cal G}=0$) shows that
the oversimplified
description of the short-distance contribution could lead
to $\sim 25\%$ systematic shift of $f_\omega$.

To infer the fraction $f_{\omega}^{+-}$
for $\pi^+\pi^-$ pairs at $Q\to 0$ from the fitted
$f_{\omega}^{\pm\pm}$
for identical charged pions, one has to take
into account its $Q$-dependence as well as the fact that,
due to a lower multiplicity of the pairs of identical
charged pions, the fraction $f_{\omega}^{\pm\pm}$
is $\sim 40\%$ higher than $f_{\omega}^{+-}$ \cite{smol}.

\section{Conclusions}
We have developed a practical formalism allowing one
to quantify the effect of a finite space-time extent
of particle emission region on the two-particle production
in continuous and discrete spectrum.
We have shown that one can usually neglect
the non-equal emission times in the pair c.m. system,
the space-time coherence and the
residual charge.
The developed formalism is in the basis of the femtoscopy
techniques allowing one to measure space-time characteristics
of particle production as well as the
low-energy strong interaction between specific particles.
We have applied it
to the problem of
lifetime measurement of hadronic atoms produced by
a high-energy beam in a thin target,
particularly, to the measurement of pionium lifetime
in the experiment DIRAC at CERN.
Based on the transport code simulations, we have
calculated so called correction factors that can be used
to take into account the finite size of the production
region by multiplying the point-like Coulomb production
cross sections of the free and bound $\pi^+\pi^-$ pairs.
We have shown that the short-distance contribution is of
minor importance for the lifetime measurement
since it leads to practically the same and nearly constant
correction factors for free and bound pairs which
cancel in the breakup probability.
The most important is the fraction $f_\omega$
of $\pi^+\pi^-$ pairs containing a pion from
$\omega$ decay and the other pion from any short-lived
source, except for pion pairs from one and the same $\omega$ decay.
Besides leading to slightly different
global shifts of the correction factors,
it also affects their $Q$- and $n$-dependence.
The resulting correction to the "point-like"
pionium lifetime composes $\sim -7.5\%$.
Assuming rather conservative
$30\%$ uncertainty in $f_\omega$,
due to the uncertainty in the $\omega$ production and
a simplified treatment of the correction
(e.g., the neglect of $f_\omega$ variation in the analyzed
$Q$-interval),
one arrives at $\sim 5\%$ uncertainty
in the extracted pionium lifetime.
It is shown that this uncertainty could be diminished
if the high statistics data on
correlations of identical charged pions
were collected in the DIRAC experiment.
The statistics required to determine $f_\omega$
to 10$\%$ and control the
finite-size effect
on the lifetime to $\sim 2\%$
composes $\sim 3\cdot 10^8$ $\pi^-\pi^-$ pairs.
The uncertainty in $f_\omega$ can be also reduced
by tuning the transport simulations with the help of
experimental data on particle (resonance) spectra and
femtoscopic correlations in proton-nucleus
collisions at the beam energy of $\sim 20$ GeV.
The lifetime uncertainty can be
essentially reduced in future experiments using the
multi-layer targets \cite{ade04} since it will be
basically determined by the uncertainty in the
calculated number $N_A$ of produced atoms only;
even for the conservative $30\%$
uncertainty in $f_\omega$,
the corresponding uncertainty in
the lifetime will be $\sim 1\%$ only.
The above estimates of the finite-size correction
to the pionium lifetime and its uncertainty
are based on a 1-dimensional fit of the
$Q$-distribution in the interval $(4,15)$ MeV/$c$.
One may expect their underestimation due to the neglected
increase of the slope of $Q$-distribution
for $Q > Q_T^{\rm cut} = 4$ MeV/$c$. On the other hand,
one may expect their reduction when extending the fit interval
down to $Q=0$ and performing a more constrained
2-dimensional fit of the $(Q_T,Q_L)$-distribution.
The effect of the finite-size uncertainty
on the breakup probability of other hadronic atoms
remains to be studied. For $\pi K$ and $\pi p$
atoms, it is expected similar to the one for
$\pi^+\pi^-$ atoms since $\sim 50\%$ decrease
of the $\omega$ contribution is compensated by
about the same decrease of the Bohr radius $|a|$
thus retaining a
similar finite-size Coulomb FSI effect
$\sim \langle r^*\rangle^{\mbox{\tiny SL}}/a$.
As for $K^+K^-$, $K^- p$ and $\bar{p}p$ atoms,
there is no $\omega$ contribution to the corresponding
hadron pairs though, due to the smaller Bohr radii,
its effect can be partly substituted
by the contribution of $\phi$-meson
and other sufficiently narrow resonances.

\section*{Acknowledgements}
The author thanks Vladimir Lyuboshitz, Leonid Nemenov,
Jan Smol\'{\i}k and Valery Yazkov for useful
discussions. This work was supported by the
Grant Agency of the Czech Republic under contracts
202/01/0779, 202/04/0793 and 202/07/0079.

\appendix
\section{Non-equal emission times}
%
%

We consider here the role of non-equal emission times
in the Bethe--Salpeter amplitude
$\psi_{\widetilde{q}}(x)={\rm e}^{{\rm i}\tilde qx/2}+
\Delta\psi ^{(+)}_{\widetilde{q}}(x)$,
where the correction
$\Delta\psi $ to the plane wave is given
in (\ref{7}).
We will consider the amplitude in the pair c.m. system,
in which
the plane wave ${\rm e}^{{\rm i}\tilde qx/2}=
{\rm e}^{-{\rm i}{\bf k}^*{\bf r}^*}$
is independent of the emission times.
First, we will prove the integral relation between the
Bethe--Salpeter amplitude and the corresponding
non-relativistic wave function, derived on the condition
${k}^{*2}\ll \mu^2$ \cite{ll82}:
\begin{equation}
\label{70} \psi ^{(+)}_{\widetilde{q}}(x) =
\int {\rm d}^3{\bf r}'
\delta_{{k}^{*}}({\bf r}^*-{\bf r}',t^*)
\psi_{-{\bf k}^*}({\bf r}')
\end{equation}
\begin{equation}
\label{70'}
\delta_{{k}^{*}}({\bf r}^*-{\bf r}',t^*)=
\frac{1}{(2\pi)^3}
\int {\rm d}^3\mbox{\boldmath $\kappa$}
{\rm e}^{-{\rm i}\mbox{\boldmath $\kappa$}({\bf r}^{*}-{\bf r}')}
\exp\!\left(-{\rm i}\frac{\mbox{\boldmath
$\kappa$}^2-{\bf k}^{*2}}{2m(t^{*})}|t^{*}|\right),
\end{equation}
where $m(t^{*}>0) = m_2$ and $m(t^*<0) = m_1$.

We start by splitting the product of the propagators
into four terms, each containing only two poles in the
complex $\kappa_0$-plane,
situated in the opposite upper and lower half-planes.
Taking into account that in the pair c.m. system ${\bf P}=0$
and that the pair energy coincides with its effective
mass: $P_0=m_{12}$, we get
\begin{equation}
\label{propag}
\eqalign{
\{(\kappa^2-m_1{}^2+{\rm i}0)[(P-\kappa)^2-m_2{}^2+{\rm i}0]\}^{-1}
\cr
\qquad
=[(\kappa_0-\widetilde{\omega}_1+{\rm i}0)(\kappa_0+
\widetilde{\omega}_1-{\rm i}0)
\cr
\qquad \qquad
\cdot
(\kappa_0-m_{12}-\widetilde{\omega}_2+{\rm i}0)
(\kappa_0-m_{12}+\widetilde{\omega}_2-{\rm i}0)]^{-1}
\cr
\qquad
=[m_{12}{}^2-(\widetilde{\omega}_1-
\widetilde{\omega}_2)^2]^{-1}
\{[(\kappa_0-\widetilde{\omega}_1+{\rm i}0)(\kappa_0+
\widetilde{\omega}_1-{\rm i}0)]^{-1}
\cr
\qquad \qquad
+[(\kappa_0-m_{12}-\widetilde{\omega}_2+{\rm i}0)
(\kappa_0-m_{12}+\widetilde{\omega}_2-{\rm i}0)]^{-1}
\cr
\qquad \qquad
-[(\kappa_0-\widetilde{\omega}_1+{\rm i}0)(\kappa_0-m_{12}+
\widetilde{\omega}_2-{\rm i}0)]^{-1}
\cr
\qquad \qquad
-[(\kappa_0+\widetilde{\omega}_1-{\rm i}0)(\kappa_0-m_{12}-
\widetilde{\omega}_2+{\rm i}0)]^{-1}\},
}
\end{equation}
where $\widetilde{\omega}_i=(m_i{}^2+
\mbox{\boldmath $\kappa$}^2)^{1/2}$.
Assuming now that the amplitude
$f^S\equiv f^S(\kappa_0,m_{12}-\kappa_0)$
is an analytical function in the complex
$\kappa_0$-plane, we can integrate over $\kappa_0$
using the residue theorem. Consider first $t^*>0$.
In this case the integration contour has to be
closed in the upper half-plane, equation (\ref{7})
then giving
\begin{equation}
\label{dpsi1}
\eqalign{
\Delta\psi ^{(+)}_{\widetilde{q}}(x) =
\frac{1}{\pi^2}m_{12}
{\rm e}^{-{\rm i}[m_{12}+(m_1{}^2-m_2{}^2)/m_{12}]t^*/2}
\int\frac{{\rm d}^3\mbox{\boldmath $\kappa$}{\rm e}^
{-{\rm i}\mbox{\boldmath $\kappa$}{\bf r}^*}}
{m_{12}{}^2-(\widetilde{\omega}_1-
\widetilde{\omega}_2)^2}
\cdot \cr
\qquad
\cdot\left[{\rm e}^{-{\rm i}\widetilde{\omega}_1t^*}f(-
\widetilde{\omega}_1,
m_{12}+\widetilde{\omega}_1)\left(\frac{1}{m_{12}+
\widetilde{\omega}_1+
\widetilde{\omega}_2}-\frac{1}{2\widetilde{\omega}_1}
\right)-
\right.
\cr
\qquad
\left.
-{\rm e}^{{\rm i}(m_{12}-\widetilde{\omega}_2)t^*}f(m_{12}-
\widetilde{\omega}_2,
\widetilde{\omega}_2)\left(\frac{1}{m_{12}-
\widetilde{\omega}_1-
\widetilde{\omega}_2+{\rm i}0}+\frac{1}{2\widetilde{\omega}_2}
\right)\right].
}
\end{equation}
Since we are interested in the limit of small
particle momenta in the pair cm. system:
${k}^{*2}\ll\mu^2$ and since the integral (\ref{dpsi1})
is dominated by
$\mbox{\boldmath $\kappa$}^2\approx {\bf k}^{*2}$,
we can use the
following non-relativistic approximations
(recall that
$\mu=m_1m_2/(m_1+m_2)$ is the reduced mass of the
two-particle system):
\begin{equation}
\label{nonrel}
\eqalign{
m_{12}\doteq m_1+m_2+\frac{{\bf k}^{*2}}{2\mu}
\cr
m_{12}+(m_1{}^2-m_2{}^2)/m_{12}\doteq 2\left(m_1+
\frac{m_2}{m_1+m_2}
\frac{{\bf k}^{*2}}{2\mu}\right)
\cr
\widetilde{\omega}_i\doteq m_i+
\frac{\mbox{\boldmath $\kappa$}^2}{2m_i}\qquad
m_{12}{}^2-(\widetilde{\omega}_1-
\widetilde{\omega}_2)^2\doteq 4m_1m_2
\cr
m_{12}-\widetilde{\omega}_1-\widetilde{\omega}_2\doteq
\frac{{\bf k}^{*2}-\mbox{\boldmath $\kappa$}^2}{2\mu}.
}
\end{equation}
Retaining in the integral (\ref{dpsi1}) only the
dominant pole term
$\sim [m_{12}-\widetilde{\omega}_1-
\widetilde{\omega}_2+{\rm i}0]^{-1}$,
we get
\begin{equation}
\label{dpsi2}
\eqalign{
\Delta\psi ^{(+)}_{\widetilde{q}}(x) = \frac{1}{2\pi^2}
\int\frac{{\rm d}^3\mbox{\boldmath $\kappa$}{\rm e}^
{-{\rm i}\mbox{\boldmath $\kappa$}{\bf r}^*}}
{\mbox{\boldmath $\kappa$}^2-{\bf k}^{*2}-{\rm i}0}
\cr
\qquad
\cdot
\exp\!\left(-{\rm i}\frac{\mbox{\boldmath $\kappa$}^2-
{\bf k}^{*2}}{2m_2}t^*\right)
f(m_{12}-\widetilde{\omega}_2,\widetilde{\omega}_2).
}
\end{equation}
Using now the equalities
$\Delta\psi ^{(+)}_{\widetilde{q}}({\bf r}^*,t^*=0)
\equiv \Delta\psi_{-{\bf k}^*}({\bf r}^*)$
and $\delta^{(3)}(\mbox{\boldmath $\kappa$}-
\mbox{\boldmath $\kappa$}')=
(2\pi)^{-3}\int {\rm d}^3{\bf r}'\cdot
\exp[i(\mbox{\boldmath $\kappa$}-
\mbox{\boldmath $\kappa$}')
{\bf r}']$, we can write:
\begin{equation}
\label{dpsi3}
\eqalign{
\Delta\psi ^{(+)}_{\widetilde{q}}(x) =
\frac{1}{2\pi^2}
\int {\rm d}^3\mbox{\boldmath $\kappa$}'
\delta^{(3)}(\mbox{\boldmath $\kappa$}-
\mbox{\boldmath $\kappa$}')
\frac{d^3\mbox{\boldmath $\kappa$}{\rm e}^
{-{\rm i}\mbox{\boldmath $\kappa$}{\bf r}^*}}
{\mbox{\boldmath $\kappa$}'^2-{\bf k}^{*2}-{\rm i}0}
\cr
\qquad \qquad
\cdot
\exp\!\left(-{\rm i}\frac{\mbox{\boldmath $\kappa$}^2-
{\bf k}^{*2}}{2m_2}t^*\right)
f(m_{12}-\widetilde{\omega}_2',
\widetilde{\omega}_2')
\cr
\qquad
= \int {\rm d}^3{\bf r}'
\int \frac{{\rm d}^3\mbox{\boldmath $\kappa$}}
{(2\pi)^3}
{\rm e}^{-{\rm i}\mbox{\boldmath $\kappa$}
({\bf r}^{*}-{\bf r}')}
\exp\!\left(-{\rm i}\frac{\mbox{\boldmath
$\kappa$}^2-{\bf k}^{*2}}{2m(t^{*})}t^{*}\right)
\cr
\qquad \qquad
\cdot
\frac{1}{2\pi^2}
\int\frac{{\rm d}^3\mbox{\boldmath $\kappa$}'{\rm e}^
{-{\rm i}\mbox{\boldmath $\kappa$}'{\bf r}'}}
{\mbox{\boldmath $\kappa$}'^2-{\bf k}^{*2}-{\rm i}0}
f(m_{12}-\widetilde{\omega}_2',\widetilde{\omega}_2')
\cr
\qquad
\equiv \int {\rm d}^3{\bf r}'
\delta_{{k}^{*}}({\bf r}^*-{\bf r}',t^*)
\Delta\psi_{-{\bf k}^*}({\bf r}'),
}
\end{equation}
where the $\delta_{{k}^{*}}$-function is given
in (\ref{70'}).
Noting that the $\delta_{{k}^{*}}$-function
in the integral (\ref{dpsi3})
acts on the plane wave ${\rm e}^{-{\rm i}{\bf k}^*{\bf r}'}$
as a $\delta$-function, we finally arrive at the
integral relation in (\ref{70}) for $t^*>0$.
The prove of this relation in the case of $t^*<0$ is
done in a similar way, the integration
$\kappa_0$-contour being now
closed in the lower half-plane. The result is the same
as in (\ref{dpsi1})-(\ref{dpsi3}),
up to the substitutions $m_2\rightarrow -m_1$
in the time-dependent phase factor and
$\widetilde{\omega}_2\rightarrow m_{12}-
\widetilde{\omega}_1$
in the arguments of the scattering amplitude $f$.

At $t^*=0$, the function
$\delta_{{k}^{*}}({\bf r}^*-{\bf r}',0)=
\delta^{(3)}({\bf r}^*-{\bf r}')$ and, at $t^*>0$,
\begin{equation}
\label{70''}
\delta_{{k}^{*}}({\bf r}^*-{\bf r}',t^*)=
\left(\frac{m_2}{2\pi {\rm i} t^*}\right)^{3/2}
\exp\!\left[i\left(\frac{k^{*2} t^*}{2m_2}+
\frac{({\bf r}^{*}-{\bf r}')^2m_2}{2 t^*}\right)\right].
\end{equation}
For negative $t^*$-values, the substitution
$m_2 \rightarrow - m_1$
has to be done in (\ref{70''}).
It is clear from (\ref{70''}) that, at small $k^*$
($k^*\ll m(t^*){r}^*/|t^*|$), the function
$\delta_{{k}^{*}}({\bf r}^*-{\bf r}',t^*)$
practically coincides with the $\delta$-function
$\delta^{(3)}({\bf r}^*-{\bf r}')$ on condition (\ref{10}).

Since the particles start to feel each other only after
both of them are created,
it is clear that a large difference in the emission times
generally leads to a suppression of particle interaction
at small ${k}^{*}$:
$|\Delta\psi ^{(+)}_{\widetilde{q}}(x)|\le
|\Delta\psi_{-{\bf k}^*}({\bf r}^*)|$;
$\Delta\psi ^{(+)}_{\widetilde{q}}(x)\rightarrow 0$ at
$|t^*|\rightarrow\infty$. Particularly instructive
is the case when one of the two particles is very heavy,
say $m_2\gg m_1$. Then the two-particle interaction is
suppressed provided the light
particle is emitted prior the emission of the heavy one
($m(t^*<0)=m_1$ in (\ref{70'})). Otherwise, the
large mass $m(t^*>0)=m_2$ prevents the suppression
even if the light particle were emitted much later
than the heavy one. Below we consider the effect of
non-equal emission times on
two-particle production in some detail.

We start with the FSI due to the short-range forces only.
Inserting the spherical wave (\ref{11}) into the
integral relation
(\ref{70}) or (\ref{dpsi3}), we get \cite{ll82}
\begin{equation}
\label{70sw}
\eqalign{
\Delta\psi ^{(+)}_{\widetilde{q}}(x)
\cr
\qquad
=
\frac{f(k^*)}{{r}^{*}}
\left\{{\rm i}\sin({k}^{*}{r}^{*})+\frac{1-{\rm i}}{2}
\left[{\rm E}_1(z_-){\rm e}^{{\rm i}{k}^{*}{r}^{*}}+
{\rm E}_1(z_+){\rm e}^{-{\rm i}{k}^{*}{r}^{*}}\right]\right\},
}
\end{equation}
where
$z_{\pm}=\left(\frac{m(t^*)}{2|t^*|}
\right)^{1/2}\left({r}^{*}\pm
\frac{{k}^{*}|t^*|}{m(t^*)}\right)$ and
${\rm E}_1(z)=\sqrt{\frac{2}{\pi}}\int\limits_0^z dy
{\rm e}^{{\rm i}y^2}$
is the Fresnel integral.
Note that the length ${k}^{*}|t^*|/m(t^*)
\equiv l_{{k}^{*}}$
can be interpreted classically, for large
${k}^{*}{r}^{*}$,
as a distance traveled by
the first emitted particle until the creation
moment of the second one.
The absolute value of the factor
$({r}^{*}\pm l_{{k}^{*}})$
in the argument $z_+$ ($z_-$) thus corresponds to
the maximal (minimal) possible distance between
the particles in their c.m. system
at the later of the two creation moments.
The effect of non-equal emission times however doesn't
reduce to the modification of the distance ${r}^{*}$,
it survives even at ${k}^{*}=0$. This effect vanishes
in the limit of small $|t^*|$, when $z_-\gg 1$,
${\rm E}_1(z_{\pm})\rightarrow (1+{\rm i})/2$ and
(\ref{70sw})
reduces to the spherical wave (\ref{11}).
In the opposite limit of large $|t^*|$, when
$\left[m(t^*)r^{*2}/(2|t^*|)\right]^{1/2}\ll 1$,
the interaction is suppressed
and the scattered wave
$\Delta\psi ^{(+)}_{\widetilde{q}}(x)$
tends to zero for arbitrary ${k}^{*}$-values.

In the simple static Gaussian model of independent
one-particle
emitters described by the amplitude (\ref{188}),
the applicability condition (\ref{10}) of the {\it equal-time}
approximation can be roughly written in the form
(\ref{10sta}).
Clearly, the latter condition is not satisfied
for very slow particles emitted by the emitters of a
long lifetime.
This is demonstrated in figures~\ref{fig3} and \ref{fig4}
for the FSI contribution
in the $\pi^0\pi^0$ correlation function.

Note that the change of the
character of the effect of non-equal times
at $v\approx 0.6$ and its increase with the
increasing velocity is not expected from condition
(\ref{10sta}).
The increase of the effect for relativistic particles
($v\rightarrow 1$) is specific
for the systems of not very large sizes and lifetimes
$\tau_0\sim r_0$, when the population of the light-cone
region ${\bf r}\sim {\bf v}t$ is not negligible.
Indeed, in this region
the arguments of the
Fresnel integrals at ${k}^*=0$ can be small even at
large $\gamma$:
$z_{\pm}\approx (\gamma m|r_L-t|/2)^{1/2}$,
leading to the modification of the spherical wave.

Consider finally the effect of non-equal emission
times on the correlations of two charged particles.
Since, at not very large $|t^*|$, the function
$\delta_{{k}^{*}}({\bf r}^*-{\bf r}')$
is close to the $\delta$-function, we can neglect
the terms of higher powers of
$({r}'/a)$ in (\ref{70}).\footnote
{
The account of these terms
is however important at large time separations to guarantee
vanishing of
the Coulomb interaction at $|t^*|\rightarrow \infty$.
}
The non-equal time correction is thus mainly
generated by the subleading term $\sim r^*/a$ and
so can be expected rather small,
similar to the case of strong FSI, where it arises from
a small finite-size contribution $\sim f/r^*$.
It concerns also the case of hadronic atoms since
the Schr\"odinger equation at a small negative energy
$-\epsilon_b=-\kappa^2/(2\mu)$ practically coincides
with that in
continuous spectrum at zero energy. As a result, for
${r}^{*}\ll \kappa^{-1}\doteq n|a|$
(n being the main atomic quantum number),
the ${r}^{*}$-dependence of the corresponding wave
functions
at a given orbital angular momentum is the same.

\section{Decay rate and normalization}
The decay rate (partial width) $\Gamma_n^\beta$
of a bound $\alpha$-channel state decay into the
$\beta$-channel
is given by the square of the wave function
$\psi_{n0}^\beta$
in (\ref{psicapp'}) (at a distance $r^*>d$),
multiplied by the product of the surface
$4\pi r^{*2}$ and the relative velocity
$v_\beta = k_\beta^*/\mu_\beta$:
\begin{equation}
\label{rate}
\Gamma_n^\beta=4\pi r^{*2}\frac{k_\beta^*}{\mu_\beta}
|\psi_{n0}^\beta|^2
= 4\pi\frac{k_\beta^*}{\mu_\alpha}|N'(n)|^2
\frac{(K^{\beta\alpha})^2}{1+(k_\beta^*K^{\beta\beta})^2}.
\end{equation}
{
Note that for two identical bosons in the channel $\beta$,
the twice as large square of the symmetrized wave function
is compensated by
twice as small surface so that the result is the same
as for two non-identical particles.
}
In the considered two-channel case, the
$\beta$-channel is the
only open one, so the decay rate coincides with the
inverse lifetime (total width) of the bound
$\alpha$-channel state
which can be calculated from the imaginary part of
the energy $E_n=-\kappa_n{}^2/(2\mu_\alpha)$:
\begin{equation}
\label{width}
1/\tau_n\equiv\Gamma_n=
-2\Im E_n\equiv 2\Re\kappa_n\Im\kappa_n/\mu_\alpha.
\end{equation}
Using (\ref{kappa2}), one has
(neglecting $\Im A^{\alpha\alpha}$ as compared with the
$\Re A^{\alpha\alpha}$ in the correction terms)
\begin{equation}
\label{reim}
\eqalign{
\Re\kappa_n=\kappa_n^c\left[1+
2\Re A^{\alpha\alpha}\kappa_n^c
+\Or\!\left(\left(2\Re A^{\alpha\alpha}\kappa_n^c
\right)^2\right)
\right]
\cr
\Im\kappa_n=2\Im A^{\alpha\alpha}(\kappa_n^c)^2
\left\{1+2[\phi(n)-1]\Re A^{\alpha\alpha}\kappa_n^c
-4\pi^2\Or\!\left(\left({\Re A^{\alpha\alpha}}/{a}
\right)^2\right)
\right\}
}
\end{equation}
\begin{equation}
\label{width'}
\Gamma_n=\frac{4}{\mu_\alpha}
\left\{1+2\phi(n)\Re A^{\alpha\alpha}\kappa_n^c
-4\pi^2\Or\!\left(\left({\Re A^{\alpha\alpha}}/{a}
\right)^2\right)
\right\}
(\kappa_n^c)^3
\Im A^{\alpha\alpha}.
\end{equation}
Using the relation
$(\kappa_n^c)^3=\pi|\psi_{n0}^{\rm coul}|^2$
and (\ref{Aaa1}) for $\Im A^{\alpha\alpha}$,
one finally gets,
in agreement with \cite{ras82,moo95}:
\begin{equation}
\label{width''}
\eqalign{
\Gamma_n=
4\pi\frac{k_\beta^*}{\mu_\alpha}|\psi_{n0}^{\rm coul}|^2
\left\{1+2\phi(n)
\frac{\Re A^{\alpha\alpha}}{n|a|}
\right.
\cr
\qquad
\left.
-4\pi^2\Or\!\left(\left(\frac{\Re A^{\alpha\alpha}}{a}
\right)^2\right)
\right\}
\frac{(K^{\beta\alpha})^2}{1+(k_\beta^*K^{\beta\beta})^2}.
}
\end{equation}
Inserting (\ref{rate}) and (\ref{width''}) into
the equality $\Gamma_n=\Gamma_n^\beta$,
one proves the relation (\ref{normapp1}) between the
normalization
factors ${\cal N}'(n)$ and $\psi_{n0}^{\rm coul}(0)$.

In the case of two or more open decay channels, the
two-channel $(\alpha,\beta)$ matrix $\hat{K}$
is no more real, particularly,
in the presence of one additional channel $j$, one has:
\begin{equation}
\label{Kaa}
\eqalign{
K^{\alpha\alpha}={\cal K}^{\alpha\alpha}+
\frac{{\rm i}k_j^* ({\cal K}^{j\alpha})^2}
{1-{\rm i}k_j^* {\cal K}^{jj}}\qquad
K^{\beta\alpha}={\cal K}^{\beta\alpha}+
\frac{{\rm i}k_j^* {\cal K}^{j\beta}{\cal K}^{j\alpha}}
{1-{\rm i}k_j^* {\cal K}^{jj}}
\cr
K^{\beta\beta}={\cal K}^{\beta\beta}+
\frac{{\rm i}k_j^* ({\cal K}^{j\beta})^2}
{1-{\rm i}k_j^* {\cal K}^{jj}},
}
\end{equation}
where ${\cal K}^{\lambda\lambda'}$ are the elements of a real
three-channel matrix ${\cal K}$.
{
Note that in the case of a two-pion system
($\alpha=\pi^+\pi^-$, $\beta=\pi^0\pi^0$),
the third channel is $j=\gamma\gamma$ so that the elements
${\cal K}^{j\beta}$ and ${\cal K}^{jj}$ can be safely
neglected. Then, only the element $K^{\alpha\alpha}$
acquires the imaginary part:
$K^{\alpha\alpha}={\cal K}^{\alpha\alpha}+
ik_j^* ({\cal K}^{j\alpha})^2$.
}

Generally, one has to account for the possible imaginary parts
of the elements of the two-channel $K$-matrix as well as,
for a possibility of a pure imaginary value of the momentum
$k_\beta^*$ in the case of a closed channel $\beta$
($k_\beta^{*2}<0$, $k_\beta^*=i (-k_\beta^{*2})^{1/2}$,
$\Gamma_n^\beta=0$).
Then
\begin{equation}
\label{Aaa2a}
\eqalign{
\Re A^{\alpha\alpha}=
\Re K^{\alpha\alpha}
-
\left\{\Im k_\beta^*\Re (K^{\beta\alpha})^2+
\Re k_\beta^*\Im (K^{\beta\alpha})^2
\right.
\cr
\qquad \qquad
\left.
+
\left|k_\beta^*\right|^2
\left[
\Re K^{\beta\beta}\Re (K^{\beta\alpha})^2+
\Im K^{\beta\beta}\Im (K^{\beta\alpha})^2
\right]\right\}
\left|1-{\rm i}k_\beta^* K^{\beta\beta}\right|^{-2}
}
\end{equation}
\begin{equation}
\label{Aaa2}
\eqalign{
\Im A^{\alpha\alpha}=
\Im K^{\alpha\alpha}
-
\left\{\Im k_\beta^*\Im (K^{\beta\alpha})^2-
\Re k_\beta^*\Re (K^{\beta\alpha})^2
\right.
\cr
\qquad \qquad
\left.
+
\left|k_\beta^*\right|^2
\left[
\Re K^{\beta\beta}\Im (K^{\beta\alpha})^2+
\Im K^{\beta\beta}\Re (K^{\beta\alpha})^2
\right]\right\}
\left|1-{\rm i}k_\beta^* K^{\beta\beta}\right|^{-2}
\cr
\qquad
 = k_\beta^*\left|\frac{K^{\beta\alpha}}
 {1-{\rm i}k_\beta^* K^{\beta\beta}}
\right|^2 \theta\!\left(k_\beta^{*2}\right)+
\sum_j k_j\left|\frac{D f_c^{j\alpha}}
 {1-{\rm i}k_\beta^* K^{\beta\beta}}
\right|^2
\cr
\qquad
=
\frac{\mu_\alpha}{4\pi|N'(n)|^2}
\left[\Gamma_n^\beta+\sum_j \Gamma_n^j
\right],
}
\end{equation}
where
$\theta(x)=1$ for $x\ge 0$, $\theta(x)=0$ for $x < 0$.
{
The second expression for $\Im A^{\alpha\alpha}$
in (\ref{Aaa2}) follows from a straightforward
though lengthy matrix algebra and, the last one -
from an obvious generalization of (\ref{rate}) using the
relation $K^{\beta\alpha}=D f_c^{\beta\alpha}$.
}
Inserting the last equality in (\ref{Aaa2}) into
(\ref{width'}), one proves (\ref{normapp1})
for the case of any number of open decay channels.

\noappendix

\pagebreak

\end{document}